\documentclass[12pt]{article}
\usepackage[greek,english]{babel}
\usepackage[LGR,T1]{fontenc}
\usepackage[latin1]{inputenc}
\usepackage{amssymb}
\usepackage{amsmath}
\usepackage{graphicx,xcolor}
\usepackage{geometry}
\usepackage{epsfig}
\usepackage{multicol}
\usepackage{array}
\usepackage{subfigure}
\usepackage{pdfpages}
\usepackage{lscape}
\usepackage{url}
\usepackage[lined,ruled]{algorithm2e}
\usepackage[nottoc, notlof, notlot]{tocbibind}
\usepackage{color}
\usepackage{rotating}
\usepackage[sectionbib,square]{natbib}

\usepackage {setspace}

\title{The Leviathan model: Absolute dominance, generalised distrust, small worlds and other patterns emerging from combining vanity with opinion propagation}
\author{Guillaume Deffuant, Timoteo Carletti, Sylvie Huet}

\begin{document}
\bibliographystyle{plainnat}
\maketitle

\abstract{We propose an opinion dynamics model that combines processes of
  vanity and opinion propagation. The interactions take place between randomly
  chosen pairs. During an interaction, the agents propagate their opinions
  about themselves and about other people they know. Moreover, each individual
  is subject to vanity: if her interlocutor seems to value her highly, then
  she increases her opinion about this interlocutor. On the contrary she tends
  to decrease her opinion about those who seem to undervalue her.  
The combination of these dynamics with the hypothesis that the opinion
propagation is more efficient when coming from highly valued individuals,
leads to different patterns when varying the parameters. For instance, for some parameters the positive opinion links between individuals generate a small world network. In one of the patterns, absolute dominance of one agent
alternates with a state of generalised distrust, where all agents have a very
low opinion of all the others (including themselves). We provide some
explanations of the mechanisms behind these emergent behaviors and finally
propose a discussion about their interest.  
}

\section{Introduction}

\textit{"Again, men have no pleasure (but on the contrary a great deal of
  grief) in keeping company where there is no power able to overawe them
  all. For every man looketh that his companion should value him at the same
  rate he sets upon himself, and upon all signs of contempt or undervaluing
  naturally endeavours, as far as he dares (...), to extort a greater value from his contemners, by damage; and from
  others, by the example."} T. Hobbes, Leviathan, chapter 13. \\
   
In this paper, our goal is to revisit Hobbes thesis using opinion dynamics
models and computer simulations. We propose a simple model including a vanity
process, inspired from the above citation, in which agents measure themselves
in the eyes of the others and retaliate against this judgement. Then, we observe the
collective patterns that emerge for different values of the parameters. 

Our approach is in the line of many others in the field of social simulation
or in sociophysics. It consists 
in making a few simple assumptions about the rules of interactions between
agents and then studying the obtained emerging behaviors. In some of such
models, the agents have binary opinions \citep{GalamReview2008,
  Sznajd-Weron2005}, while 
in others the opinions are continuous \citep{deffuant2000mixing,Deffuant_2006,fortunato2004universality,huet2008rejection,PUB00029104,Urbig2008,lorenz2007continuous,Gargiulo2010EPL} (see \citet{castellano2009statistical}
for a review). Our model is closer to recent ones which include a set of
affinities between agents, leading to emerging networks
\citep{PhysRevE.76.066105,ACS.14.1}.     

In the proposed model we assume that each agent can have a continuous opinion about
every other agent and we truncate it to remain between -1 and +1. In the initial state, we suppose
 that the agents don't have any opinion about the others. The agents
interact by randomly chosen pairs and two different processes apply. The first
one supposes that during any interaction, each agent propagates
 her opinions about herself, about her interlocutor and about several
randomly chosen other agents that she knows. In this propagation, highly valued agents are more influential. The second process represents a vanity effect: an agent likes to be highly valued by the others, thus she increases her opinion on those who value her well. On the contrary she decreases her opinion about those who
undervalue on her. These assumptions are inspired by Hobbes, but also by more recent experiments and observations from social-psychologists \citep{FeinSpencer1997,Leary01052006,Buckley200414,Srivastava2005,StephanMaiano2007,WoodForest2011}. Moreover, we suppose that the access to the opinion of the
others is not perfect: people may not express exactly what they think and the
listener may misinterpret these expressions. To take this into account in the model, the
propagated opinions are distorted by some noise. 

Despite its simpliciy, the model shows a surprising variety of patterns when changing the parameters. We identified the following five major ones: 
\begin{itemize}
	\item Equality. Each
          agent has a positive opinion about herself, she is connected by strong positive mutual opinions
          with a small set of agents and has very negative opinions about all the others. All agents have a similar number of positive (and negative) links. For some parameters network of positive links shows the characterisitics of small world networks. 
    \item Elite. The pattern shows two categories of agents: the elite and second category agents. The elite agents have a positive self-opinion and are strongly supported by a friend, but they have a very negative opinion of all the other elite agents and of all the second category agents. The second category agents have a very negative self opinion, they have a very negative opinion of all the other second category agents and their opinion about the elite agents is moderate.
		\item Hierarchy. All agents share a similar opinion about every other agent (called reputation) and the reputations are widely spread between $-1$ and $+1$. There are more agents of low reputation than of high reputation: this gives the image of a classical hierarchy with a wide basis and progressively shrinking when going up to the top.      		
		\item Dominance. As in the hierarchy pattern, all agents share a similar opinion about every other agent,  but a single agent has a high reputation while most of the other agents have a very low reputation.
				\item Crisis. Each agent has a very negative opinion of all the others and of herself.
\end{itemize}

Moreover, we observe also mixtures of these patterns in a same run, in particular crisis and dominance patterns alternate in some simulations. These mixtures of patterns add significantly to the richness of the model. We believe that this richness cannot be explored in details in a single paper. Therefore, our goal is rather to draw a global view of the main features of this model and to derive the main theoretical explanations of some of them. Doing so, we identify several issues that would require more detailed studies.
 
The paper is organised as follows. We firstly describe in details the dynamics of the model. Then we describe the
main patterns obtained and some maps of their presence in the parameter space. We then propose
some theoretical or qualitative explanations of these patterns. Finally, we propose a discussion
about the potential use of the model in philosophical or sociological debates. 

\section{The dynamics of the model}
This section describes the model in details and particularly its evolution rules representing opinion propagation and vanity. 

We consider a set of $N$ agents, each agent $i$ is characterised by
her list of opinions about the other agents and about herself:
$(a_{i,j})_{1 \leq i,j \leq N}$. 

We assume $a_{i,j}$ lie between $-1$ and $+1$, or it is equal to {\em nil} if the
agent $i$ never met $j$ and nobody talked to $i$ about $j$ yet.
At the initialisation, we suppose that the agents never met, therefore all their opinions are set to {\em nil} .
 
The individuals interact by uniformly randomly drawn pairs $(i, j)$ and at
each encounter, we apply both processes: opinion propagation and vanity.

Let us remark that we always keep the opinions between $-1$ and $+1$, by
truncating them to $-1$ if their value is below $-1$ after the interaction, or to $+1$ if their value is above $+1$. Moreover, in the
following, we consider that one iteration, i.e. one time step $t\rightarrow
t+1$, is $N/2$ random pair interactions (each individual interacts $N$ times on average during one iteration).

\subsection{Opinion propagation}

Let us assume that agents $i$ and $j$ have been drawn. During an encounter, we
suppose that agent $j$ propagates to $i$ her opinions about herself
($j$), about $i$, and about $k$ agents randomly chosen among her
acquaintances. Moreover, we suppose that if $i$ has a high opinion of $j$,
then $j$ is more influential.   

This hypothesis is implemented by introducing a propagation coefficient,
denoted $p_{i,j}$, which is based on the difference between the opinion of $i$ about $j$ ($a_{i,j}$) and the opinion $i$ about herself ($a_{i,i}$). It uses the logistic function with parameter $\sigma$. If $a_{i,j}=nil$ ($j$ is unknown to $i$), we assume that $i$ has a neutral opinion about $j$ and we set $a_{i,j}\leftarrow 0$. Let us also observe
that, at the initialisation an agent has no opinion about
herself ($a_{i,i}=nil$), before she takes part of a first encounter, thus
we also set $a_{i,i}\leftarrow 0$. 
Then we compute the propagation coefficient $p_{i,j}$, which rules the intensity of the opinion propagation from $j$ to $i$:
\begin{equation}
 p_{i,j} =  \frac{1}{1 + exp\left(-\frac{a_{i,j} - a_{i,i}}{\sigma}\right)}\, . 
 \end{equation} 

\begin{figure}
 	\centering
 		\includegraphics[width=8 cm]{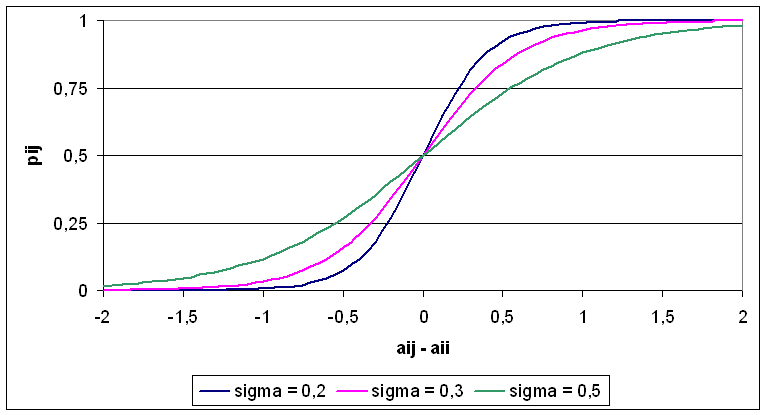} 
 	\caption{Examples of variations of the propagation coefficient $p_{i,j}$ when $a_{i,j} - a_{i,i}$ varies, and for three values of parameter $\sigma$.}
 	\label{fig:propaCoef}
 \end{figure} 	
 
Figure \ref{fig:propaCoef} represents the value of $p_{i,j}$ when the difference $a_{i,j} - a_{i,i}$ varies (between -2 and +2), for three different values of parameter $\sigma$. One can observe that $p_{i,j}$ tends to $1$ when $a_{i,j} - a_{i,i}$ is
close to 2 ($i$ values $j$ higher than herself), and tends to $0$ when it is
close to -2 ($i$ values $j$ lower than herself). The influence of $j$ on $i$
is then expressed as follows ($\rho$ is a parameter of the model ruling the
importance of opinion propagation): 
\begin{equation}
a_{i,i} \leftarrow  a_{i,i} + \rho p_{i,j} ( a_{j,i}- a_{i,i} +
\texttt{Random}(-\delta,\delta) )\, ,
\end{equation}
and
\begin{equation}
a_{i,j} \leftarrow  a_{i,j} + \rho p_{i,j} ( a_{j,j}- a_{i,j} +
\texttt{Random}(-\delta,\delta) )\, .
\end{equation}

Where we denoted by \texttt{Random}$(-\delta, \delta)$ a
  uniformly distributed random number between $- \delta$ and $+\delta$, that
  can be seen as a noise that distorts the perception that $i$ has about
$j$'s opinions. The parameter $\delta$ rules the amplitude of this noise. 

Moreover, $j$ propagates her opinion about (at most) $k$ of her
acquaintances. More precisely, 
let $n_j$ be the number of acquaintances of $j$ different from $i$ (number of individuals $q$
such that $a_{j,q}$ is not {\em nil} and $j \neq i$)\footnote{We also tested without checking that the acquaintance is different from $i$, and did not notice any major change in the emergent patterns of the model.}. We choose at random with reinsertion
$\min(k, n_j)$ agents among $j$'s acquaintances 
(i.e. an acquaintance of $j$ can be selected several times, while other
are not selected).

The propagation to $i$ of $j$'s opinion about $q$ is expressed thus by:
\begin{equation}
a_{i,q} \leftarrow  a_{i,q} + \rho p_{i,j} ( a_{j,q}- a_{i,q} +
\texttt{Random}(-\delta,\delta) )\, .
\end{equation}
And this will be repeated for every $q$ selected at random.

In the interaction, we apply the influence of $j$ on $i$ and then the reciprocal one of $i$ on $j$. Indeed, we suppose that the interactions take place in both ways (from $i$ to $j$ and from $j$ to $i$), the order being randomly chosen.

\subsection{Vanity dynamics}

The second ingredient of the model is the dynamics representing agent's vanity. This
dynamics expresses that agents tend to reward the agents that value them more
positively than they value themselves and to punish the ones that value them
more negatively than they value themselves. 

We assume that the modification of $i$'s opinion about $j$ is
simply linear with the difference between the
opinion of $i$ about herself and the opinion of $j$ about $i$ (slightly
modified randomly): 
\begin{equation}
a_{i,j} \leftarrow a_{i,j} + \omega ( a_{j,i} - a_{i,i} +
\texttt{Random}(-\delta,\delta))\, .
\end{equation}
If $i$ has a lower (resp. higher) self opinion than her perception of the opinion
$j$ has about her ($i$), then $a_{i,j}$ is increased
(resp. decreased). Parameter $\omega$
rules the importance of the vanity dynamics. 
We also assume that the intensity of the vanity effect is independent from the opinions. Indeed, 
it is possible that one forgives more easily a disappointment coming from a
loved person because precisely of this affection. But on the other hand, it is
well known that hatred can also easily come from disappointed love. Since
there is no clear intuition in one way or the other, we made the simplest
assumption that the vanity effect is linear with the difference of opinions. 

\subsection{Summary}
\label{summary}
The Algorithm~\ref{Algo:Iteration} describes one iteration:  $N/2$ random pairs of individuals are drawn, with reinsertion, and we
suppose that each individual 
influences the other during the encounter. Algorithm~\ref{Algo:Interaction}
codes the interaction with the
two aspects of the dynamics: opinion propagation and vanity. 
\begin{algorithm}
\SetKwFunction{Random}{Random}
\SetKwFunction{Interaction}{Interaction}
\BlankLine
\For{$N/2$ times}{
	$i \leftarrow$ \Random($1, N$) \;
	$j \leftarrow$ \Random($1,N,\neq i$) \;
	\Interaction($i,j$) \;
	\Interaction($j,i$) \;
}
\label{Algo:Iteration}
\caption{Iteration}
\end{algorithm}

\begin{algorithm}
\SetKwFunction{Random}{Random}
\SetKwFunction{KnownAgents}{KnownAgents}
\BlankLine
\BlankLine
\lIf {$a_{i,i}$ = nil} {$a_{i,i} \leftarrow$ 0} \;
\lIf {$a_{i,j}$ = nil} {$a_{i,j} \leftarrow$ 0} \;
\lIf {$a_{j,i}$ = nil} {$a_{j,i} \leftarrow$ 0} \;
\lIf {$a_{j,j}$ = nil} {$a_{j,j} \leftarrow$ 0} \;
		$p_{i,j}$ $\leftarrow $ $\frac{1}{1 + exp\left(-\frac{a_{i,j} - a_{i,i}}{\sigma}\right)}$ \tcp*[r]{Computing propagation coefficient} \;
		$a_{i,i}$ $\leftarrow$ $ a_{i,i} + \rho p_{i,j} ( a_{j,i}- a_{i,i} + $ \Random(-$\delta$,$\delta$) ) \tcp*[r]{$j$  propagates $a_{j,i}$} \;
		$a_{i,j} \leftarrow a_{i,j} + \rho p_{i,j} ( a_{j,j}- a_{i,j} + $ \Random(-$\delta$,$\delta$) ) \tcp*[r]{$j$ propagates $a_{j,j}$} \;
		\For(\tcp*[r]{$n_j$ number of individuals known by $j$}){$\min(k, n_j)$ times }{
			$q \leftarrow \Random(\KnownAgents(j), \neq i, \neq j)$  \tcp*[r]{$q$ random known by  $j$} \;
			\lIf {$a_{i,q}$ = nil} {$a_{i,q} \leftarrow$ 0} \;
			$a_{i,q} \leftarrow a_{i,q} + \rho p_{i,j}( a_{j,q}- a_{i,q} + $ \Random(-$\delta$,$\delta$) ) \tcp*[r]{$j$ propagates $a_{j,p}$} \;
		} 	
		$a_{i,j} \leftarrow a_{i,j} + \omega ( a_{j,i} - a_{i,i} +$ \Random($-\delta$, $\delta$)  ) \tcp*[r]{	Vanity dynamics, $i$ modifies $a_{i,j}$} \;
\label{Algo:Interaction}
\caption{Interaction($j$,$i$)}
\end{algorithm}

Summarizing the parameters of the model are thus:
\begin{itemize}
	\item $N$, number of agents; 
	\item $\rho$, ruling the intensity of the opinion influence;
	\item $\omega$, ruling the intensity of the vanity;
	\item $k$, number of acquaintances about which the pair of agents
          discuss in the opinion influence; 
	\item $\delta$, intensity of noise perturbing the evaluation of
          other's opinions; 
	\item $\sigma$, ruling the slope of the logistic function determining
          the propagation coefficients; 
\end{itemize}

\subsection{Representations of the population state}
\label{representations}
We use two different representations of the population state:
\begin{itemize}
\item Matrix representation: the opinion list of each agent
  is represented as the row of a $N\times N$ square matrix. The element $a_{i,j}$ from line $i$ and column $j$ is the opinion of agent $i$ about agent $j$. We use colours to code for the opinions: blue for negative and red for positive opinions with light colours meaning that the absolute value is close to 0. This
  representation provides all the information about the state of the
  population at a given time step, but it is sometimes difficult
  to interpret. 
	\item Network representation: we represent the agents as
          nodes of a network, in a 2D space and we draw links
          between two agents, only when one of the agents has a positive opinion about the other. The position in the 2D space of the
          nodes representing the agents is obtained by a dynamical
          algorithm, 
          which at each interaction moves the nodes in order to get
          a distribution where the distances between
          the nodes are as close as possible to the values of a
            simple function depending on the sum of the relative agents'
            opinions. The function yields a large value when two agents 
          have negative opinions about each other, and
          reciprocally, a small value when their opinions about each other are positive. When the sum of the opinions is close to 0, the function yields a medium value. The colour of the links is yellow when the link is close to 0 and it get close to red when the value of the link is close to 1. This
          representation does not show all the information about the state, because we do not represent the negative links nor the assymmetry of the links.
          However, some features of the emergent structures appear more easily with this representation. 
\end{itemize}  
In most cases, we propose both representations for the same state.

\section{The observed patterns and their sensitivity to parameters}

This section describes the five main prototypical patterns that we use to describe the behaviour of the model. Some patterns of the model are mixtures of these main ones; they are described in a following section.

\subsection{The main patterns}  

\subsubsection{Equality}  

\begin{figure}
 	\centering
 	\begin{tabular}{ccc}
 		\includegraphics[width=5 cm]{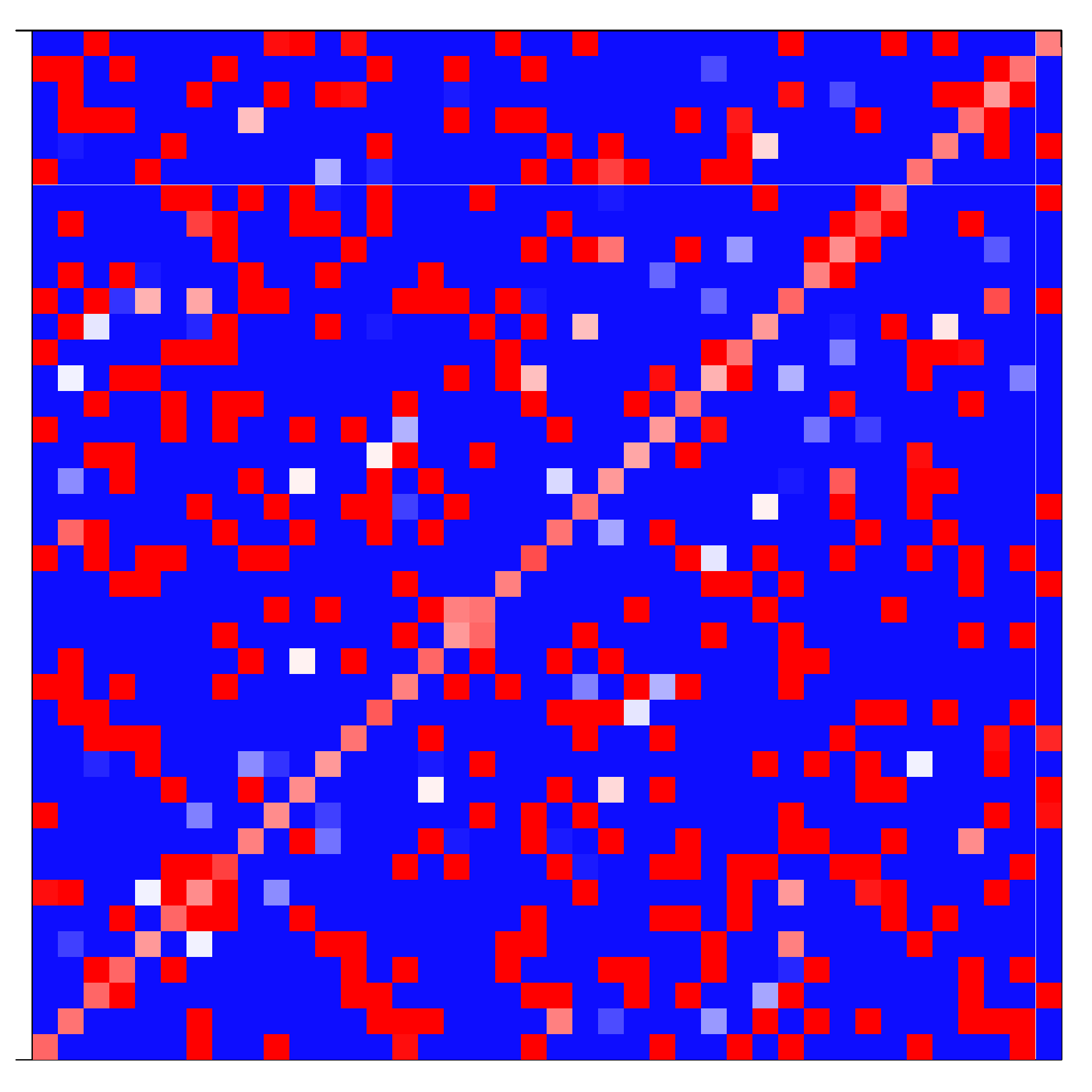} & 	\includegraphics[width= 5 cm]{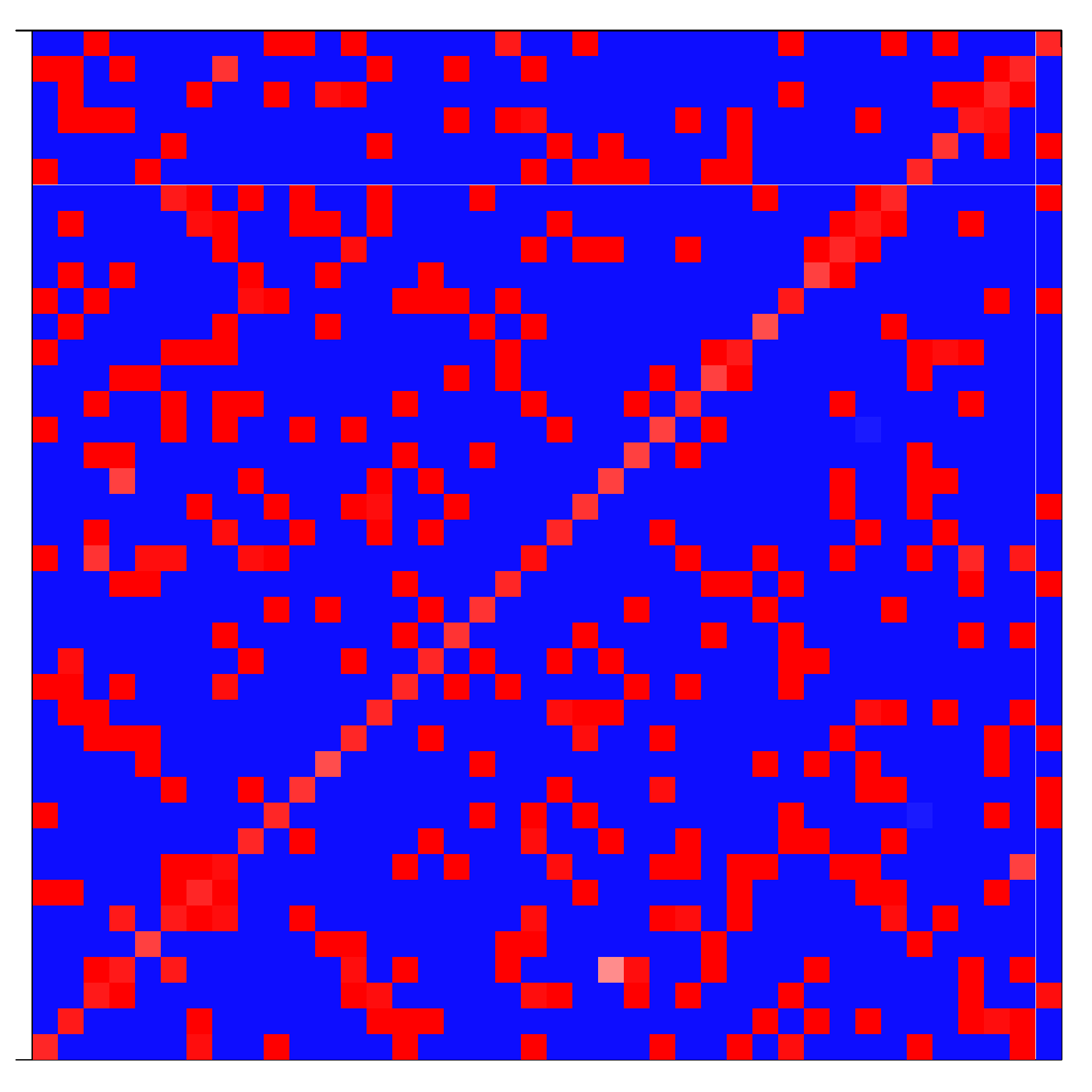} & \includegraphics[width= 5 cm]{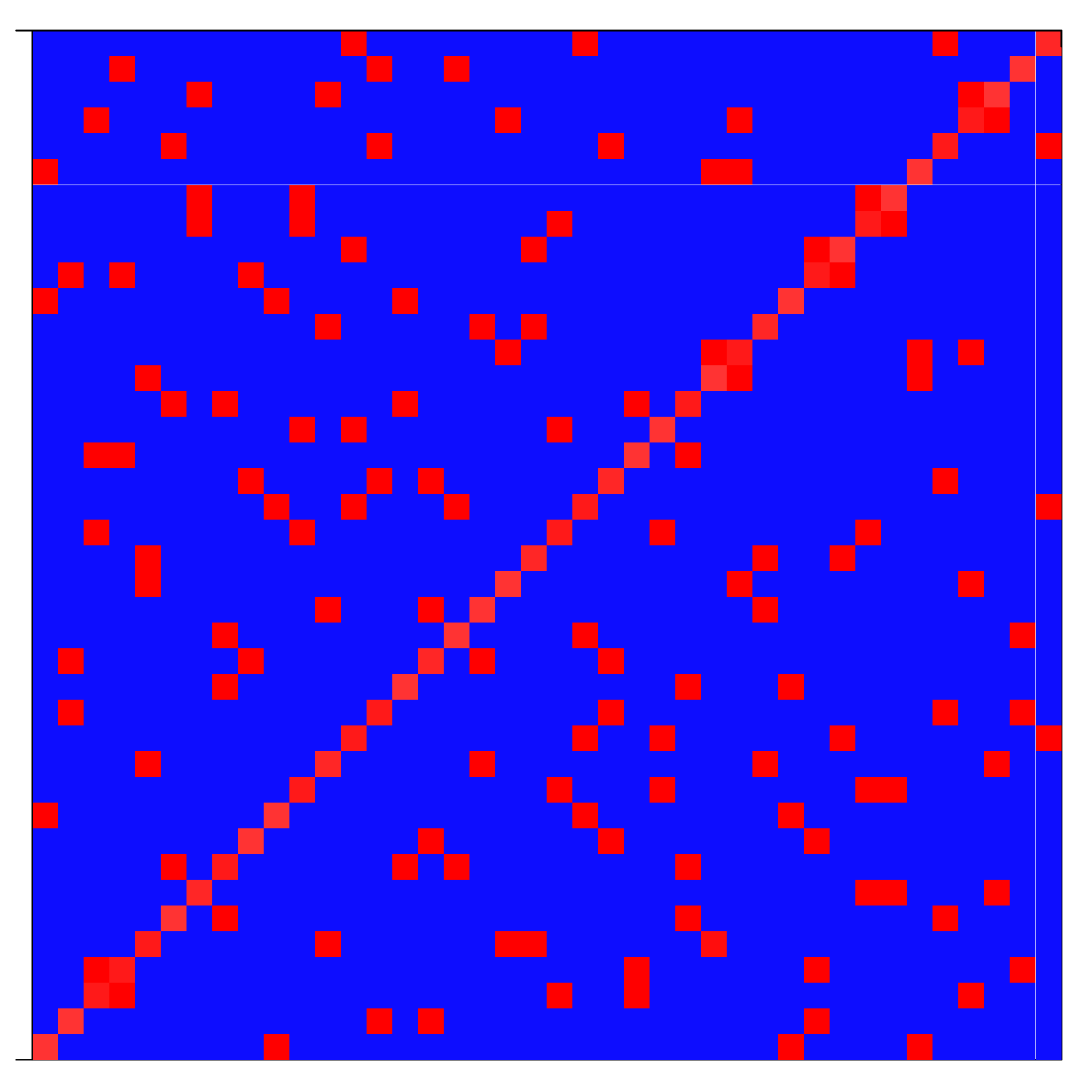}\\
 	 	\includegraphics[width= 5 cm]{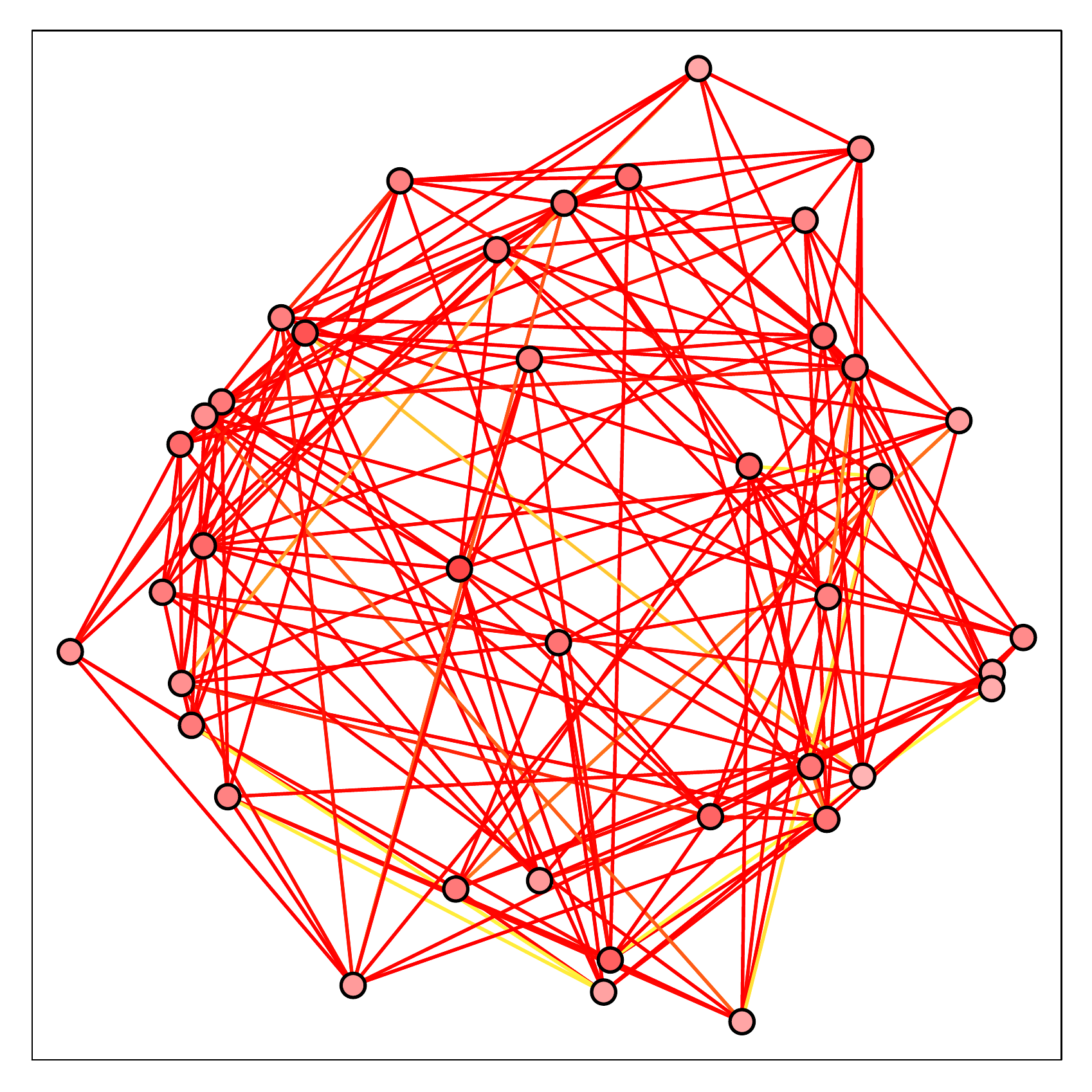} & \includegraphics[width= 5 cm]{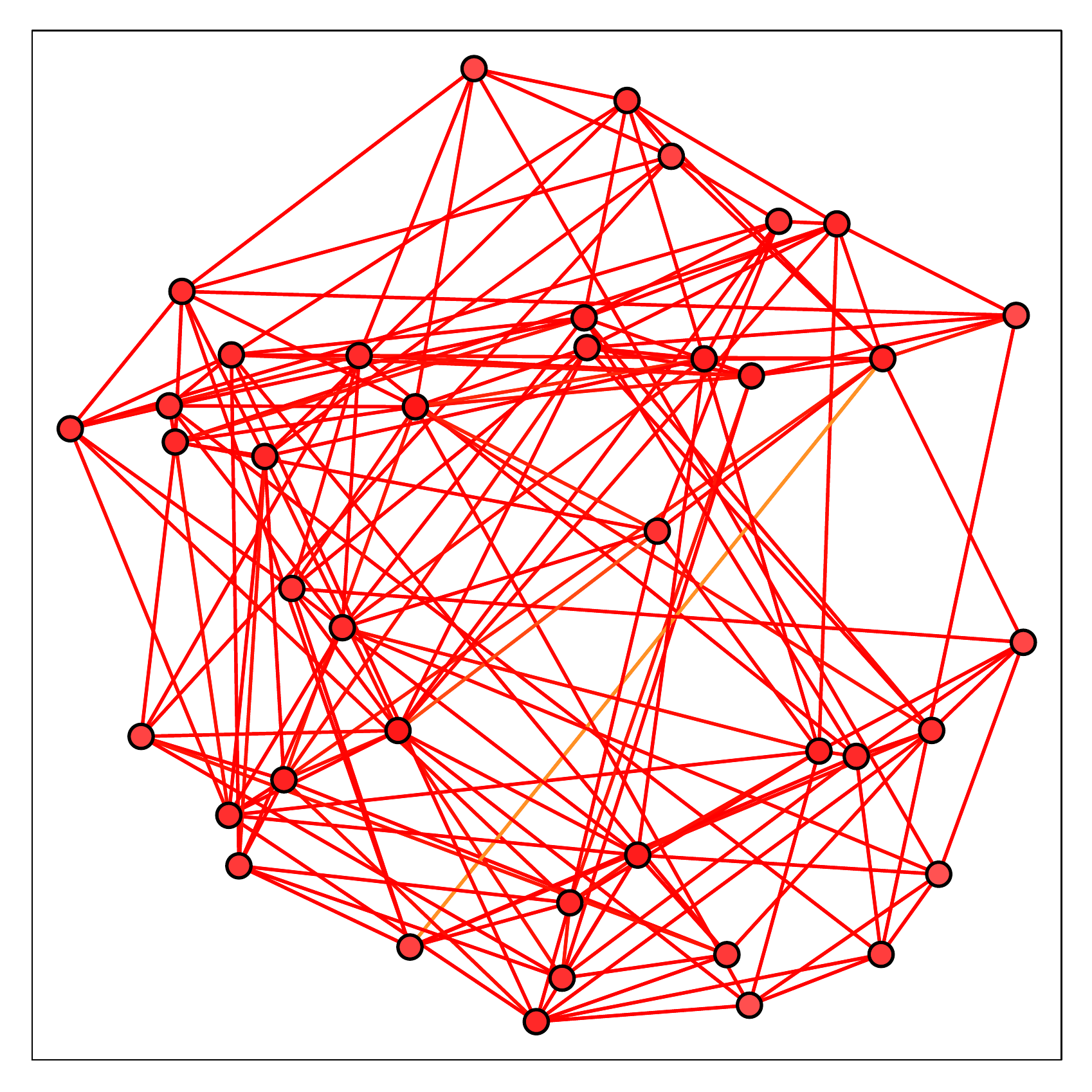}  & \includegraphics[width= 5 cm]{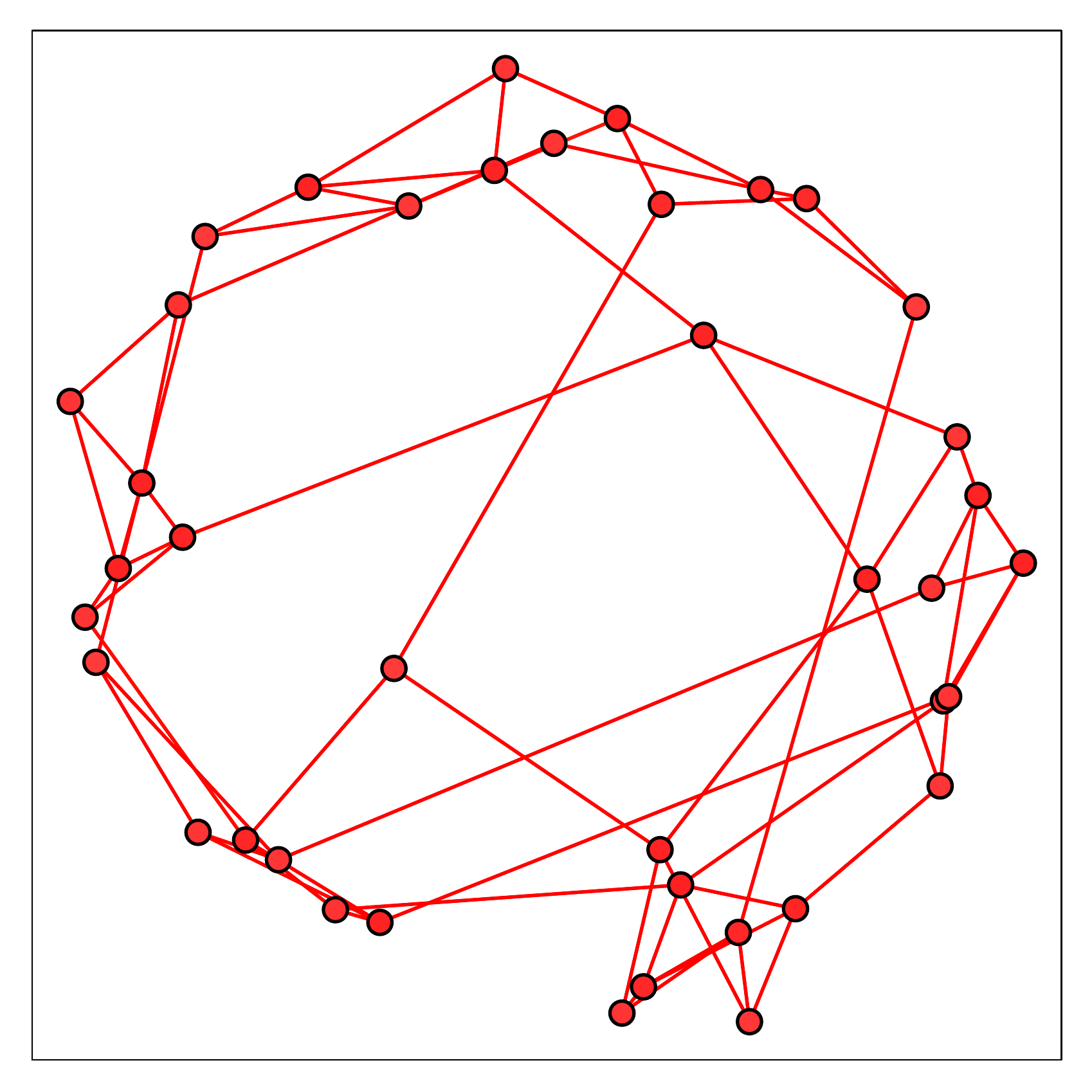} \\
 	 	$t$ = 500 & $t$ = 1000 & $t$ = 50000\\
 	\end{tabular}
 	\caption{Example of equality pattern. Parameters are: $N = 40$$\omega = 0.3$, $\rho
          = 0.01$, $\sigma = 0.35$, $k = 5$, $\delta = 0.2$. The final average degree is $3.4$, the clustering coefficient is $0.25$, whereas in a random network with the same number of nodes and average degree, the clustering coefficient is $0.08$ (average over 10 replicas). The mean shortest path is $3.05$ which is smaller than the mean shortest path of the equivalent random network ($5.02$ average on 10 replicas). See
          Section~\ref{representations} for general explanation about the 
          representations.} 
 	\label{fig:smallWorld}
 \end{figure}

Figure~\ref{fig:smallWorld} shows, using both
matrix and network representations, three states of the model for $N = 40$, $\omega = 0.3$, $\rho = 0.01$, $\delta = 0.2$, $\sigma = 0.35$, $k = 5$. One can observe two phases before reaching a stable state:
\begin{itemize}
	\item In a few hundred iterations, the model reaches a state in which the opinion matrix is symmetric, with only extreme
values ($-1$ or $+1$). Each agent separates the population between two groups:
the ones she hates and the ones she loves. The first group is larger than
the second (more blue squares in the matrix representation at $t = 1000$). Another
feature is striking: the opinions on the diagonal are significantly positive:
the agents have a good opinion of themselves (though not as good as their
opinion about their loved ones).  
\item During a few thousand iterations, many positive
connections are progressively destroyed before reaching
a stable state (we tested this stability during several hundred thousand iterations). Looking at the network representation of the stationary state (iteration
$50000$), we observe the presence of small connected sets made by $3$ or $4$ individuals. The sets
tend to be located far from each other (on a circle), indicating that each
connected set has a negative opinion about the others (as shown in the matrix representation). 
\end{itemize}
When doing other runs with the same parameters, we always get a final stable state with a similar number of friends in the networks. This pattern is typically obtained when the vanity is dominating over the opinion propagation (we describe the positions of the patterns in the parameter space later). For some parameters, the network of friends shows the properties of small world networks (this is the case of the one shown on Figure \ref{fig:smallWorld}). We call this pattern "`equality"' because all the agents are in a similar situation.

\subsubsection{Elite}  
\begin{figure}
 	\centering
 	\begin{tabular}{ccc}
 		\includegraphics[width=5 cm]{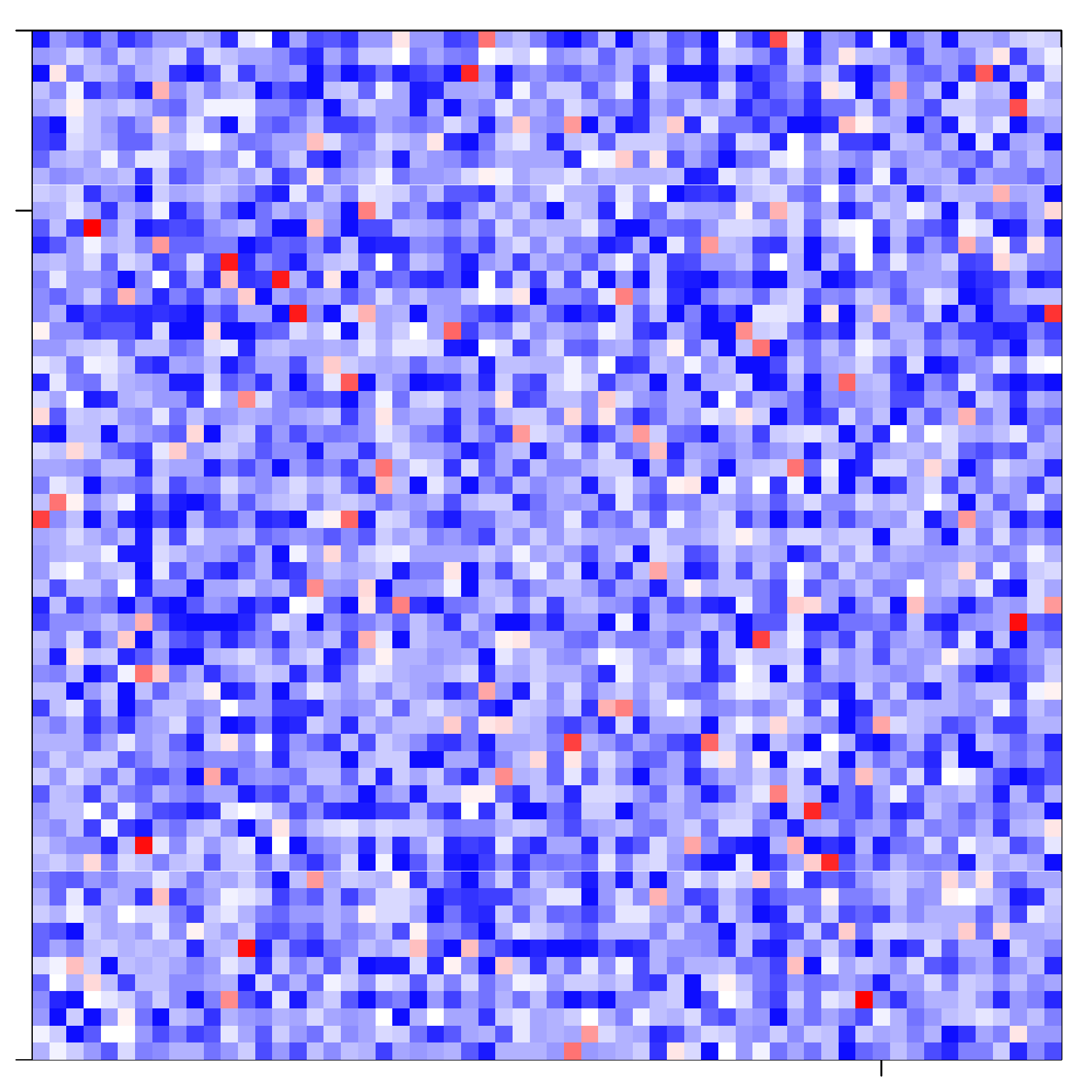} & 	\includegraphics[width= 5 cm]{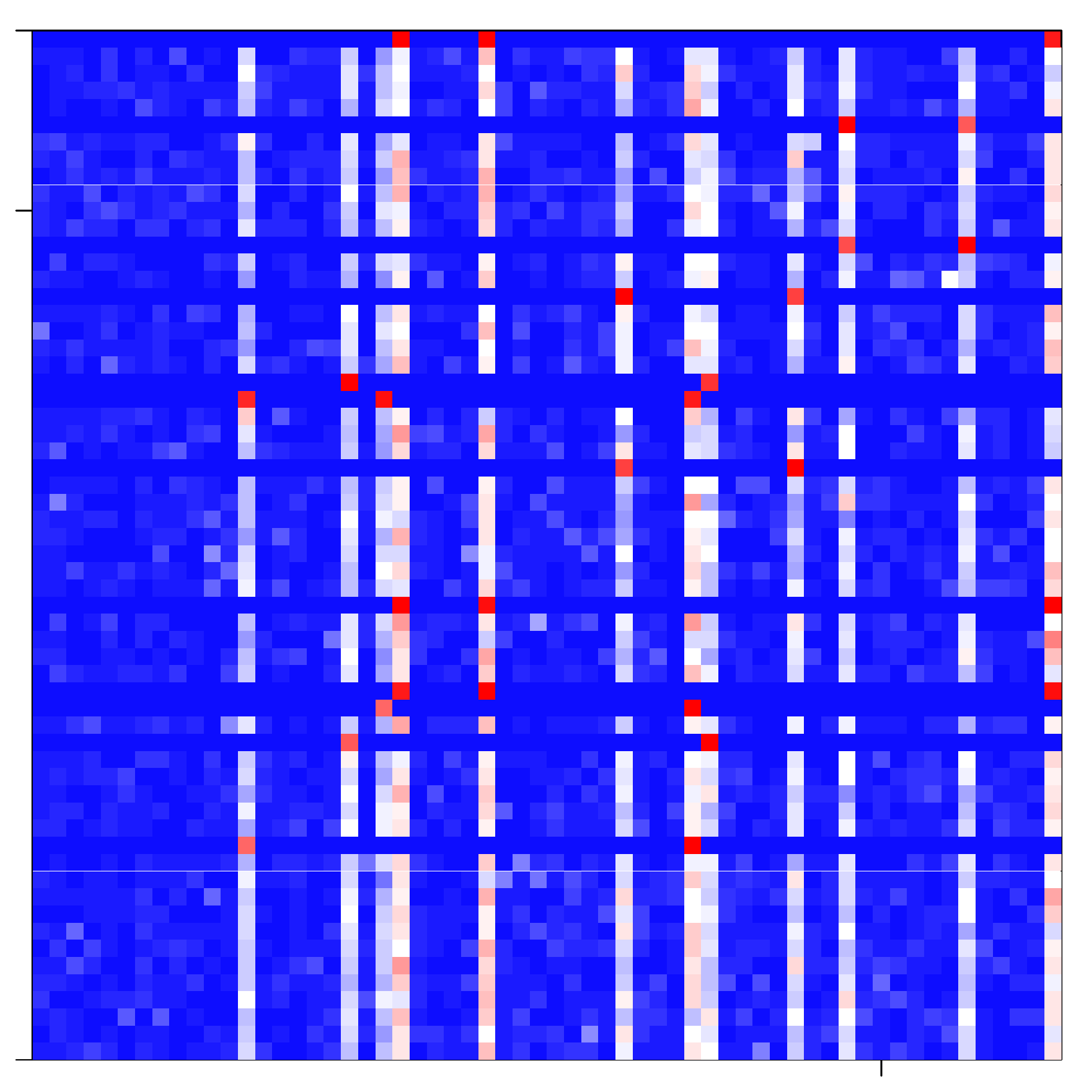}& 	\includegraphics[width= 5 cm]{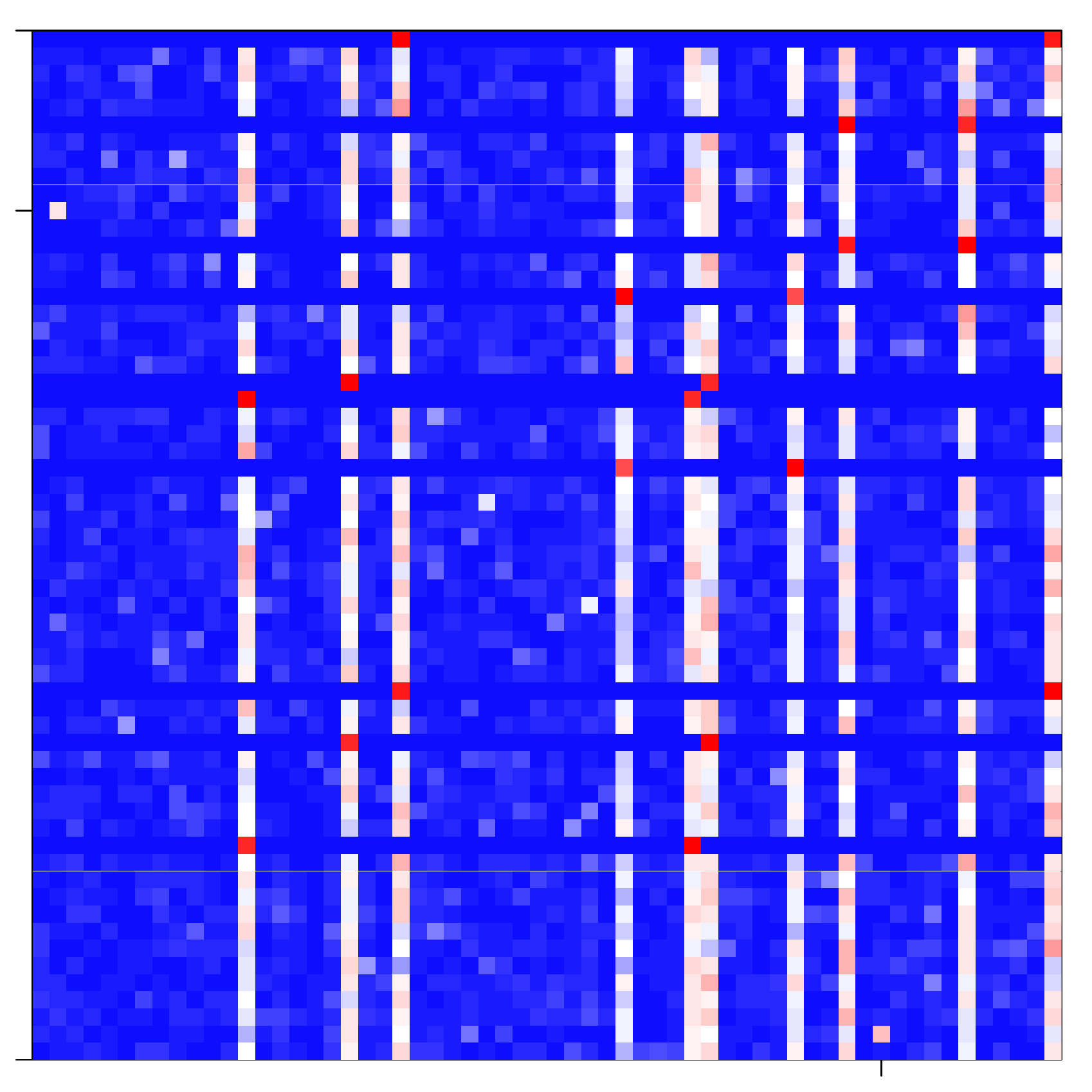} \\
 			\includegraphics[width=5 cm]{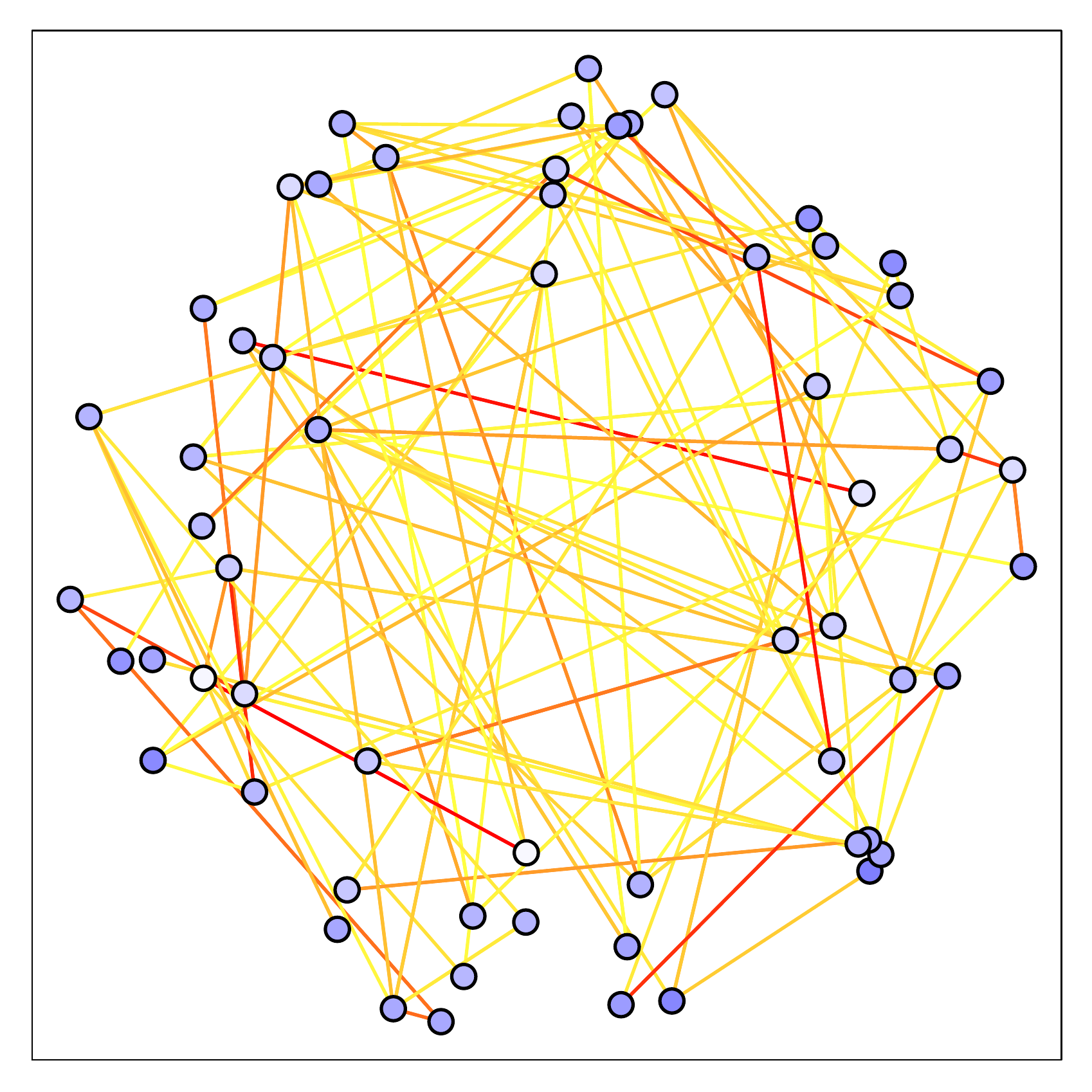} & 	\includegraphics[width= 5 cm]{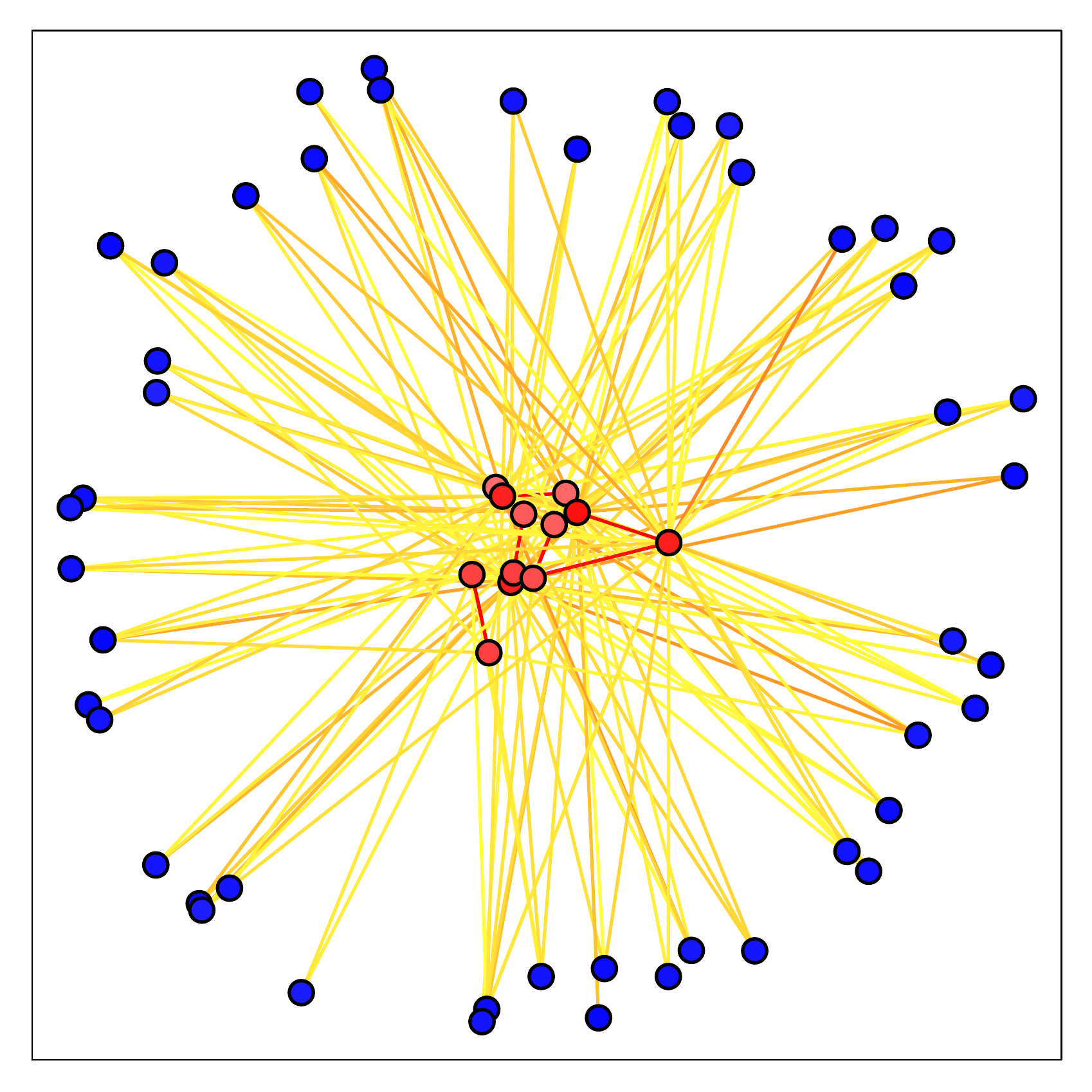}& 	\includegraphics[width= 5 cm]{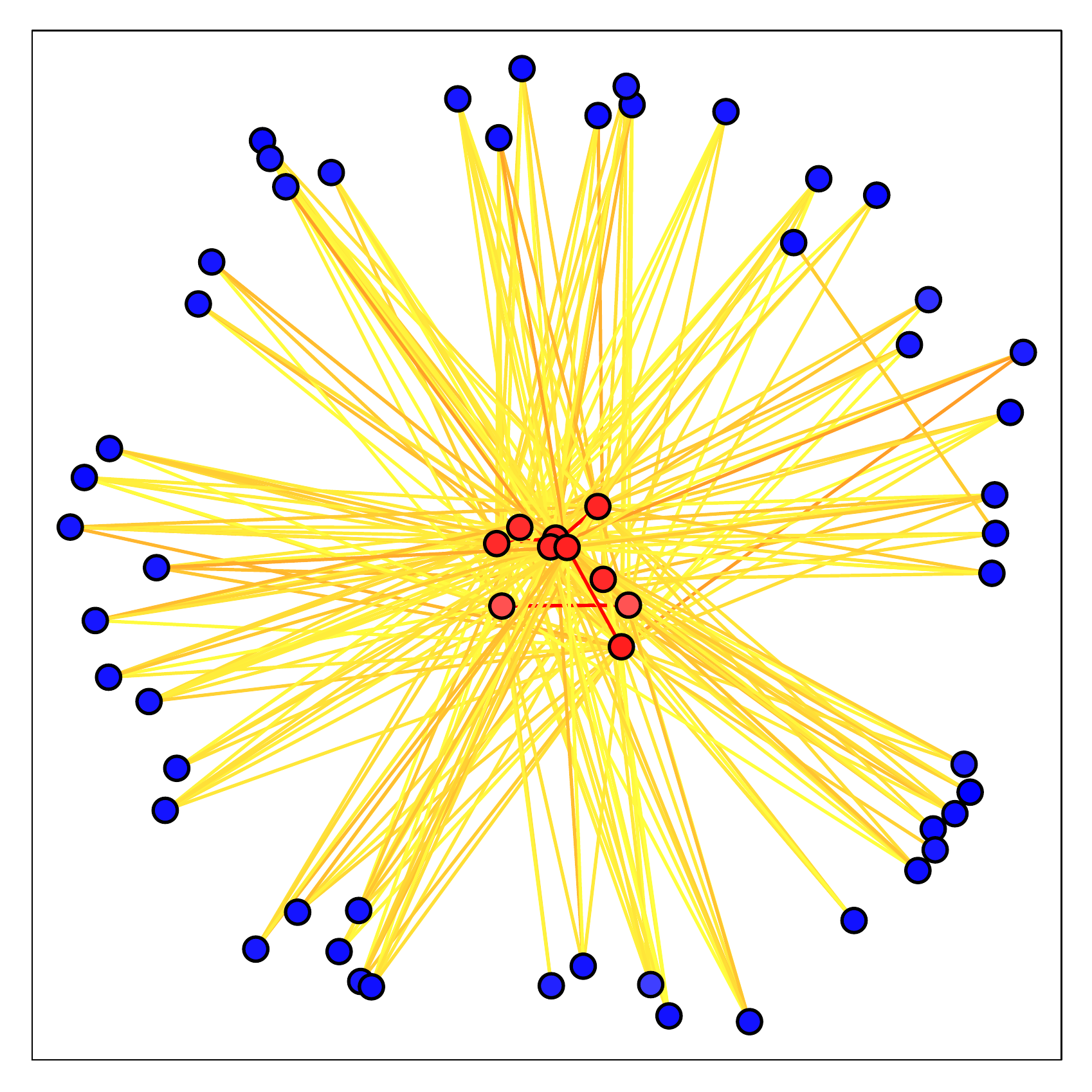} \\
 		$t$ = 1000 & $t$ = 5000  & $t$ = 50000 \\
 	\end{tabular}
 	\caption{Example of elite pattern. $N = 60$, $\omega = 0.3$ and $\rho = 0.1$, $\delta = 0.2$, $\sigma = 0.3$, $k = 5$. Two types of agents appear, the elite agents (in the middle of the network representation) have a positive self-opinion. The second category agents have a negative self-opinion. See section \ref{representations} for general explanation about the representations.}
 	\label{fig:vanityPattern2}
 \end{figure}

In this pattern, shown on Figure \ref{fig:vanityPattern2}, we note that after a while, there are two categories of agents:
\begin{itemize}
	\item elite agents have a very positive self-opinion, they have a good friend who is also in the elite, but all the other members of the elite are their foes. These appear as red nodes in the centre of the network representation for $t = 5000$ and $t = 50000$. In the matrix representation, they correspond to the red squares of the diagonal.
	\item second category have a very negative self-opinion, they have a very negative opinion the other second category agents, and they have a moderate opinion of the members of the elite (which can be moderately negative or moderately positive according to the parameters). These agents are represented as blue nodes on a circle in the network representation of Figure \ref{fig:vanityPattern2}. In this case, the opinions of the second category agents about the elite are moderately positive (yellow links). These links appear as partial light columns in the matrix representation. These columns are cut with blue squares which correspond to the negative opinions that second category agents have about each other.
\end{itemize}
  
We notice that the evolution from the initial state is also very different from the one of the equality pattern. Indeed, we see that after 1000 iterations, the matrix is less symmetric than in the equality pattern after the same number of iterations. Then we observe that the structure with the two categories of agents is already reached at $t = 5000$. This structure is rather stable, it has not changed much at $t = 50000$. However, one can obtain configurations where the elite agents often become second category and vice versa, with different values of the parameters.

\subsubsection{Hierarchy}

\begin{figure}
 	\centering
 	\begin{tabular}{ccc}
 		\includegraphics[width=5 cm]{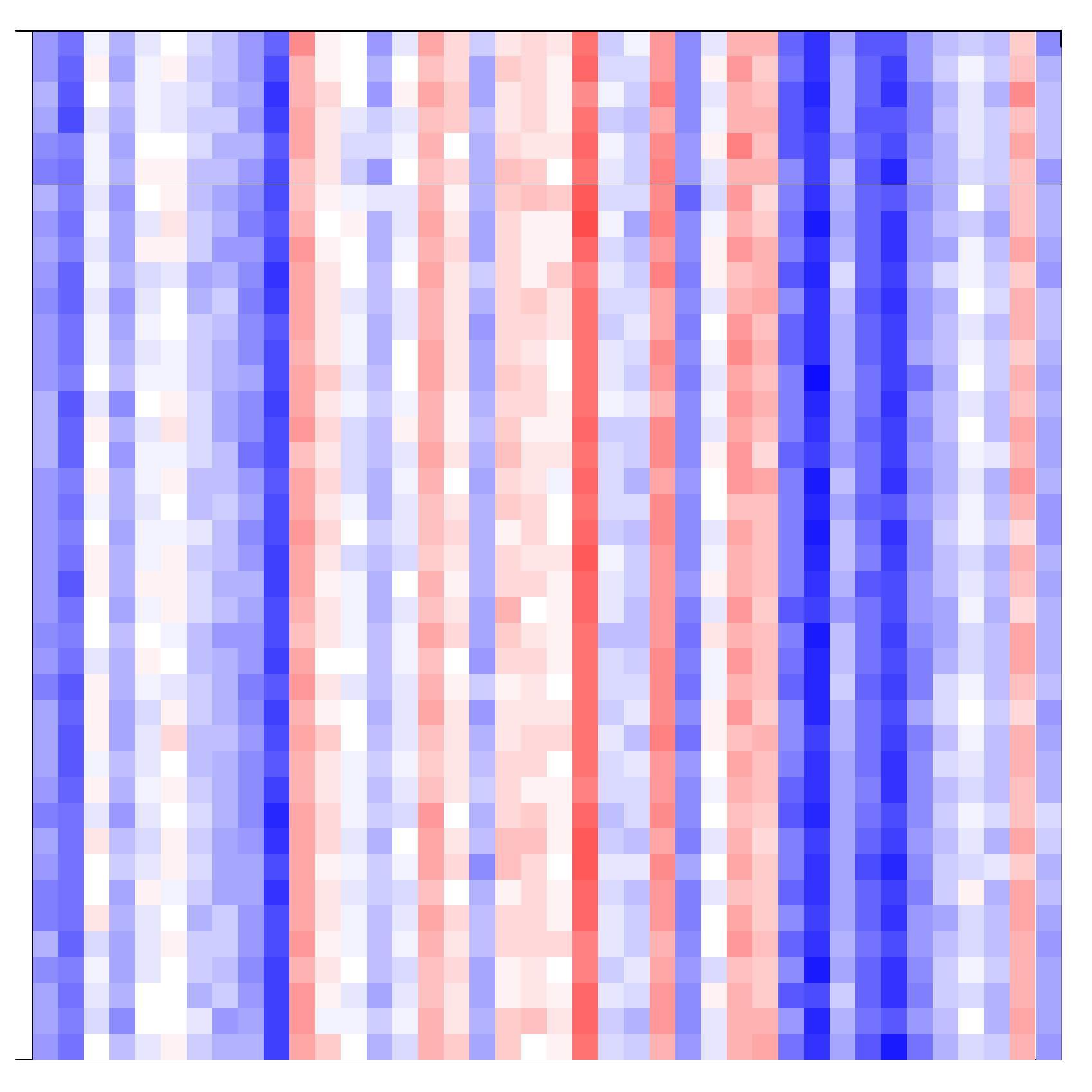} & 	\includegraphics[width= 5 cm]{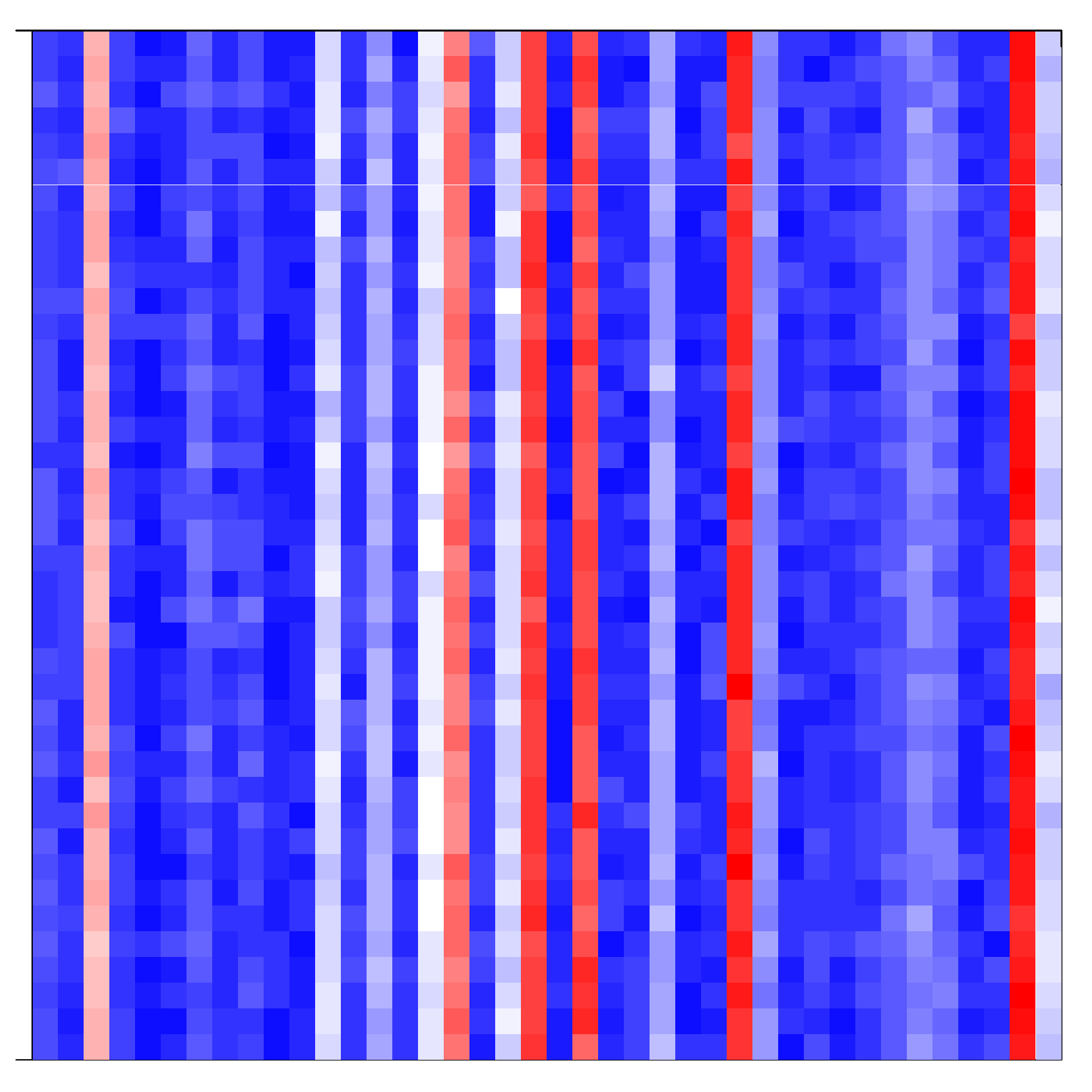}& 	\includegraphics[width= 5 cm]{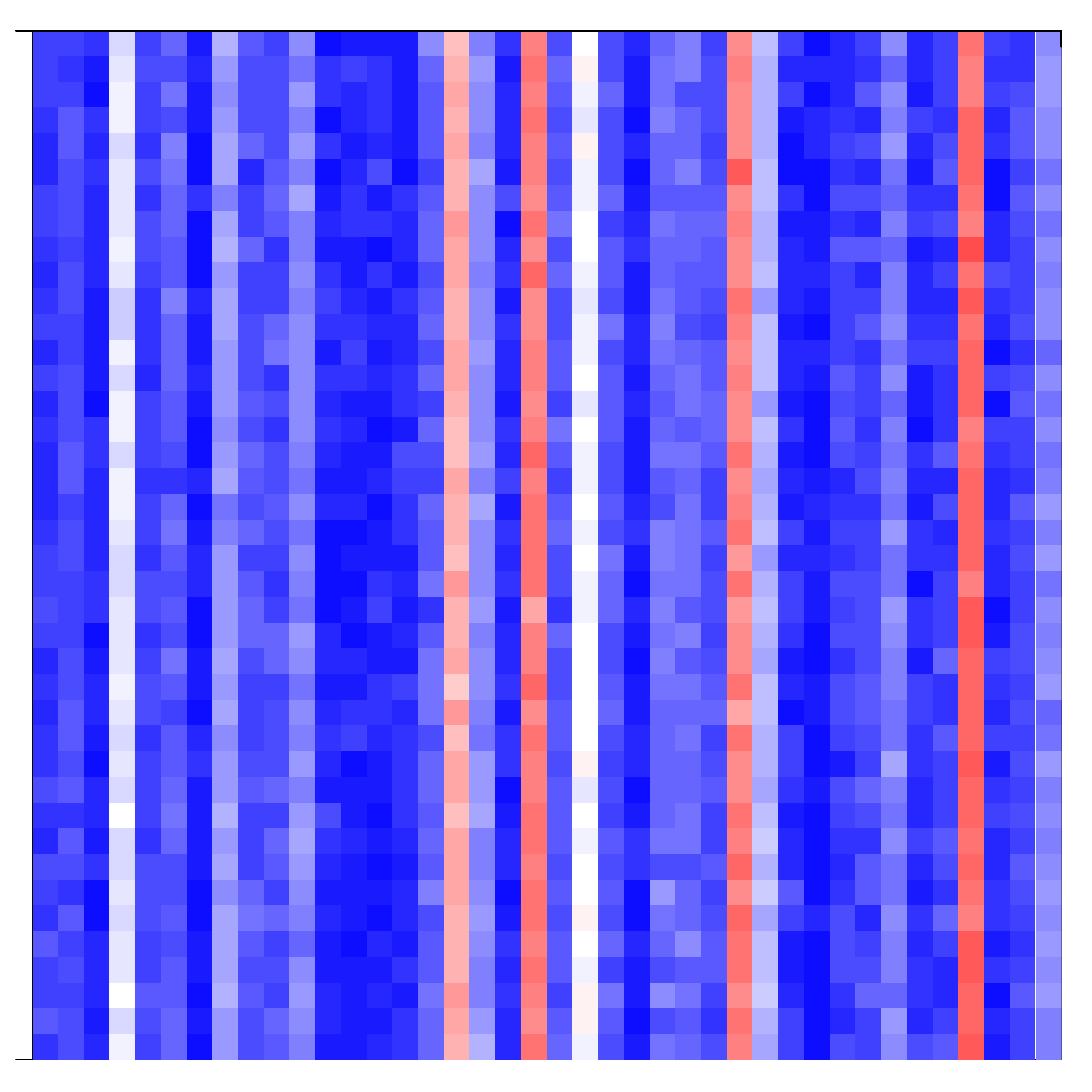} \\
 		\includegraphics[width=5 cm]{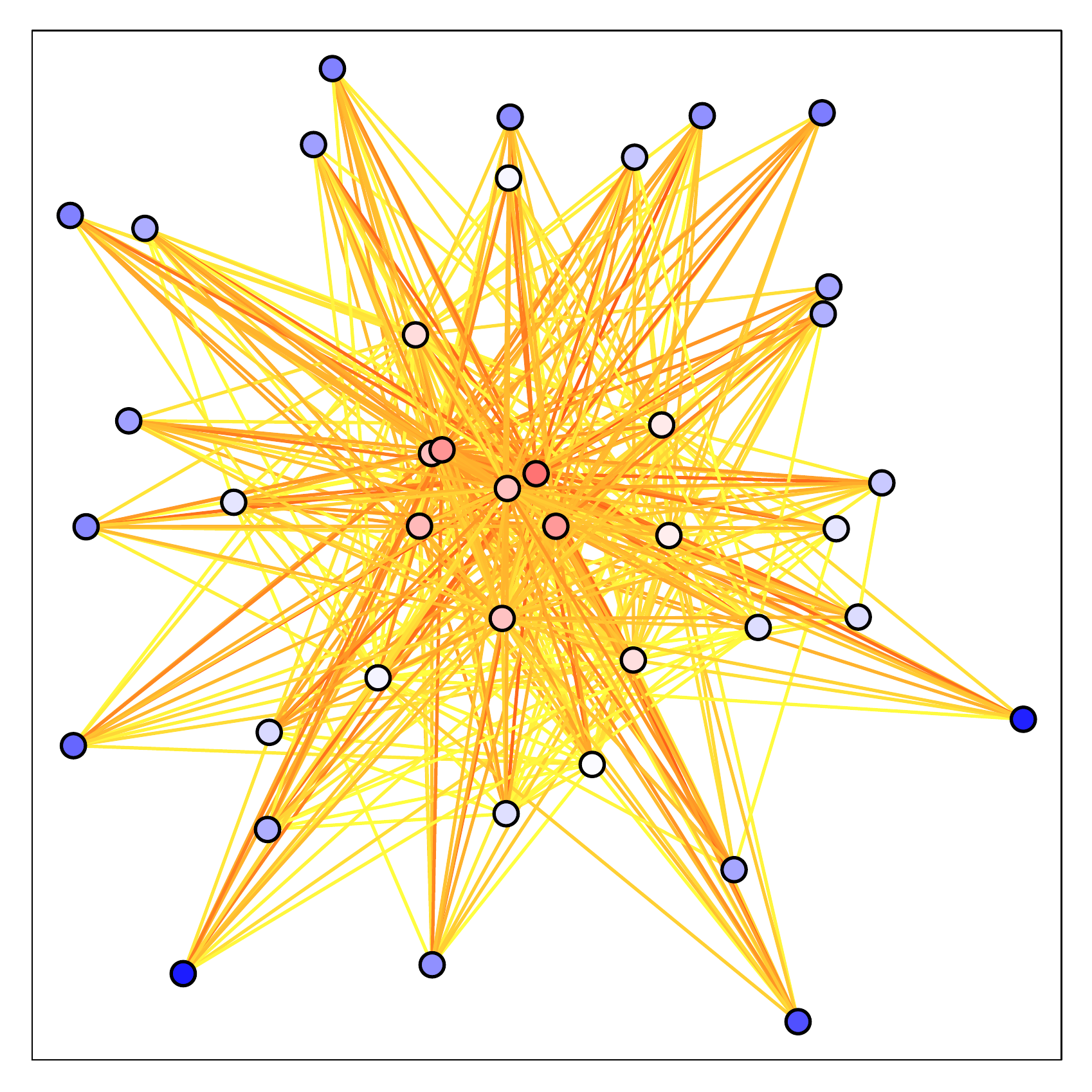} & 	\includegraphics[width= 5 cm]{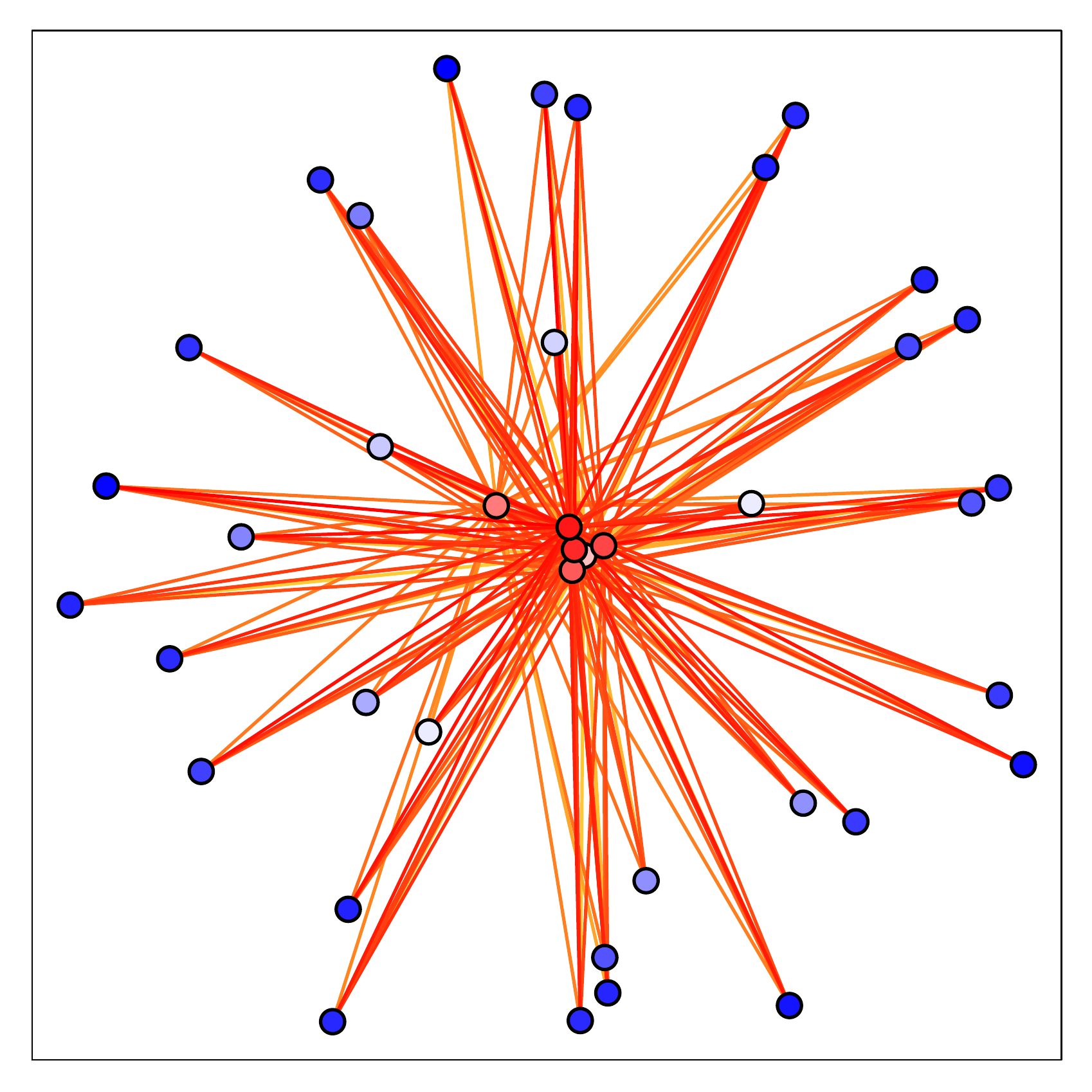}& 	\includegraphics[width= 5 cm]{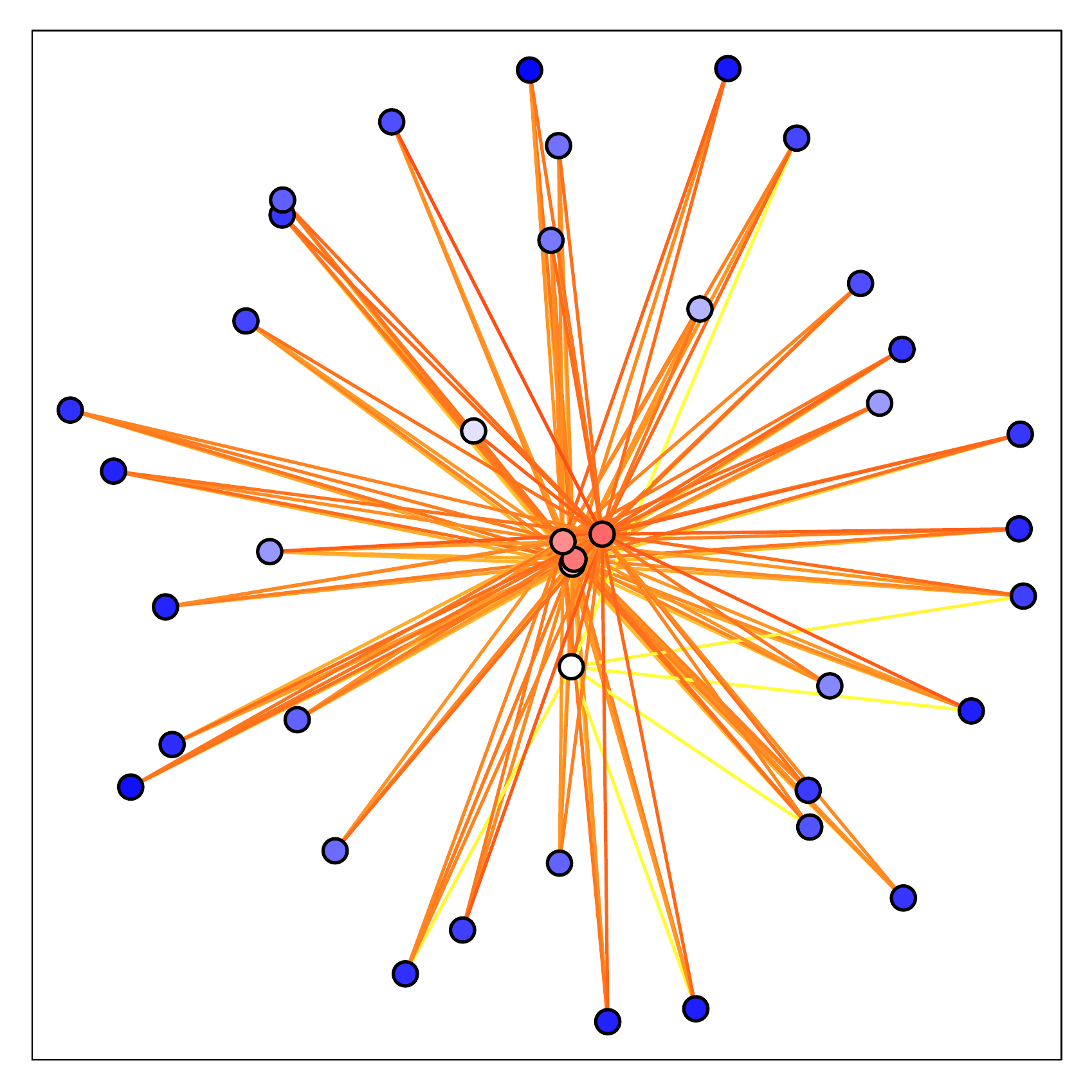} \\
 		$t$ = 5000, $\bar{a} = -0.22$  & $t$ = 20000, $\bar{a} = -0.53$,  & $t$ = 50000, $\bar{a} = -0.64$, 
 	\end{tabular}
 	\caption{Example of hierarchy pattern. $\omega = 0.2$, $\rho = 0.5$, $N = 40$, $\delta = 0.2$, $k = 10$, $\sigma = 0.3$. See section \ref{representations} for general explanation about the representations. $\bar{a}$ is the average opinion at the considered iteration. We observe that it is significantly negative.}
 	\label{fig:stableHierarchy}
 \end{figure}

When the opinion propagation gets stronger than in the previous examples, all
agents tend to have the same opinion about each other agent.  This is
shown for instance on Figure~\ref{fig:stableHierarchy} obtained using $N = 40$, $\omega
= 0.2$, $\rho = 0.5$, $\delta = 0.2$, $\sigma = 0.3$, $k = 10$. In the matrix representations, this is visualized by
cells with similar colors in the same column. As explained more formally
later, this agreement of opinions takes place because of the strength of the
averaging dynamics in the opinion propagation. We call the reputation $r_i$ of agent $i$ the average of the opinions about this agent. The random noise added to the opinion
propagation introduces some random fluctuations of the opinions around the
average value of the reputation and also some fluctuations of this
average value. 

Moreover, if we observe the distribution of the reputations, it appears that
the number of agents is decreasing with the reputation. This gives an image of a
hierarchy of reputations with a large basis and a progressively shrinking
number of agents while the reputation is increasing. The average opinion is significantly negative, reflecting the higher number of low reputation agents in the population. It can be observed on
the matrix representation that agents with the highest reputation are continuously
changing: they can drop after a while
to the basis of the hierarchy and conversly an agent of the lowest reputation
and vice versa.  

Moreover, the distribution itself is not fully stable. We observe that, from times to times we observe a few agents that have a reputation close to 1, whereas all the other reputations are close to -1. The highest reputation agents generally rapidly loose their high reputation which decrease to medium levels (0.5 or even a bit lower), and then the pattern with high reputations appears again. For instance, the highest reputation at $t = 20000$ is higher reputation than at $t = 50000$.

\subsubsection{Dominance}
	
\begin{figure}
 	\centering
 	\begin{tabular}{ccc}
 		\includegraphics[width=5 cm]{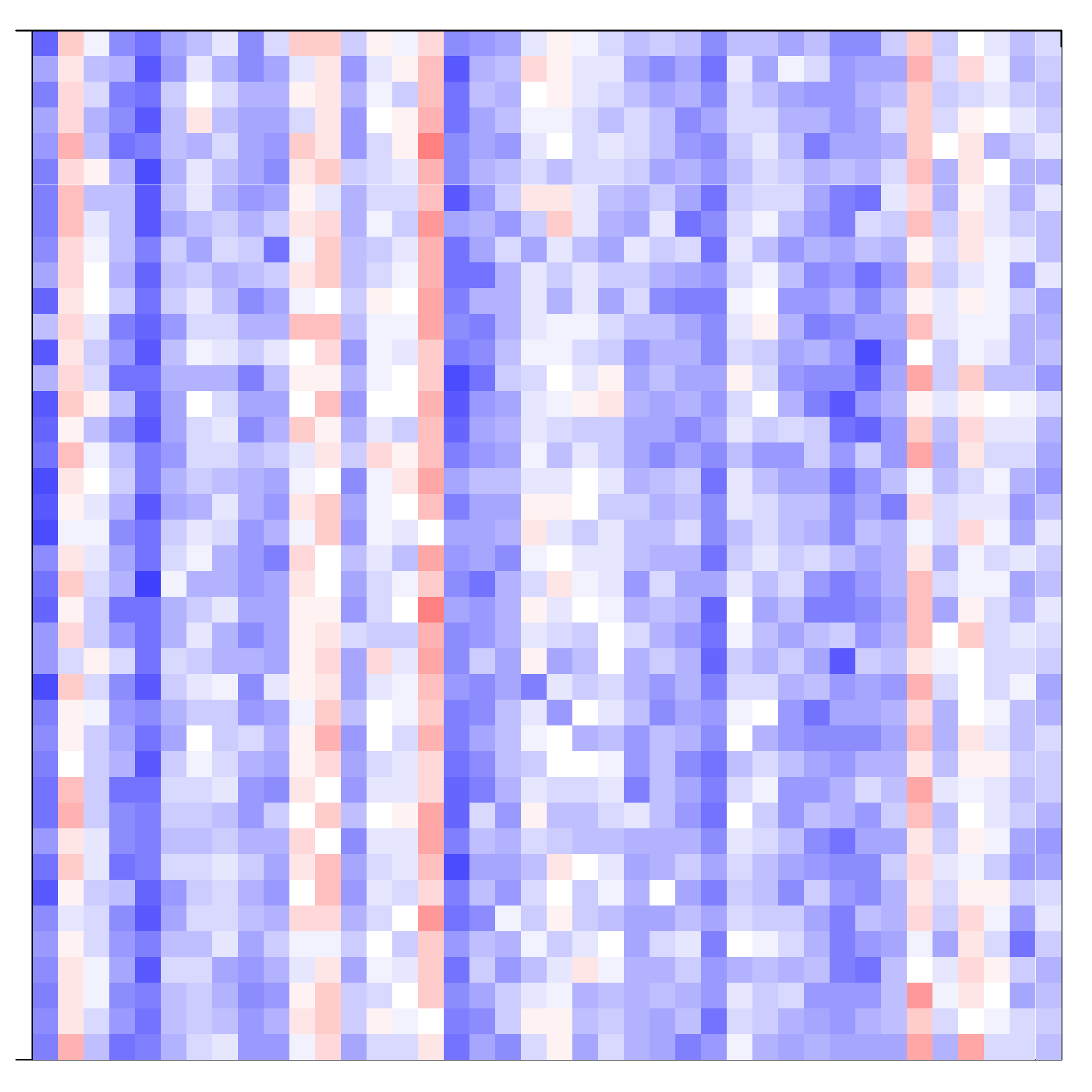} & 	\includegraphics[width= 5 cm]{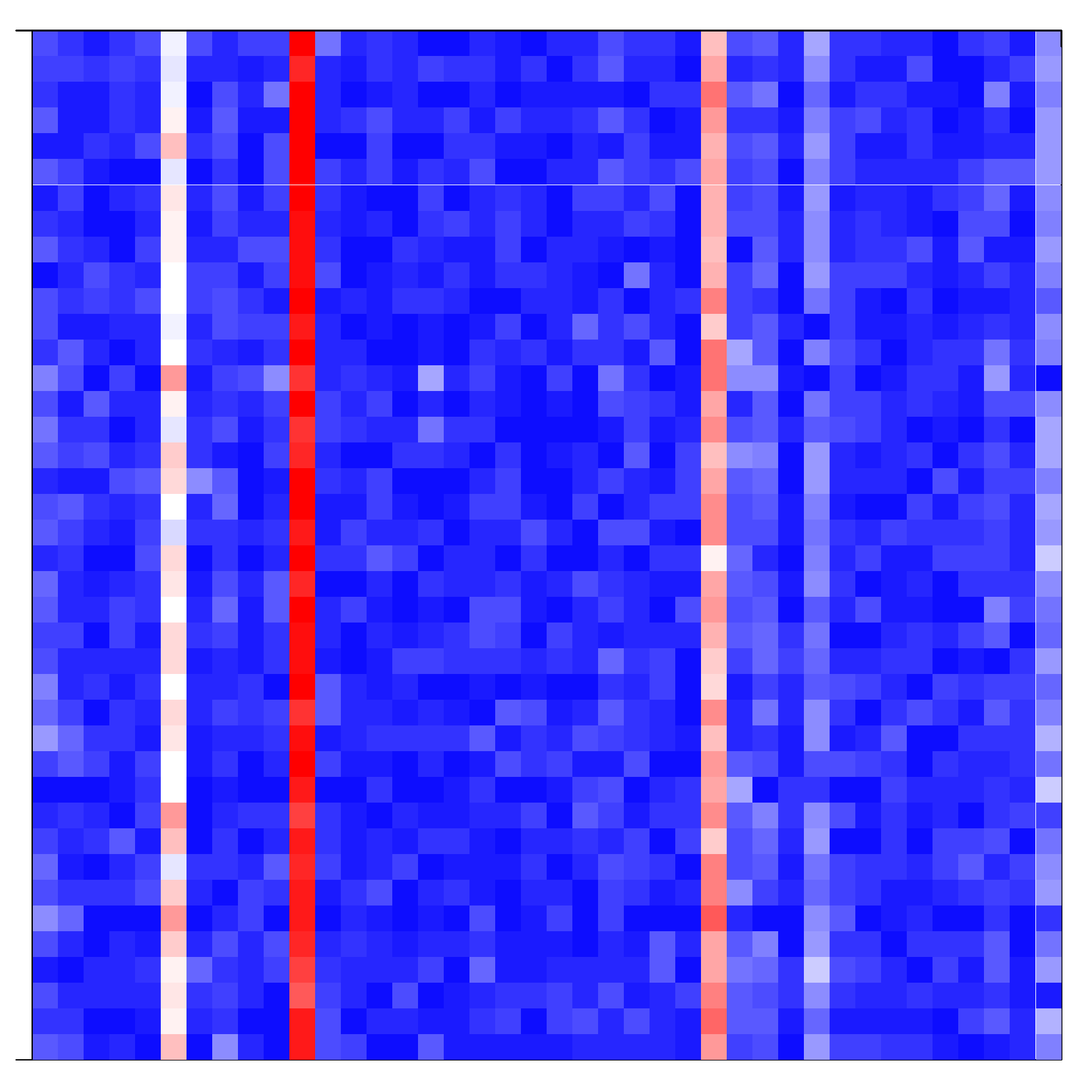}& 	\includegraphics[width= 5 cm]{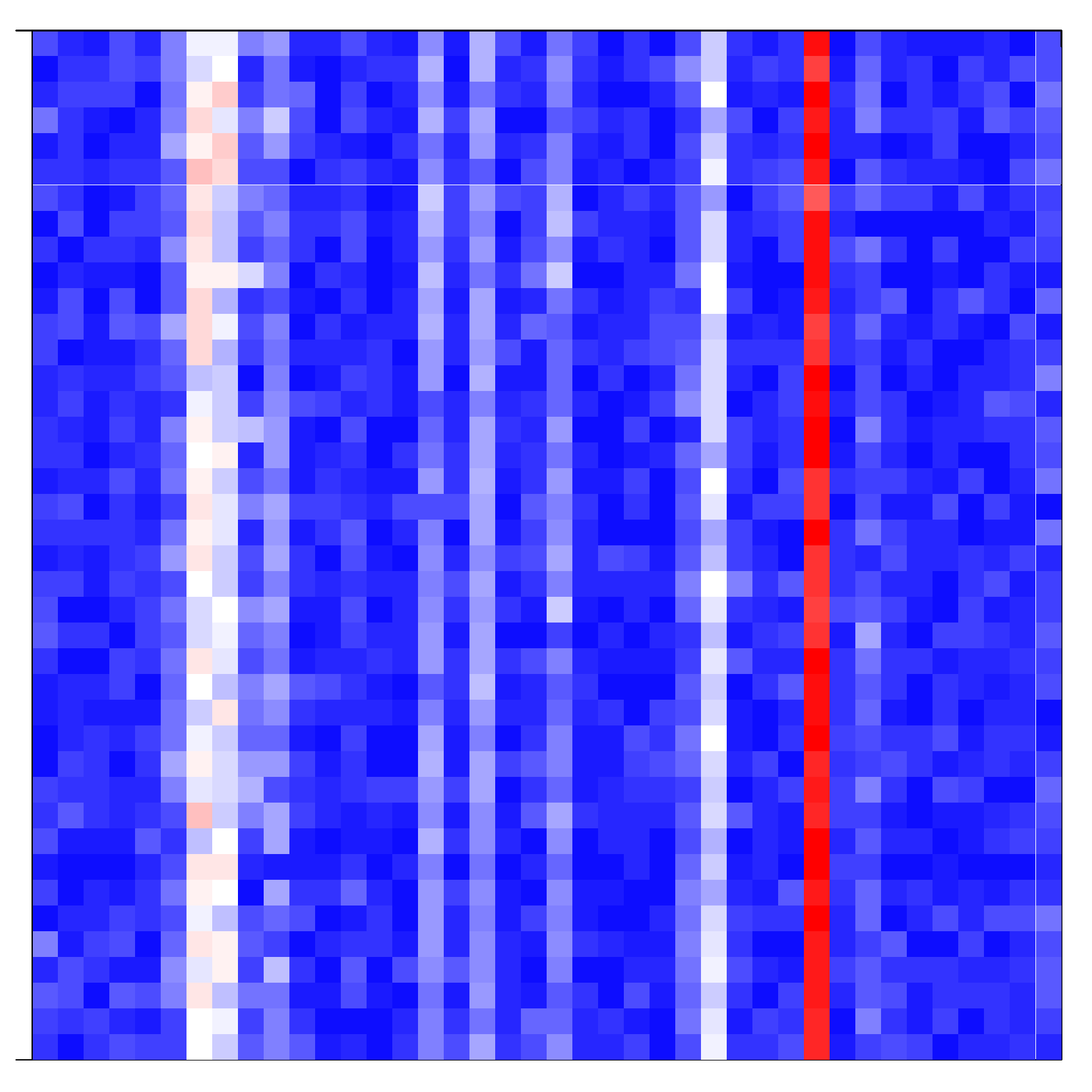} \\
 		\includegraphics[width=5 cm]{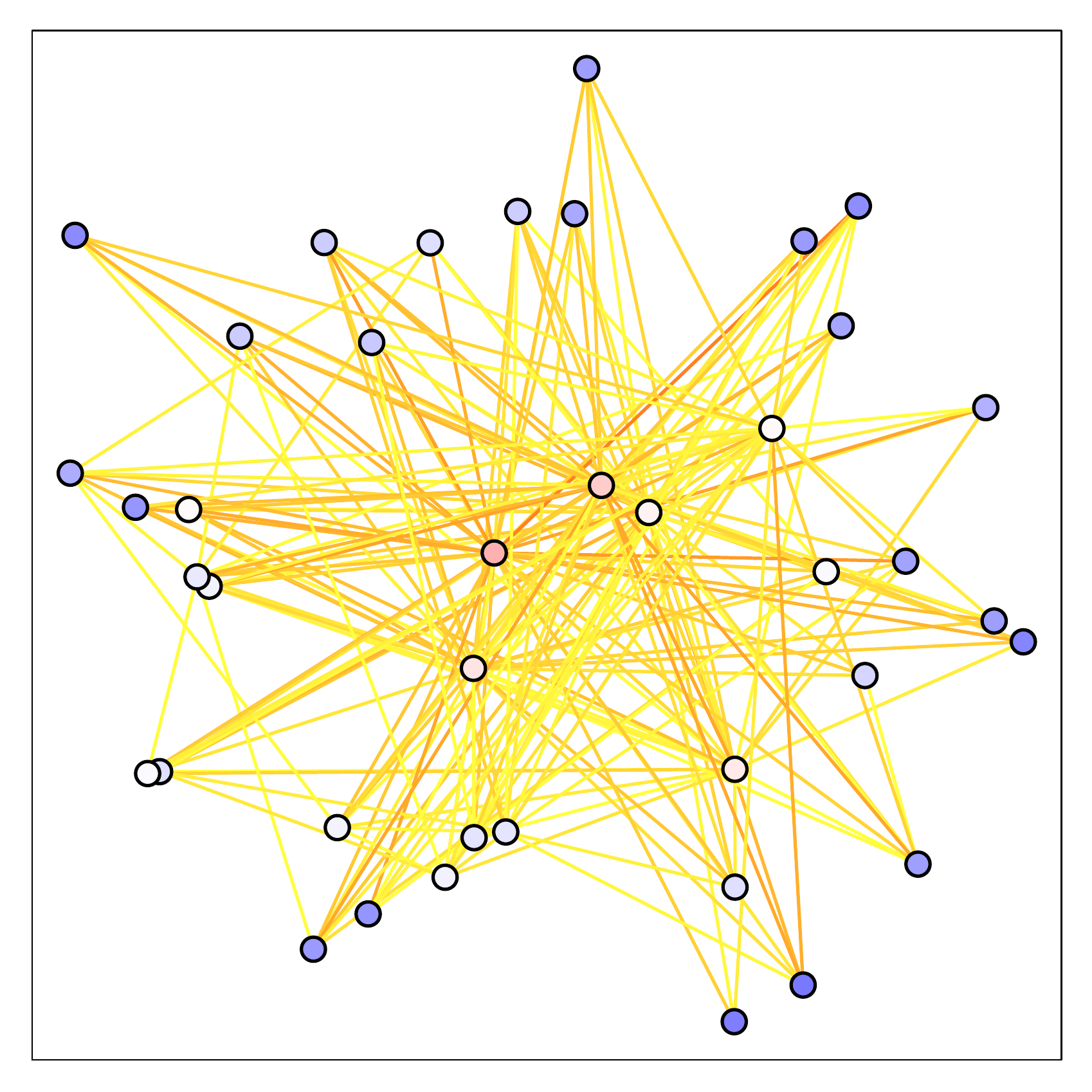} & 	\includegraphics[width= 5 cm]{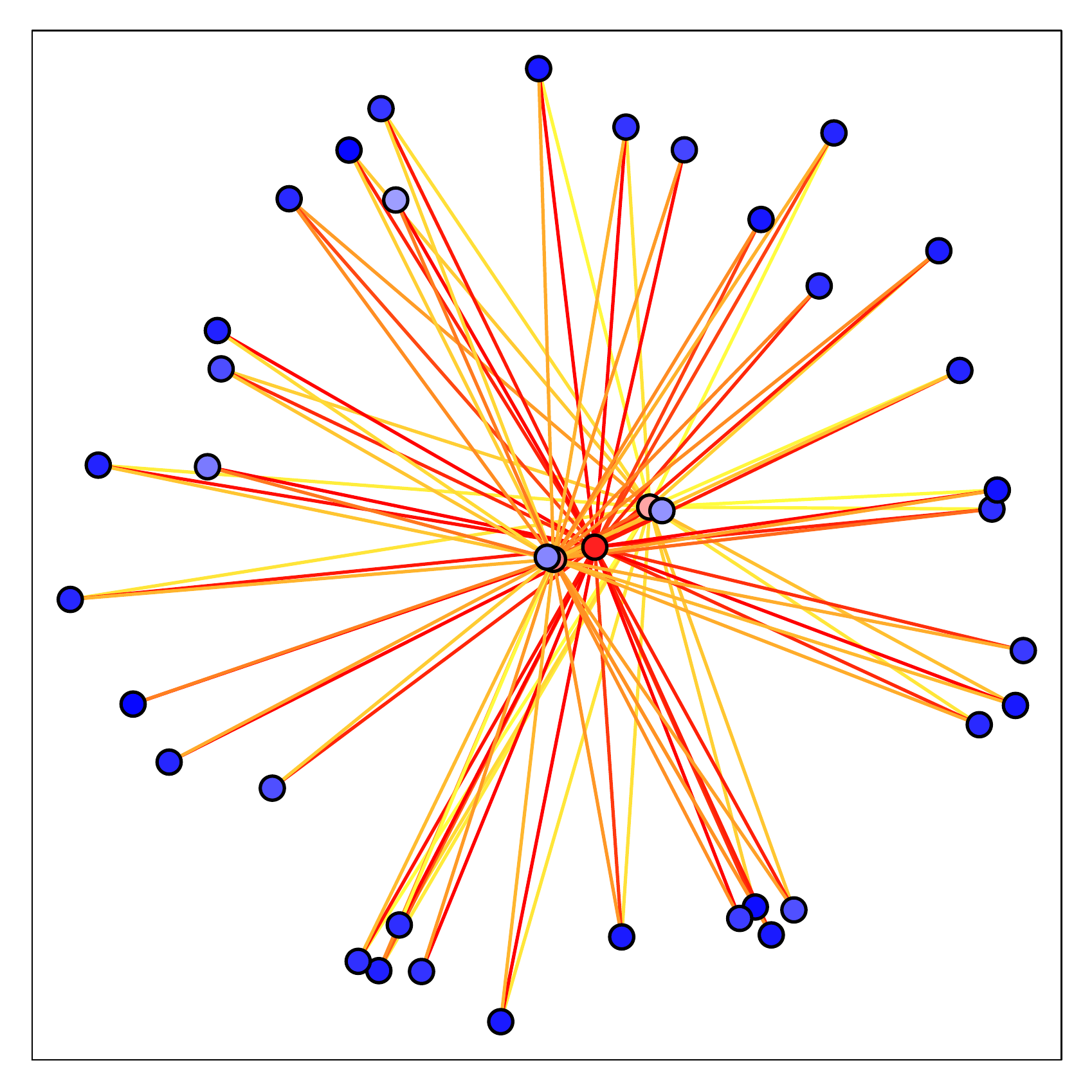}& 	\includegraphics[width= 5 cm]{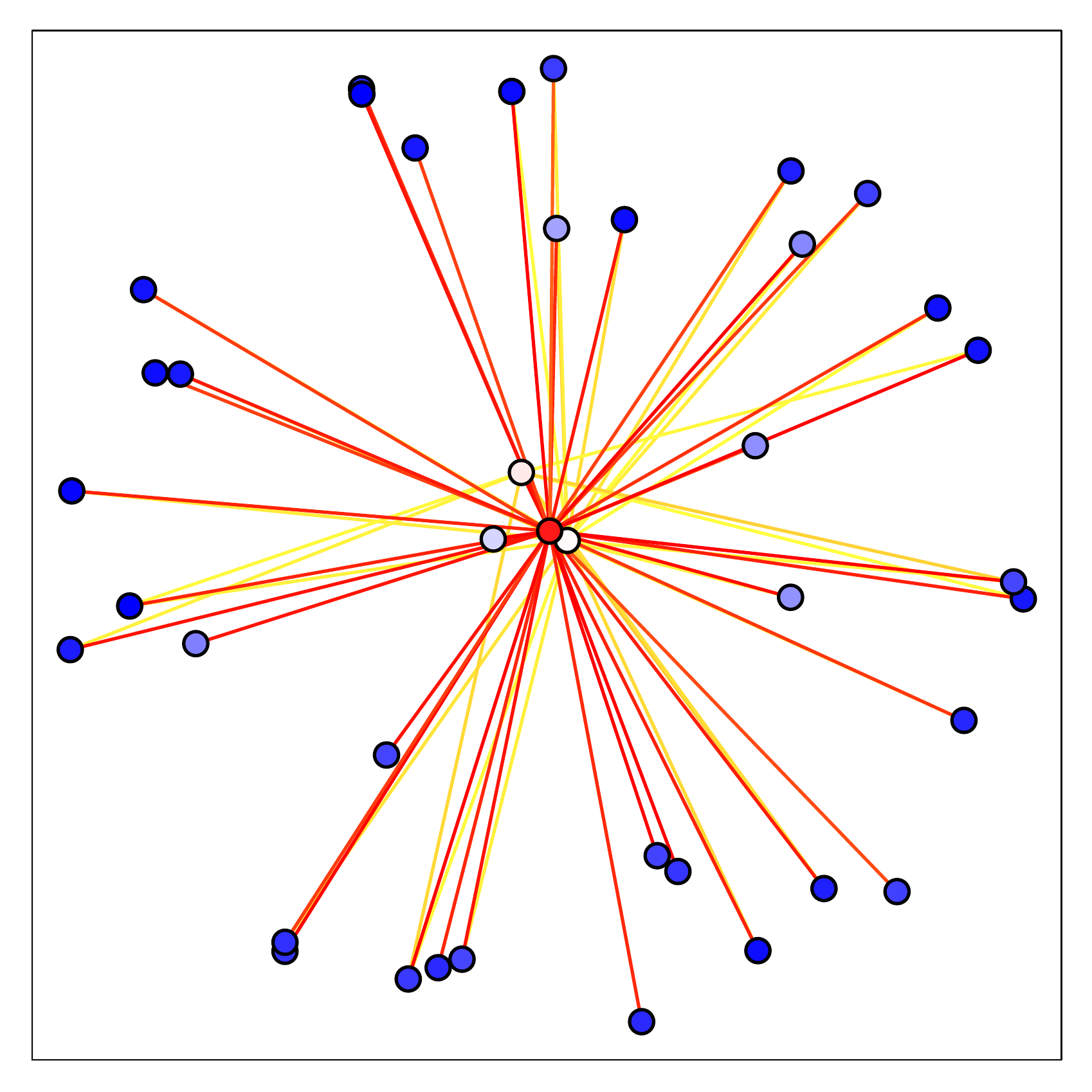} \\
 		$t$ = 1000, $\bar{a} = -0.23$  & $t$ = 10000, $\bar{a} = -0.74$,  & $t$ = 50000, $\bar{a} = -0.71$, 
 	\end{tabular}
 	\caption{Example of dominance pattern. $\omega = 0.4$, $\rho = 0.8$, $N = 40$, $\delta = 0.2$, $k = 2$, $\sigma = 0.3$. See section \ref{representations} for general explanation about the representations. $\bar{a}$ is the average opinion.}
 	\label{fig:dominance}
 \end{figure}

For some values of the parameters, we observe one agent who is highly valued by all the others (the one corresponding to the red column in the matrix representation and the red node at the centre of the network representation in Figure \ref{fig:dominance} for $t = 10000$ and $t = 50000$), whereas most of the other agents are homogeneously viewed negatively. Hence one can consider that the agent with the highest reputation strongly dominates the group. This dominating agent (called afterwards the leader for short) is metastable; after a while the leadership is taken by another agent. For instance during the simulation illustrated by Figure \ref{fig:dominance}, the leader changed 3 times in 50000 iterations. The transition between the leaders takes place when a second agent gets a reputation as high as the one of the current leader, then there is a competition between these two agents and only one remains the leader. In this example a leader maintains her dominance for typically more than 10000 iterations. There is no transition phase between the leader and her successor, as we can observe in an other pattern (mixed pattern described later).

\subsubsection{Crisis}

For some values of the parameters, after a while, all the opinions remain negative (often below $-0.5$). This state corresponds to a situation where each agent is the enemy of all the others and even of herself. In the matrix representation, all the squares are blue, and in the network representation, the nodes are not connected and they tend to be located on a circle because the algorithm tends to maximise the distance each agent has with all the others. In some cases, this state of general distrust is not stable and can lead to other patterns such as absolute dominance or elite (the mixed patterns are described in more details later).

 \begin{figure}
 	\centering
 	\begin{tabular}{ccc}
 		\includegraphics[width=5 cm]{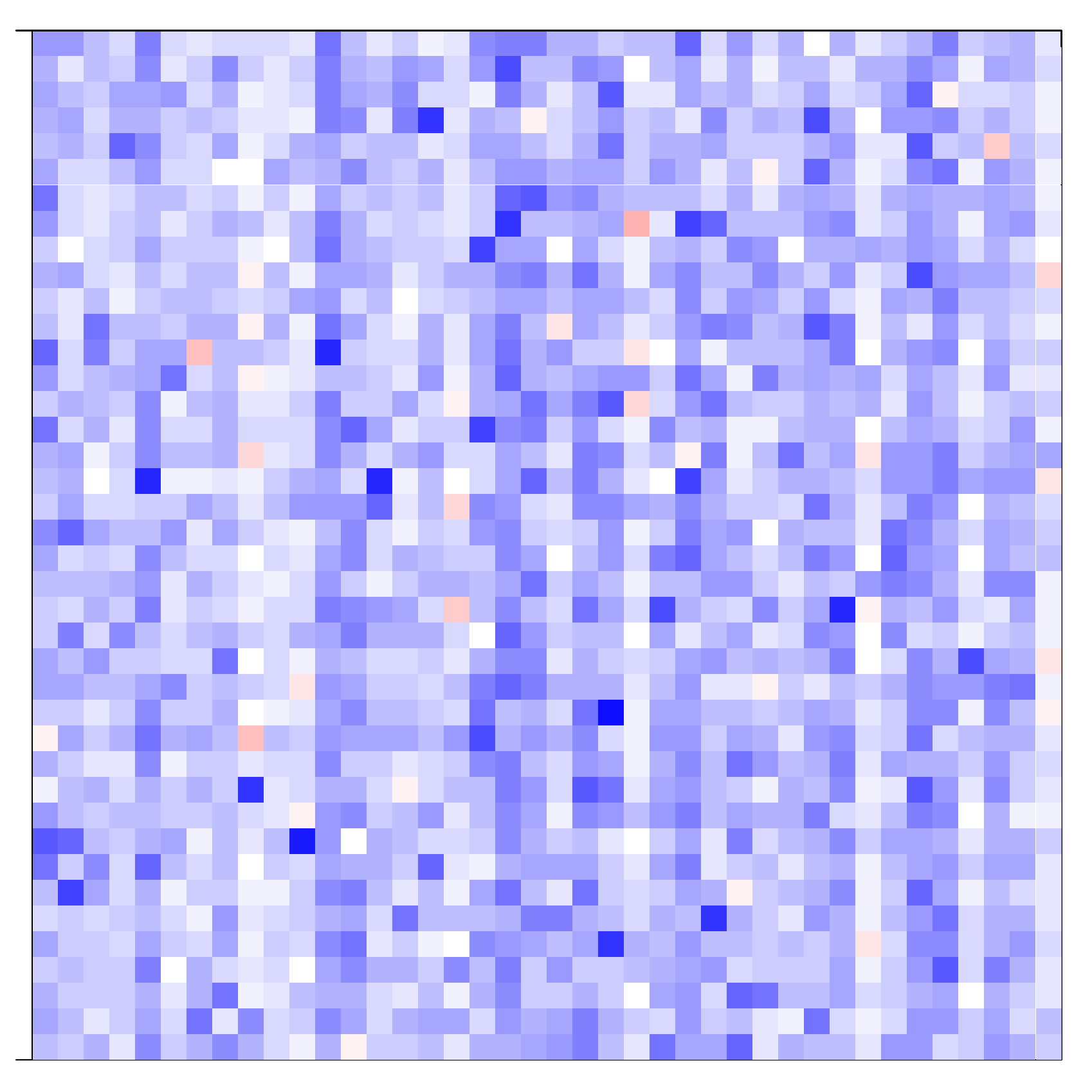} & 	\includegraphics[width= 5 cm]{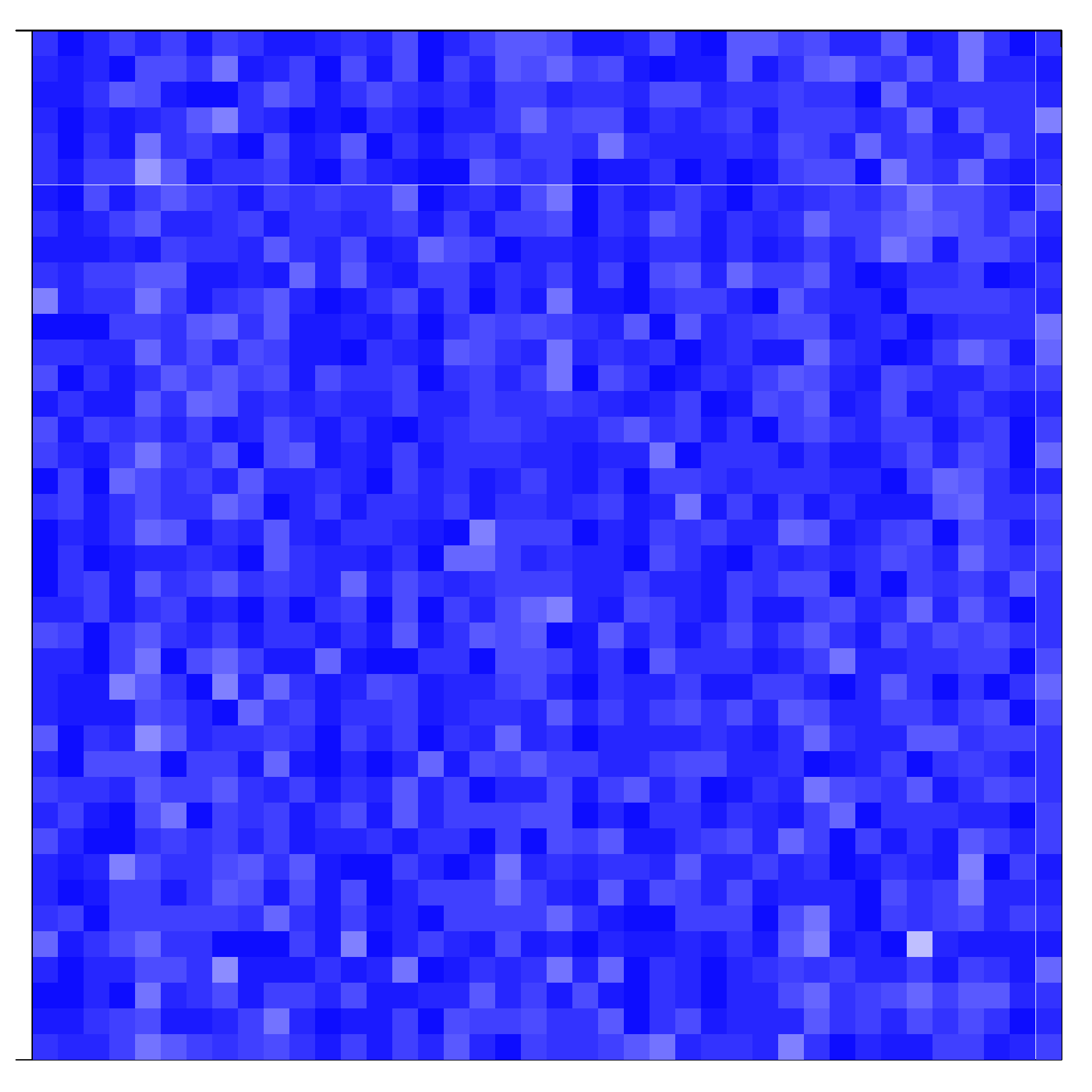}& 	\includegraphics[width= 5 cm]{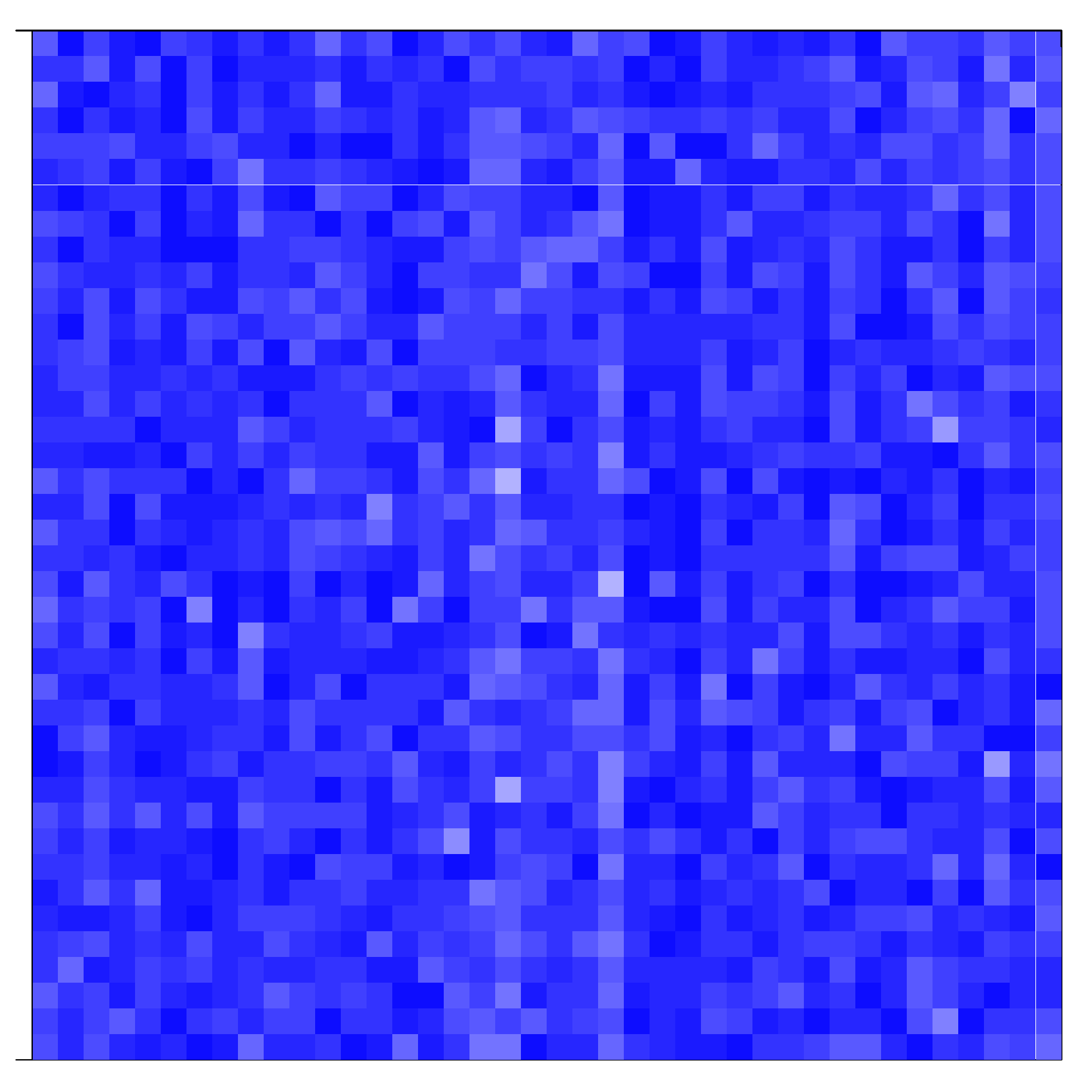} \\
 		\includegraphics[width=5 cm]{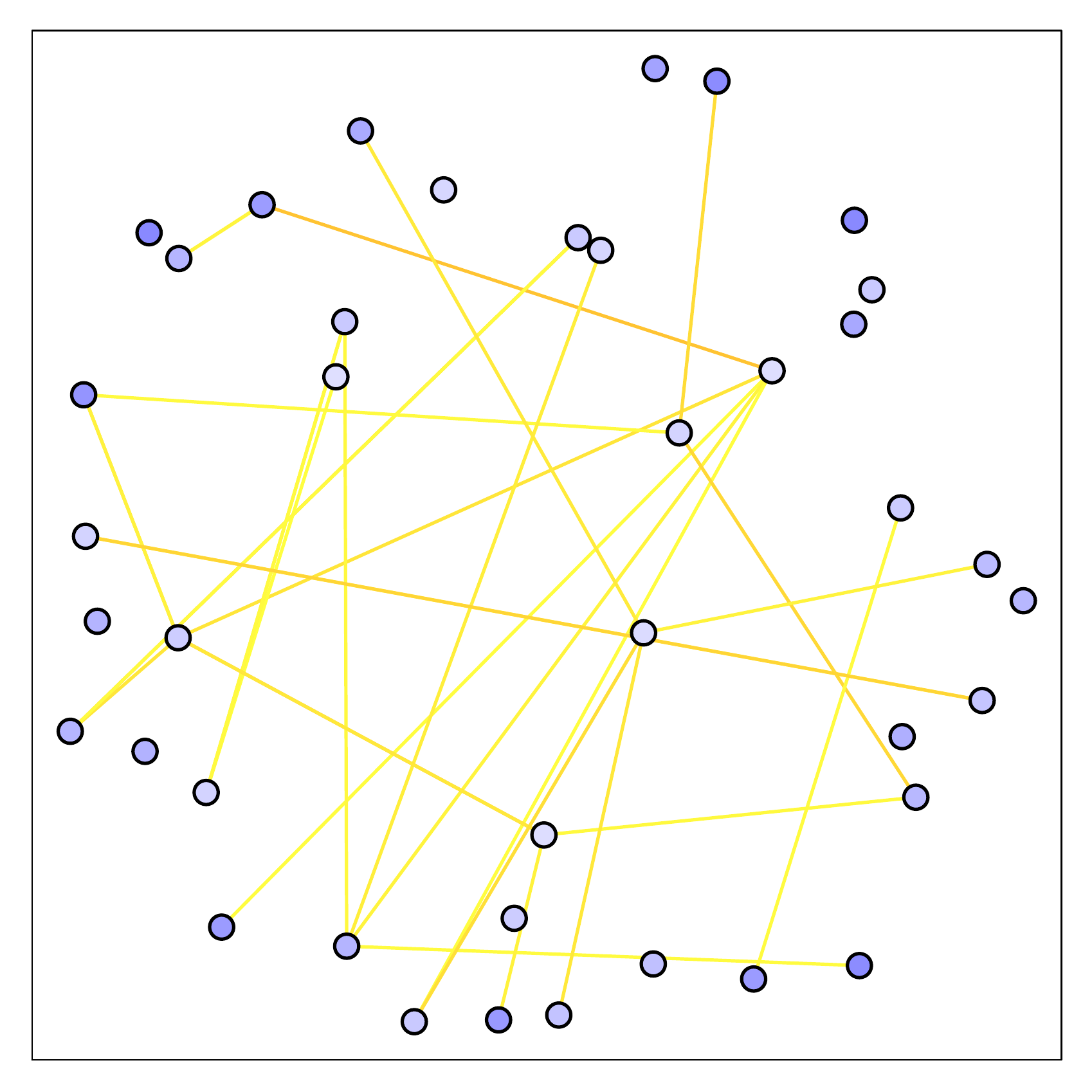} & 	\includegraphics[width= 5 cm]{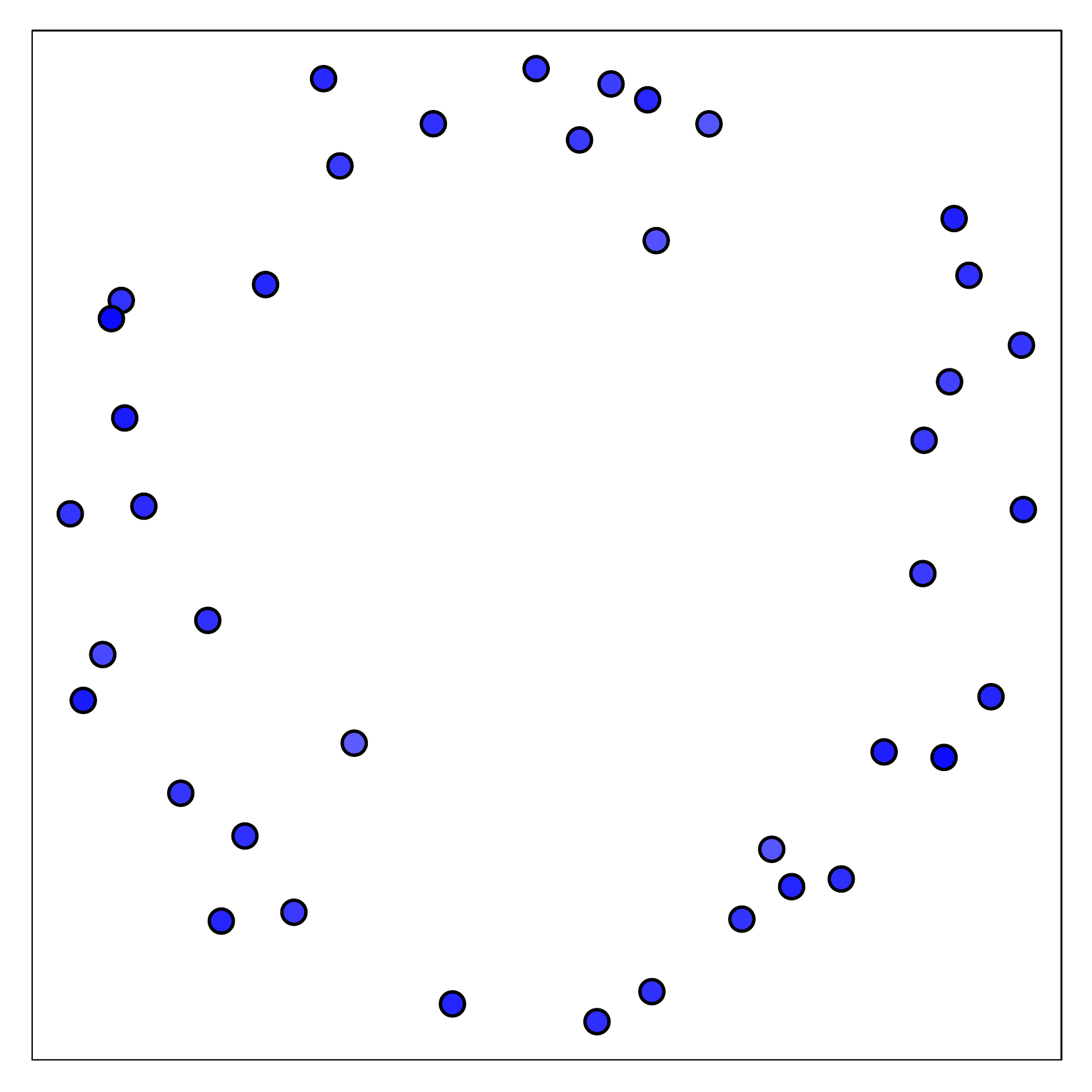}& 	\includegraphics[width= 5 cm]{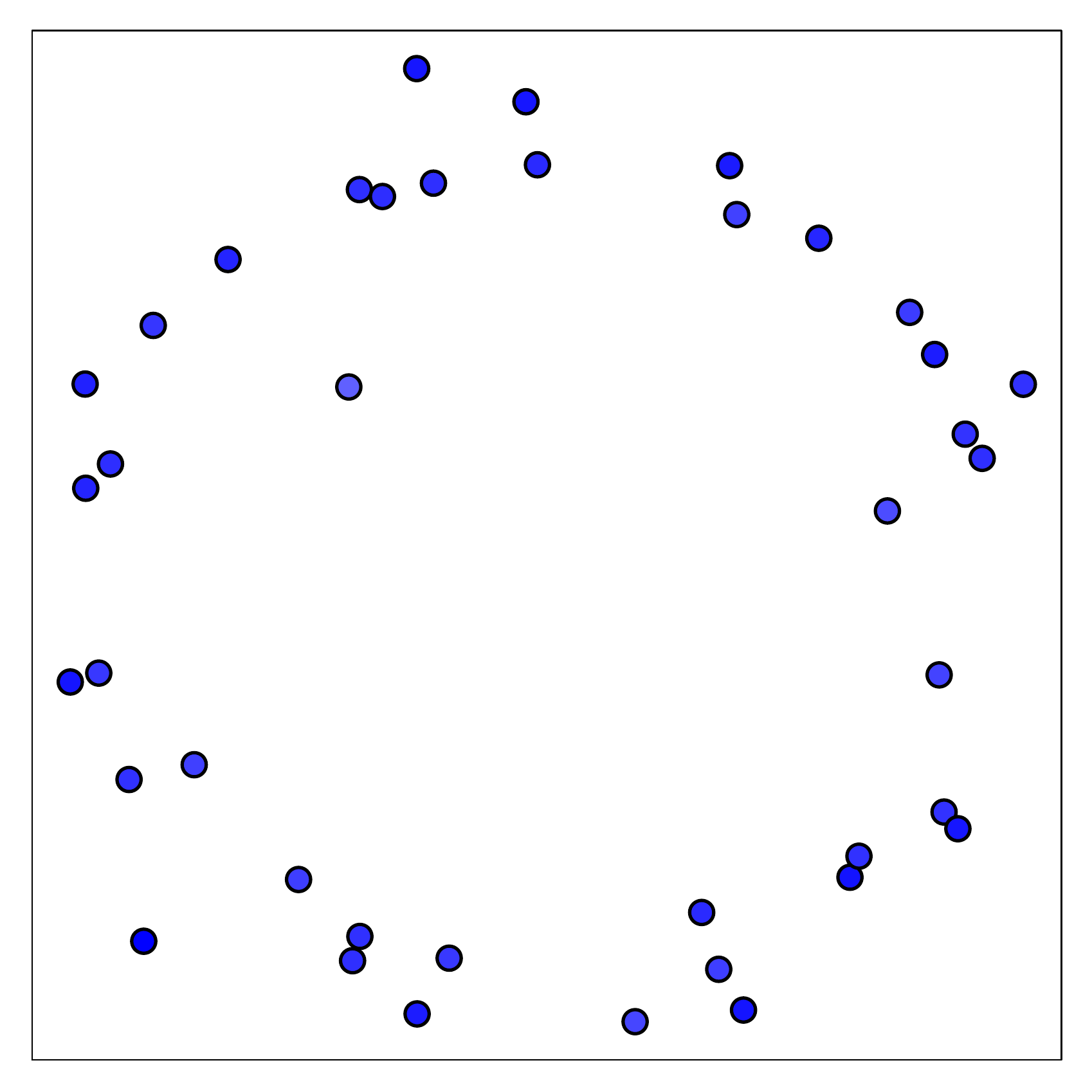} \\
 		$t$ = 1000, $\bar{a} = -0.28$  & $t$ = 5000, $\bar{a} = -0.82$,  & $t$ = 50000, $\bar{a} = -0.83$, 
 	\end{tabular}
 	\caption{Example of crisis pattern. $\omega = 0.4$, $\rho = 0.35$, $N = 40$, $\delta = 0.2$, $k = 2$, $\sigma = 0.5$. See section \ref{representations} for general explanation about the representations. $\bar{a}$ is the average opinion which is very negative for this pattern.}
 	\label{fig:crisis}
 \end{figure}

\subsection{Maps of patterns in the parameter space}
\subsubsection{Principle of the experiments}
The model includes 6 parameters and it is difficult to make an exhaustive study in the complete parameter space. We decided to study with more attention the influence of parameters $\rho$ and $\omega$ because they rule the respective influence of vanity and opinion propagation which are the main components of the model. We decided to fix $\delta = 0.2$ in this first analysis of the model, because it seems to us that the influence of this noise parameter is the easiest to interpret. Therefore, in the following, we propose maps in the space $\rho$ and $\omega$, both varying from $0.05$ to $1$ with a step of $0.05$, leading to $400$ pairs of values, for a small number of selected values of $N$, $k$ and $\sigma$. For each couple of parameter values $\rho$ and $\omega$, we run the model for 210 000 iterations (one iteration corresponding to $N/2$ random pair interactions), and we repeat this for 30 replicas. From iteration 10 000 to 210 000 every 100 iteration, we classify the state of the population into one of the 5 possible patterns: equality, elite, hierarchy, dominance, crisis. We do not consider the first 10 000 iterations because they are a transitory state that is not meaningful for the pattern. The rules for this classification are as follows:

\begin{itemize}
	\item \textit{smallWorld}: One opinion of the individual with the highest reputation is higher than 0.5, and one opinion about this individual is lower than -0.5 and one opinion about the individual with the lowest reputation is higher than 0 ($i_M$ such that  $r_{i_M} = max_i (r_i)$, $i_{m}$ such that $r_{i_m} = min_i (r_i)$, $\exists j$ such that $a_{i_{M},j} > 0.5$ and $\exists k$ such that $a_{k,i_{M}} < -0.5$ and $\exists l$ such that $a_{l,i_{m}} > 0$);
	\item \textit{elite}: One opinion of the individual with the highest reputation is higher than 0.5, and one opinion about this individual is lower than -0.5 and all the opinions about the lowest reputation individual are negative ($\exists j$ such that $a_{i_{M},j} > 0.5$ $\exists k$ such that $a_{k,i_{M}} < -0.5$ and $\forall l, a_{l,i_{m}} < 0$);
	\item \textit{hierarchy}: The highest reputation is above 0, and there are more than one reputation between the highest reputation and the highest reputation minus 0.5 ($r_{i_{M}} >0$  and $\#\left\{i \text{ such that } r_i > r_{i_{M}} - 0.5 \right\} > 1 $);
	\item \textit{dominance}: The highest reputation is above 0, and there is at most one reputation which is between the highest reputation and the highest reputation minus 0.5 ($r_{i_{M}} >0$, and $\#\left\{i \text{ such that } r_i > R - 0.5 \right\} \leq 1 $);
	\item \textit{crisis}: All opinions are below -0.5, and all reputations below 0 ($a_{i,j} \leq -0.5$, for $1 \leq i, j \leq N$  and $r_i \leq 0$ for $1 \leq i, j \leq N$).	
\end{itemize}

Then, for the 30 replicas, we compute the percentages of presence of each pattern during the runs. These percentages are used for computing the maps that are presented in the following sections: We show the sign corresponding to one pattern as soon as this pattern appears in more than 20\% of considered iterations. 

\subsubsection{Effect of $k$ and $\sigma$ for $N = 40$.}

Figure \ref{fig:patMapN40} shows the maps of these patterns for the following values: 
\begin{itemize}
	\item $N = 40$, $k = 10$, $\sigma = 0.3$;
	\item $N = 40$, $k = 10$, $\sigma = 0.5$;
	\item $N = 40$, $k = 2$, $\sigma = 0.3$; 
	\item $N = 40$, $k = 2$, $\sigma = 0.5$;
\end{itemize}
This first set of experiments shows the respective influence of parameters $k$ and $\sigma$ for a given value of $N$. The figure can be conveniently interpreted by considering a radius, starting on the $\rho$ axis and turning around the point $(\rho = 0, \omega = 0)$ until it reaches the $\omega$ axis. Following this rotation, the patterns are found in the same order in the four maps (hierarchy, dominance, crisis, elite and smallWorld) with important overlaps of two  and sometimes of three patterns (we study some of these mixted patterns in the next sections). The hierarchy occurs therefore when there is a low vanity and a high opinion propagation, whereas the smallWorld pattern occurs for a high vanity and a low opinion propagation. The other patterns appear for stronger coupling between the two processes.

Moreover, we note that the area covered by each pattern vary significantly when $k$ and $\sigma$ vary. More precisely, increasing $k$, the number of acquaintances talked about, increases significantly the sectors of hierarchy and dominance, and decreases the sectors of smallWorld and elite. The effect on the crisis pattern is not clear from these experiments. From the dynamics of the model, we know that increasing $k$ increases the opinion propagation which has similar effects to increasing $\rho$, which is confirmed by the observation that the limits between the sectors of different patterns tend to turn upward when $k$ is growing. We also observe that increasing $\sigma$ increases the sector of smallWorld and hierarchy and decreases the sector of elite.

 \begin{figure}
 	\centering
 	\begin{tabular}{cc}
 	\includegraphics[width=9 cm]{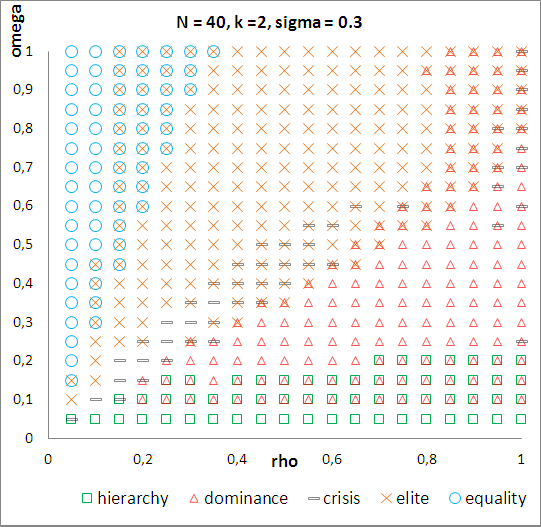} & 	\includegraphics[width= 9 cm]{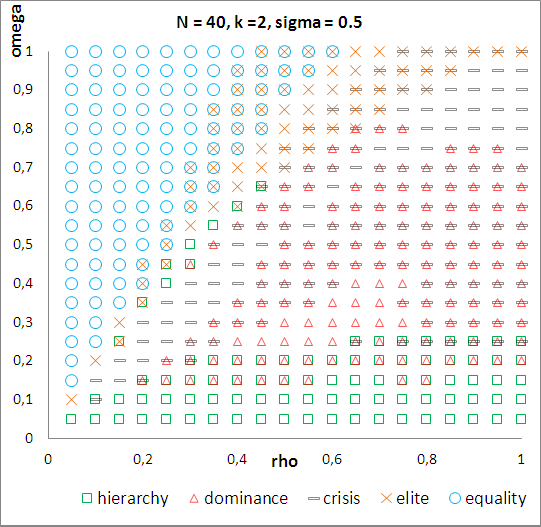}\\
 		\includegraphics[width=9 cm]{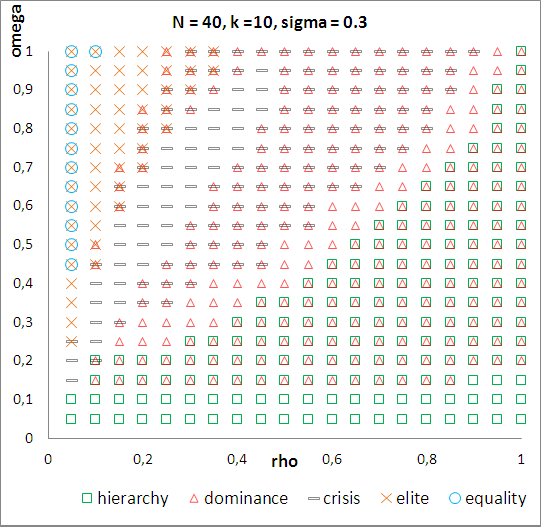} & 	\includegraphics[width= 9 cm]{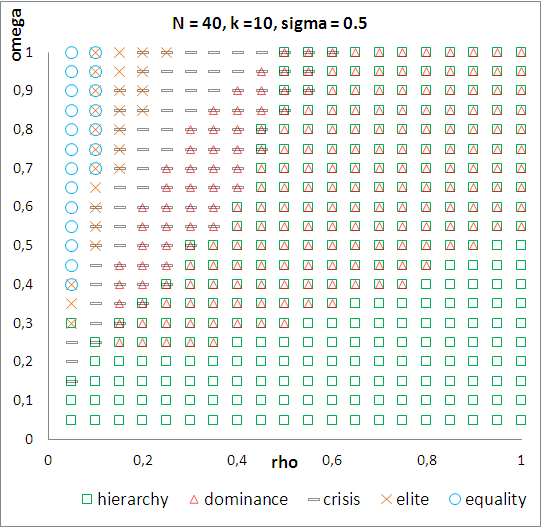} 
 		 	\end{tabular}
 	\caption{Maps of patterns for $N = 40$, $\delta = 0.2$, different values of $k$ (number of acquaintances talked about in the opinion propagation) and $\sigma$ (parameter ruling the slope of the propagation coefficient function). A pattern is visualised on the map when it appears in more than 20\% of the considered iterations in the 30 replicas, whenever two or three patterns are present in more than 20\% of replicas, two or three symbols are superposed. (see text for more details).}
 	\label{fig:patMapN40}
 \end{figure} 	

\subsubsection{Maps of patterns for $N = 10$ and $N = 100$.}

In this section, we try to get first observations on the effect of the size of the population on the sectors of patterns in the parameter space.

\begin{itemize}
	\item $N = 10$, $k = 2$, $\sigma = 0.3$;
	\item $N = 10$, $k = 2$, $\sigma = 0.5$;
	\item $N = 100$, $k = 10$, $\sigma = 0.3$;
	\item $N = 100$, $k = 10$, $\sigma = 0.5$;
\end{itemize}

This second set of experiments shows that the maps are also organised in sectors of patterns in the same order for different values of the number of agents.  We observe that the crisis pattern does not appear for $N = 10$ whereas it is more present for $n = 100$ than for $N = 40$. It seems that when the population reaches a crisis state, it takes more time to get out of it when the population is large. For $N = 10$ the dominance pattern is covering a larger sector than for the other values of $N$. This is probably due to the very simple criterion to distinguish between dominance and hirerachy, which should take the size of the population into account in a more refined version.

\begin{figure}
 	\centering
 	\begin{tabular}{cc}
 	\includegraphics[width=9 cm]{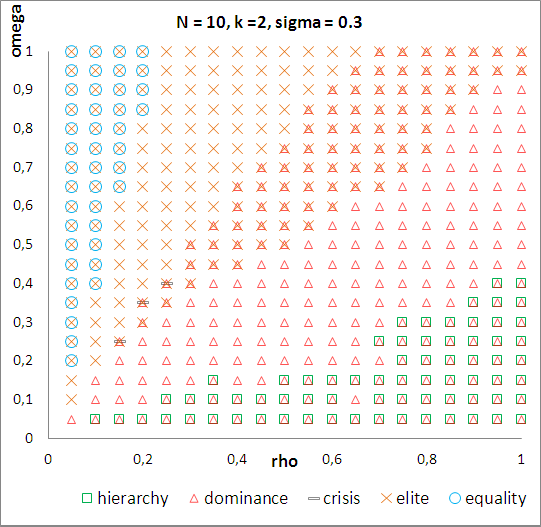} & 	\includegraphics[width= 9 cm]{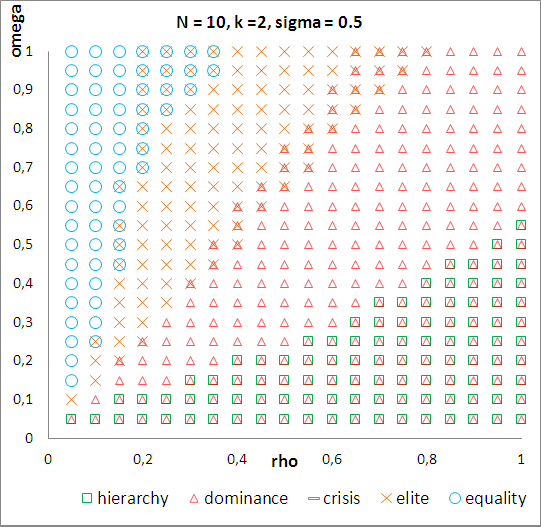}\\
 		\includegraphics[width=9 cm]{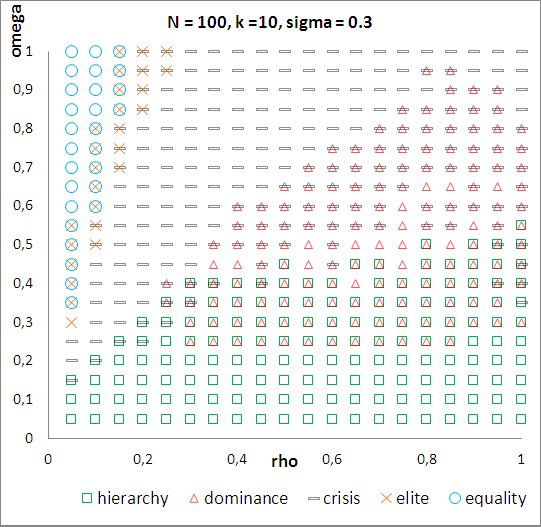} & 	\includegraphics[width= 9 cm]{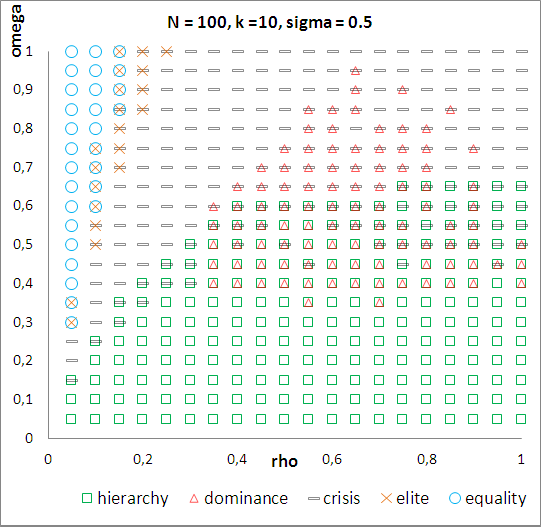} 
 		 	\end{tabular}
 	\caption{Maps of patterns for $\delta = 0.2$, $N = 10, k = 2$ and $N = 100, k =10$, for $\sigma = 0.3$ or $\sigma = 0.5$. A pattern is visualised when it appears in more than 20\% of the considered iterations in the 30 replicas for $N = 10$ or 10 replicas for $N = 100$, whenever two or three patterns are present in more than 20\% of replicas, two or three symbols are superposed(see text for more details).}
 	\label{fig:patMapN40}
 \end{figure} 	

\subsubsection{Stability of the patterns over the replicas.}
We do not think that it is very useful to provide the maps of variations of the patterns over the replicas. Globally, we observe that the main patterns for which vanity dominates (equality) or for which opinion propagation dominates (hierarchy) are very stable. The mixed patterns are the less stable, particularly around the diagonal.
We suspect that this variability, at least in some cases, could nevertheless be misleading. Indeed, for instance in the case of the mixed crisis-dominance pattern, the length of episodes of dominance or of crises can vary significantly with time. An interesting property to test for these patterns is the ergodicity of the system: running it during a very long time would be equivalent to make many replicas. If this ergodicity was verified, we would have to evaluate the necessary time to run the model in order to get a good view of the distribution of events. We consider that this work is out of the scope of this first paper. We now present two examples of mixed patterns.

\subsubsection{First example of mixed pattern: equality and elite}
\begin{figure}
 	\centering
 	\begin{tabular}{ccc}
 		\includegraphics[width=5 cm]{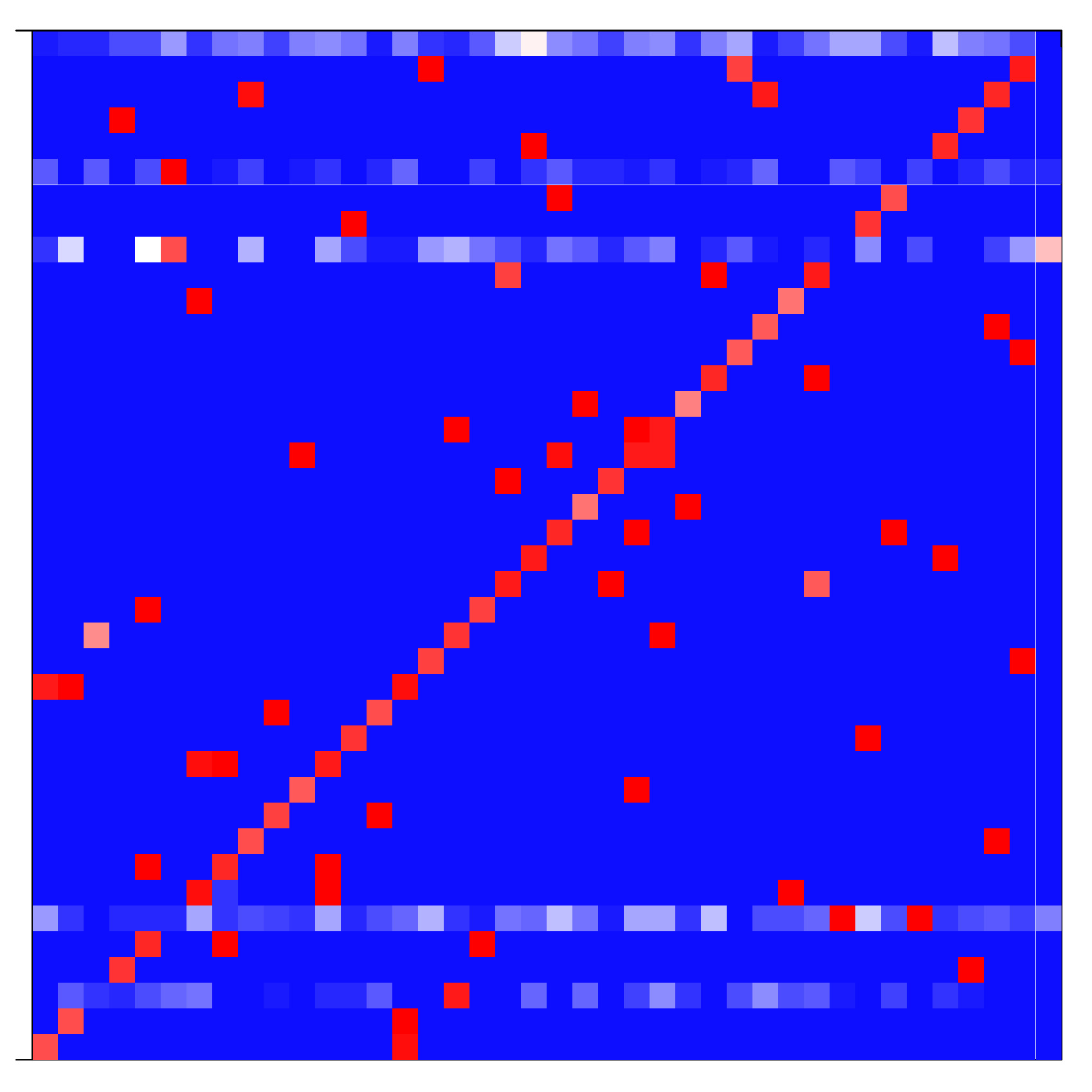} & 	\includegraphics[width= 5 cm]{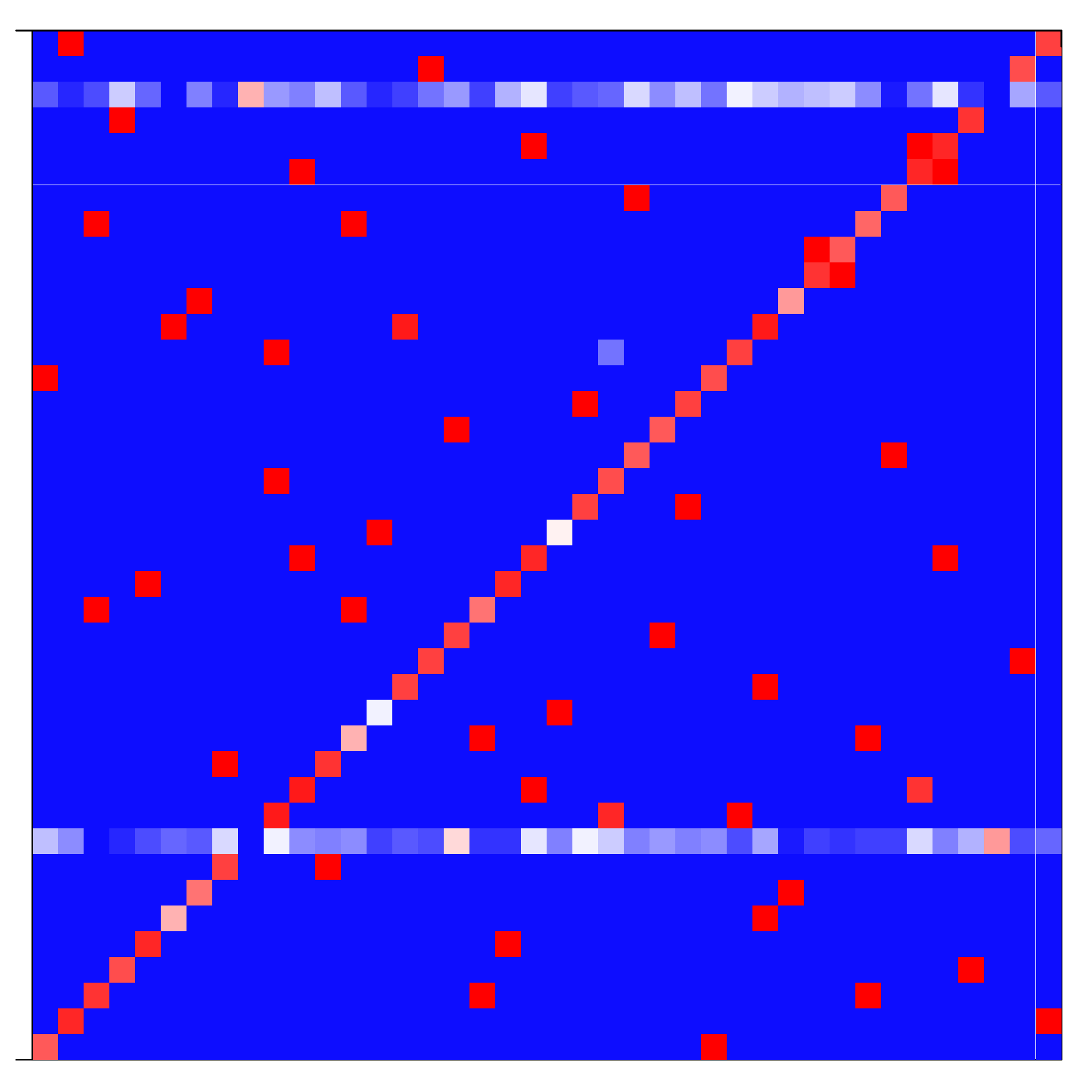}& 	\includegraphics[width= 5 cm]{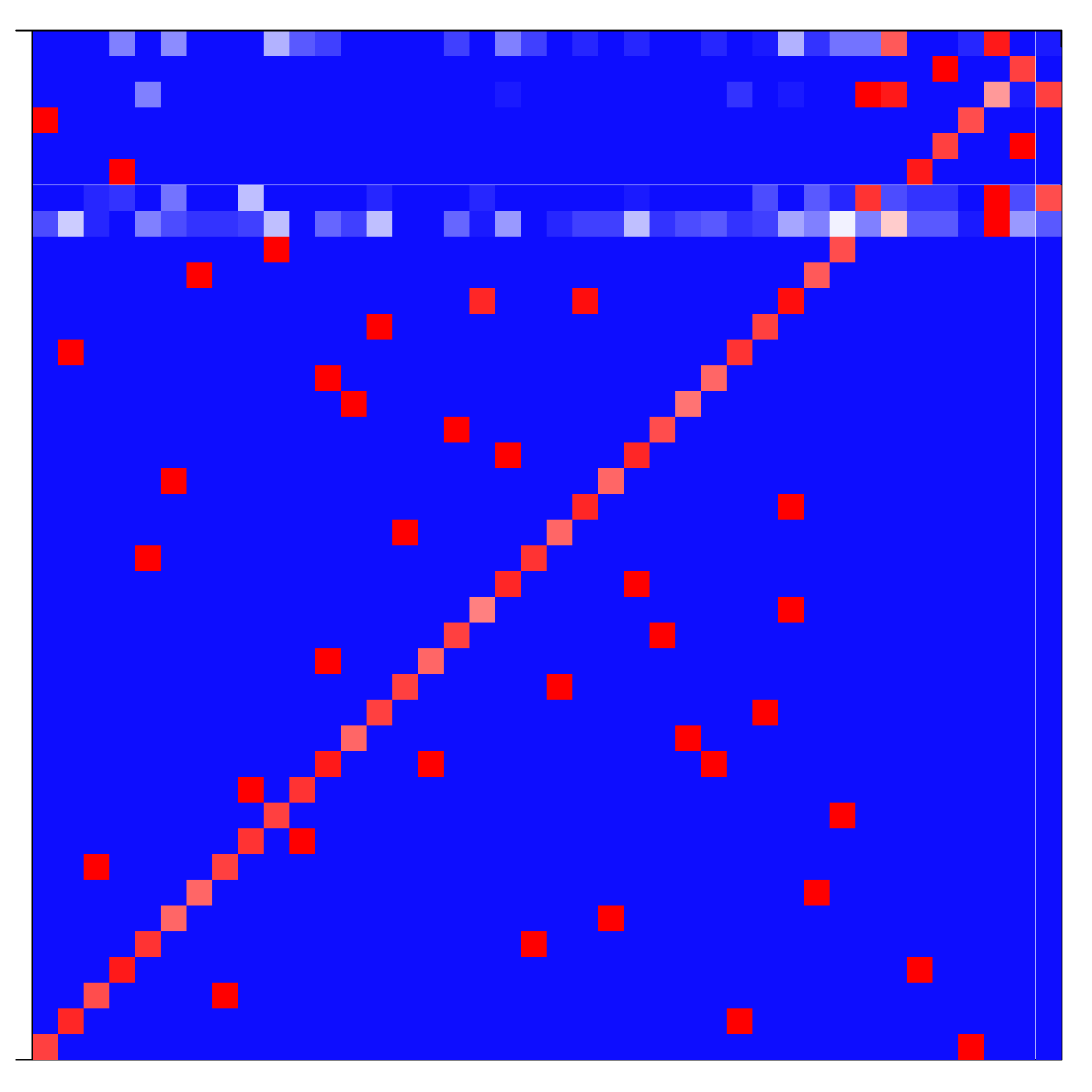} \\
 			\includegraphics[width=5 cm]{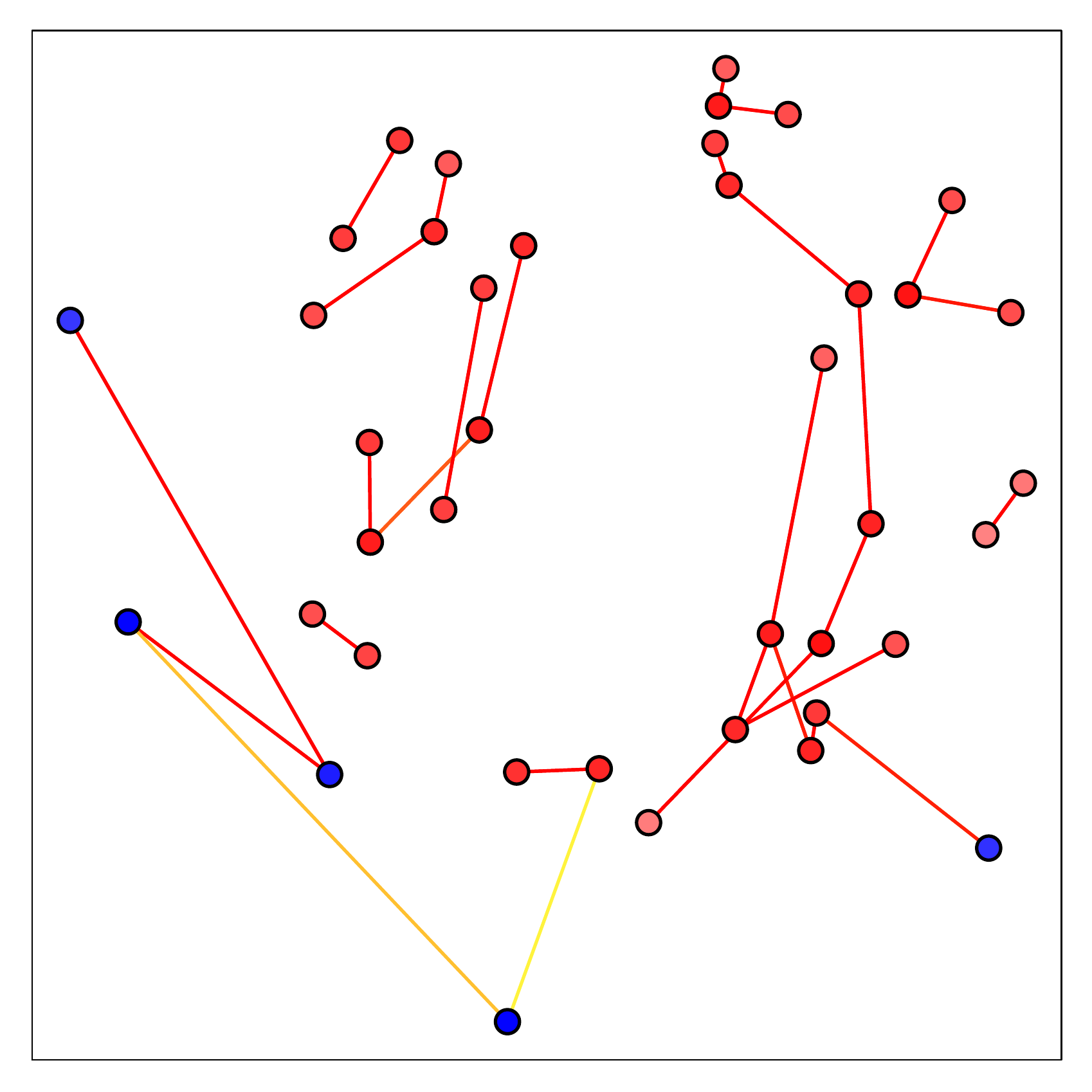} & 	\includegraphics[width= 5 cm]{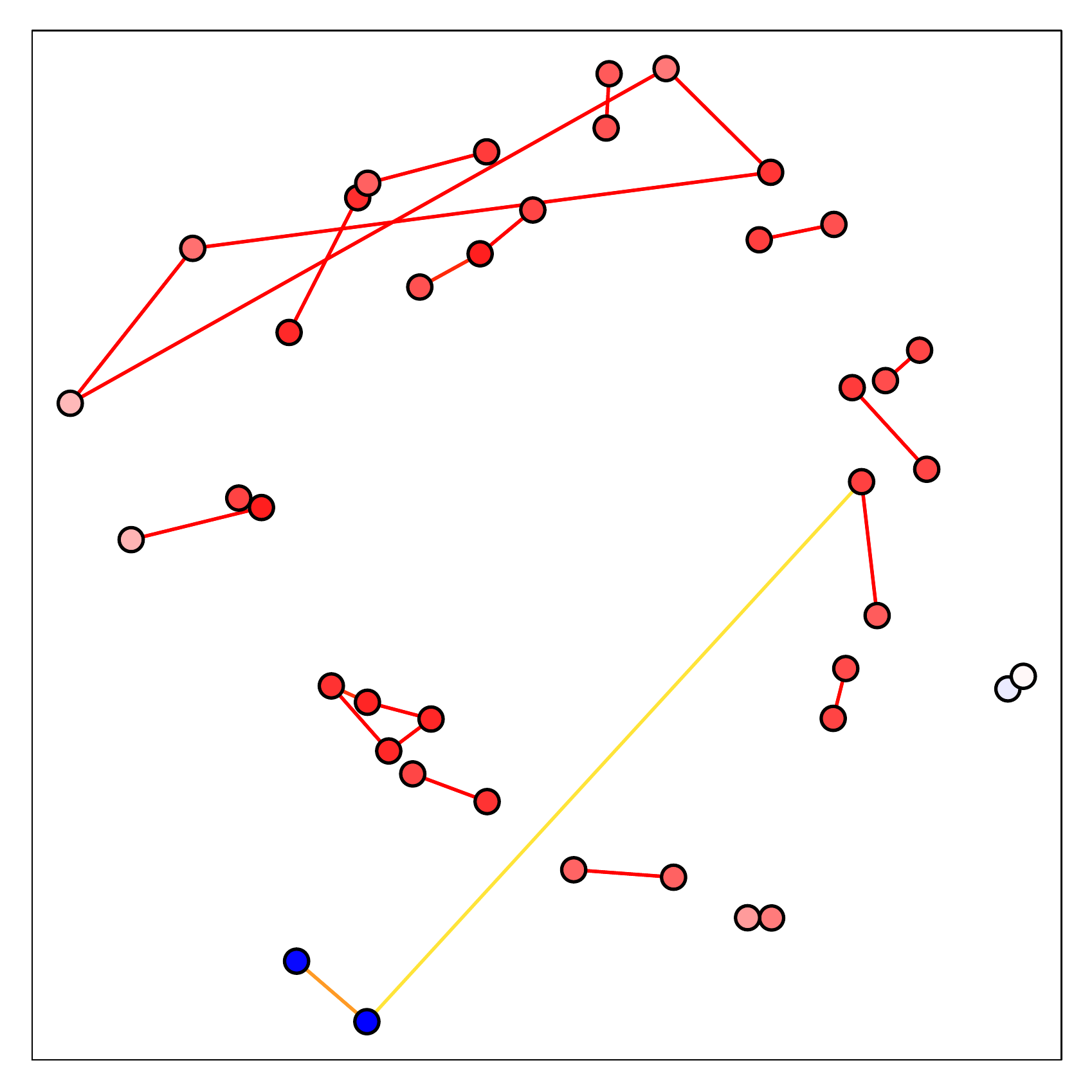}& 	\includegraphics[width= 5 cm]{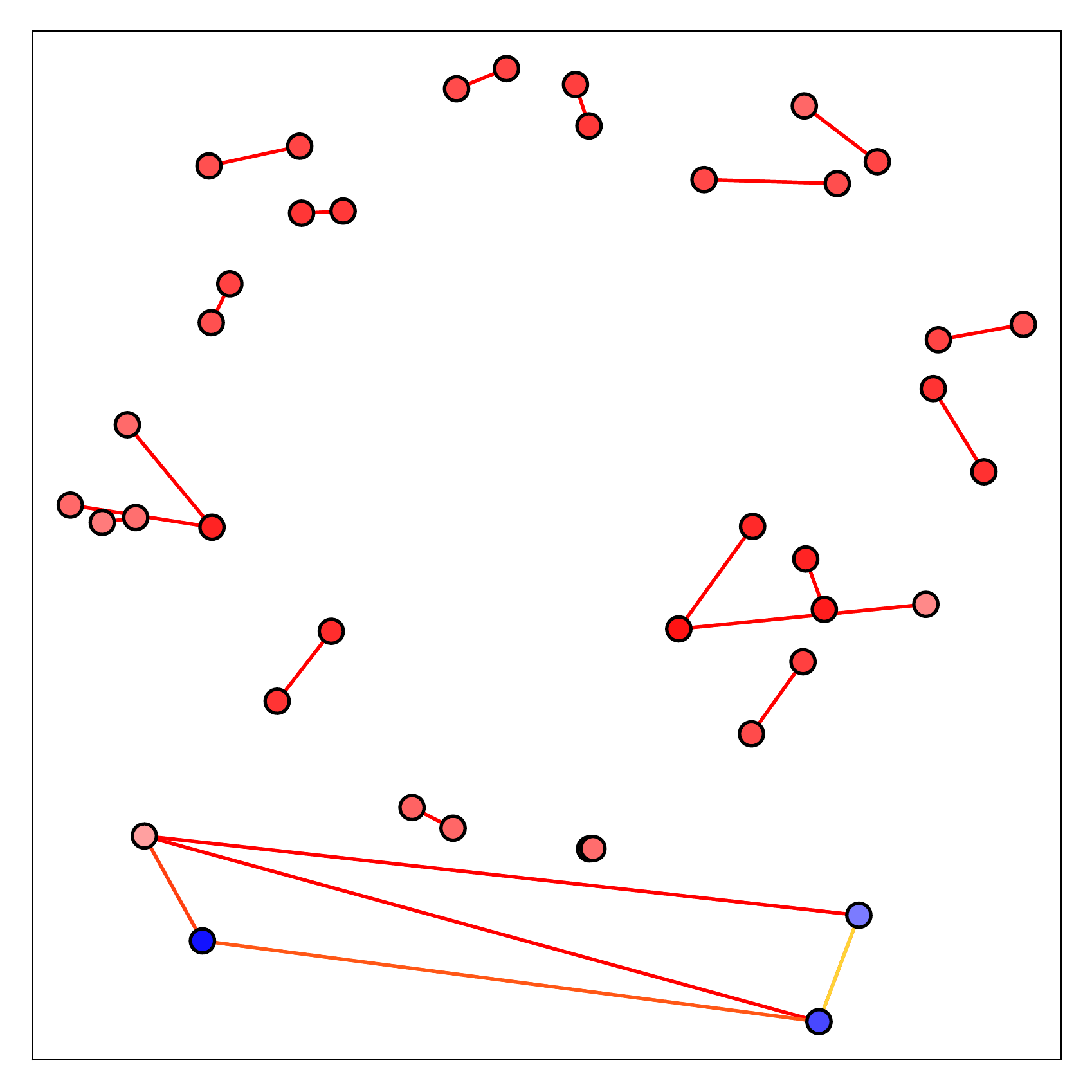} \\
 		$t$ = 1000 & $t$ = 15000  & $t$ = 50000 \\
 	\end{tabular}
 	\caption{Example of mixed pattern between equality and elite. $N = 40$, $\omega = 0.8$ and $\rho = 0.15$, $k = 2$, $\sigma = 0.3$, $\delta = 0.2$. Several individuals, with a low self opinion appear, and the matrix is no more symmetric with respect to the diagonal (horizontal lines appear). See section \ref{representations} for general explanation about the representations.}
 	\label{fig:eliteSW}
 \end{figure}

When the weight of the opinion propagation increases, the equality pattern is modified
as shown on Figure~\ref{fig:eliteSW}, where $\omega = 0.8$, $\rho = 0.15$, in the case $N = 40$, $k = 2$, $\sigma = 0.3$. We observe that the
matrix is no longer symmetric, a few 
horizontal lines appear with a majority of slightly
negative values, the value of the diagonal for these lines being
negative. Moreover, these lines are not stable, their number and positions
vary in time. 
When looking at the network representation, we observe several blue nodes in addition to the red nodes which are connected together. These correspond to the second category agents in the elite pattern. However, contrary to the elite pattern, the second category  agents tend to establish connections
between them and then, they tend to increase
their self opinion, and they may come back to the elite (not shown
on the figure). Meanwhile, other elite agents can get a negative self-opinion and then her opinion about all the others become less negative, then loose all their friends and become second category. Globally the number of agents of each ring fluctuates around a constant
average value over the time.   This pattern can be considered as transitory between equality and elite.
The model satisfies the rules of classification for elite when one agent of the second category is disconnected from all the others. This rule may appear too simple because in this case, there are more elite than second category agents, which is not the idea we have in general of an elite. This could be refined in further studies of the model.

\subsubsection{Second example of mixed pattern: dominance and crisis}

\begin{figure}
 	\centering
 	\begin{tabular}{ccc}
 	\includegraphics[width=5 cm]{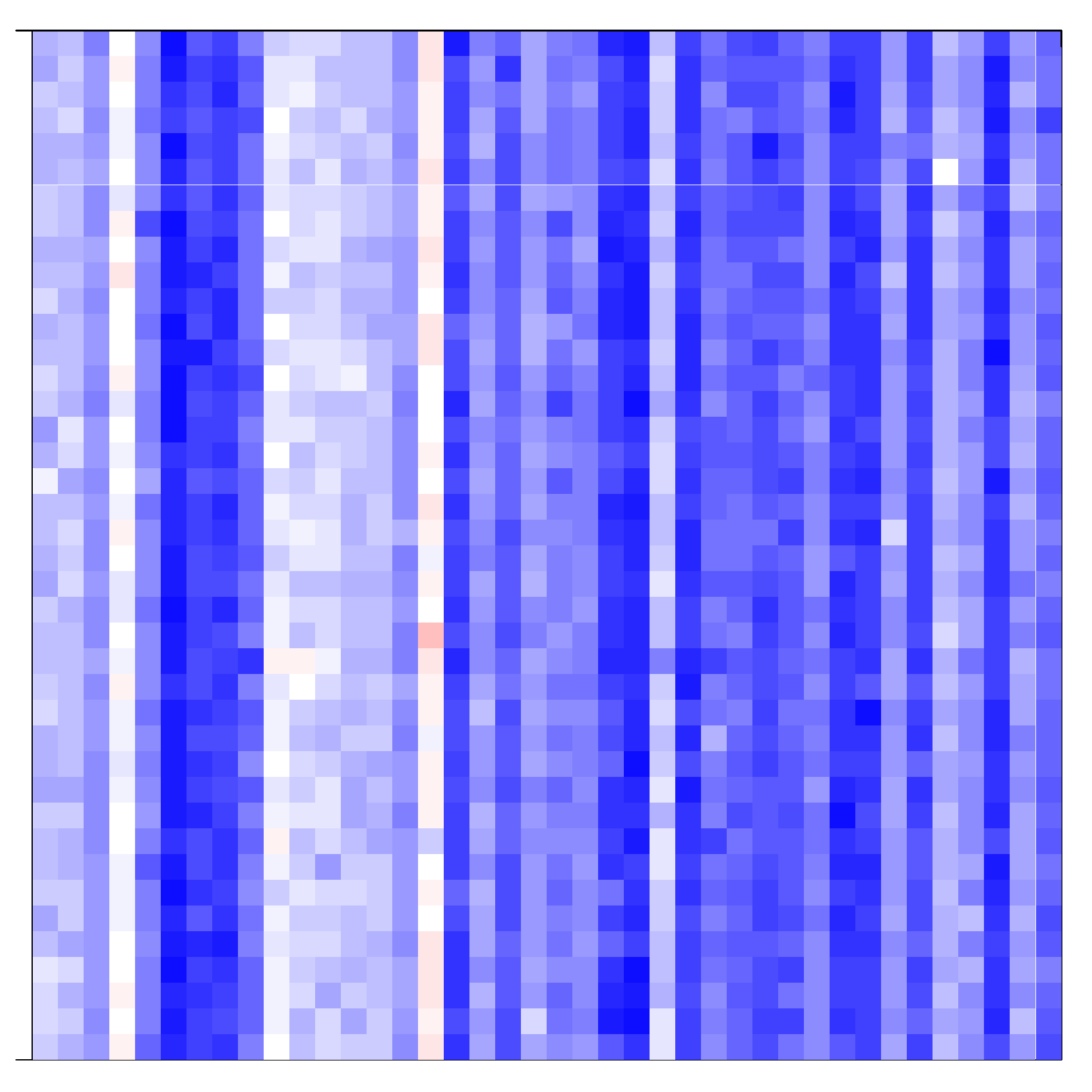} & 	\includegraphics[width= 5 cm]{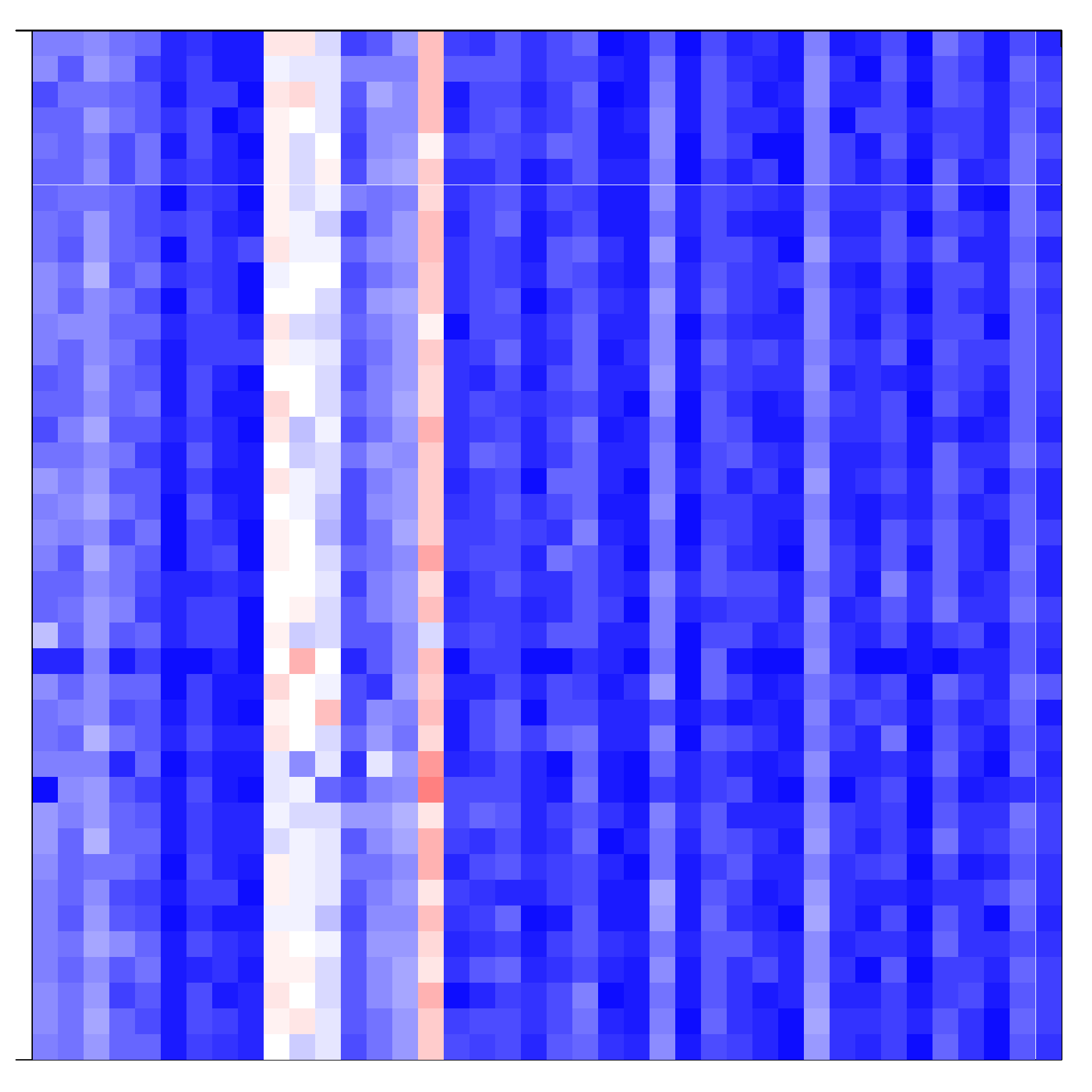}& 	\includegraphics[width= 5 cm]{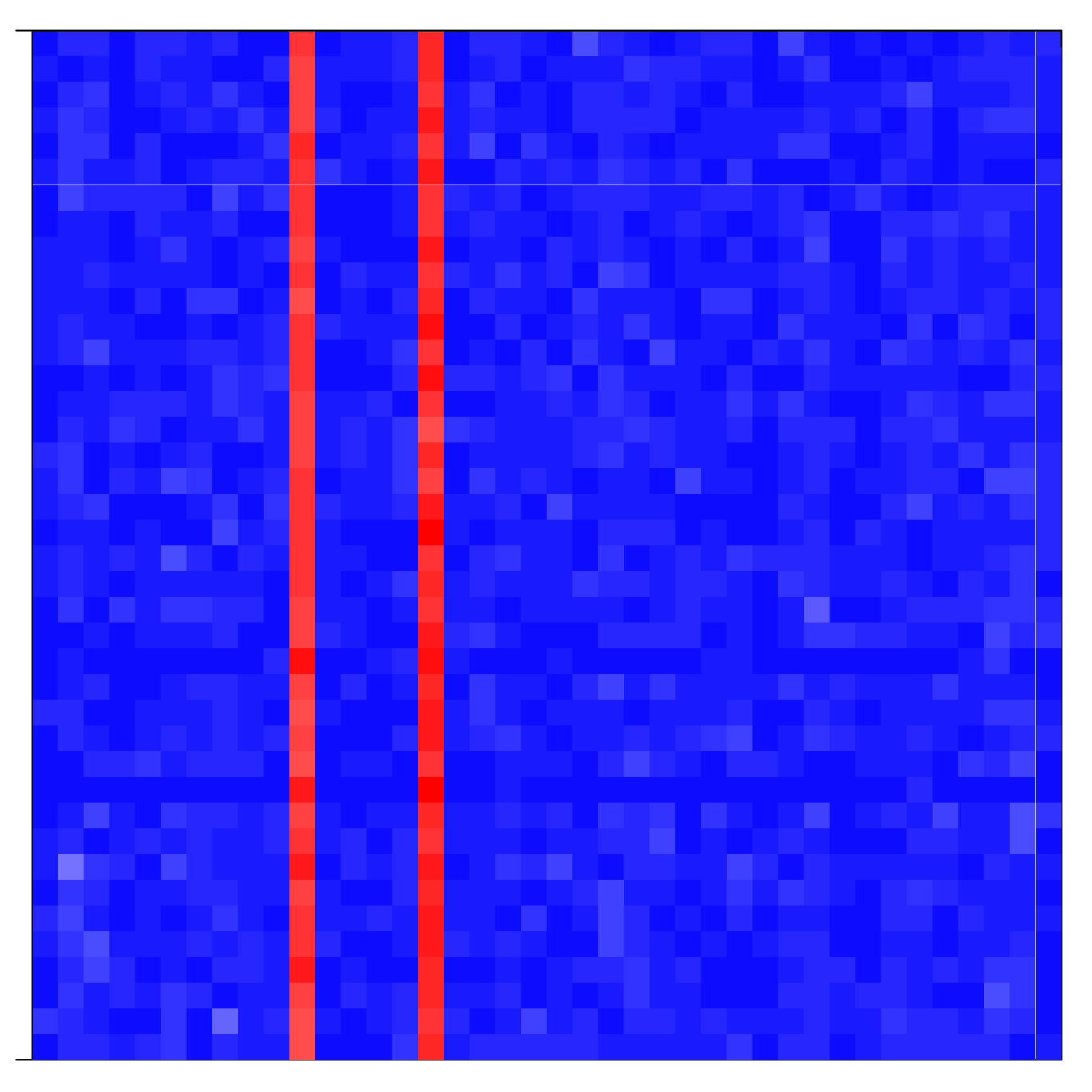} \\
 		\includegraphics[width=5 cm]{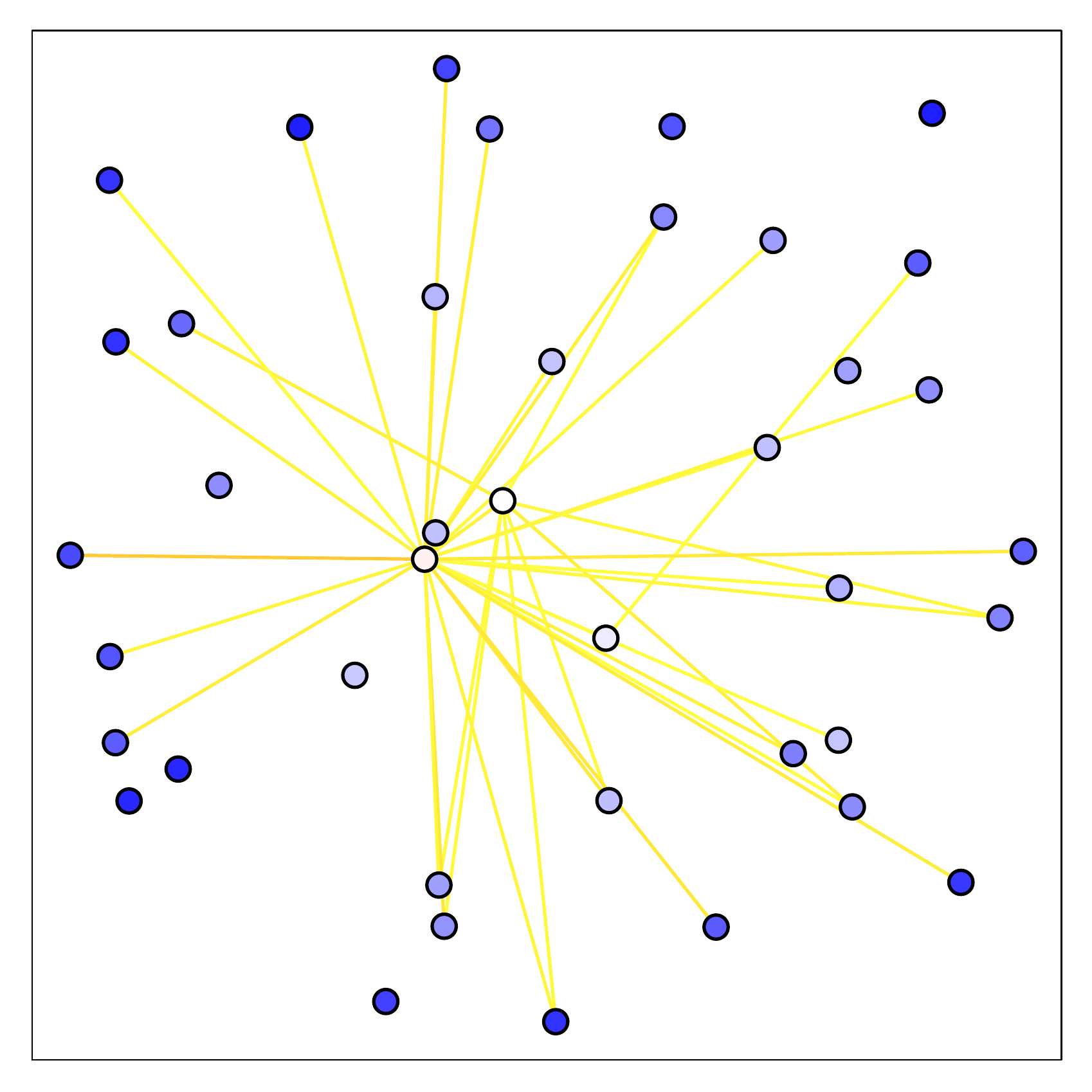} & 	\includegraphics[width= 5 cm]{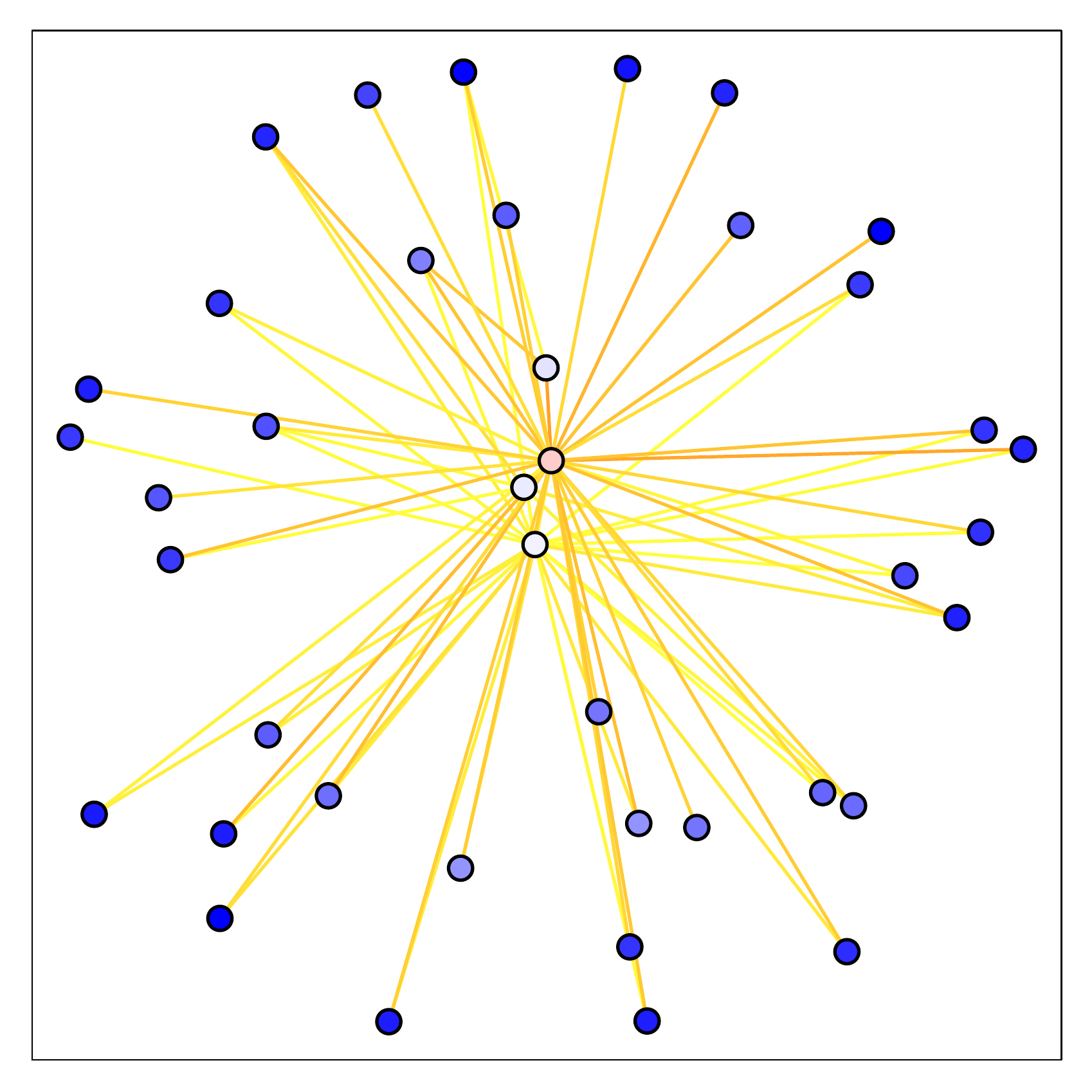}& 	\includegraphics[width= 5 cm]{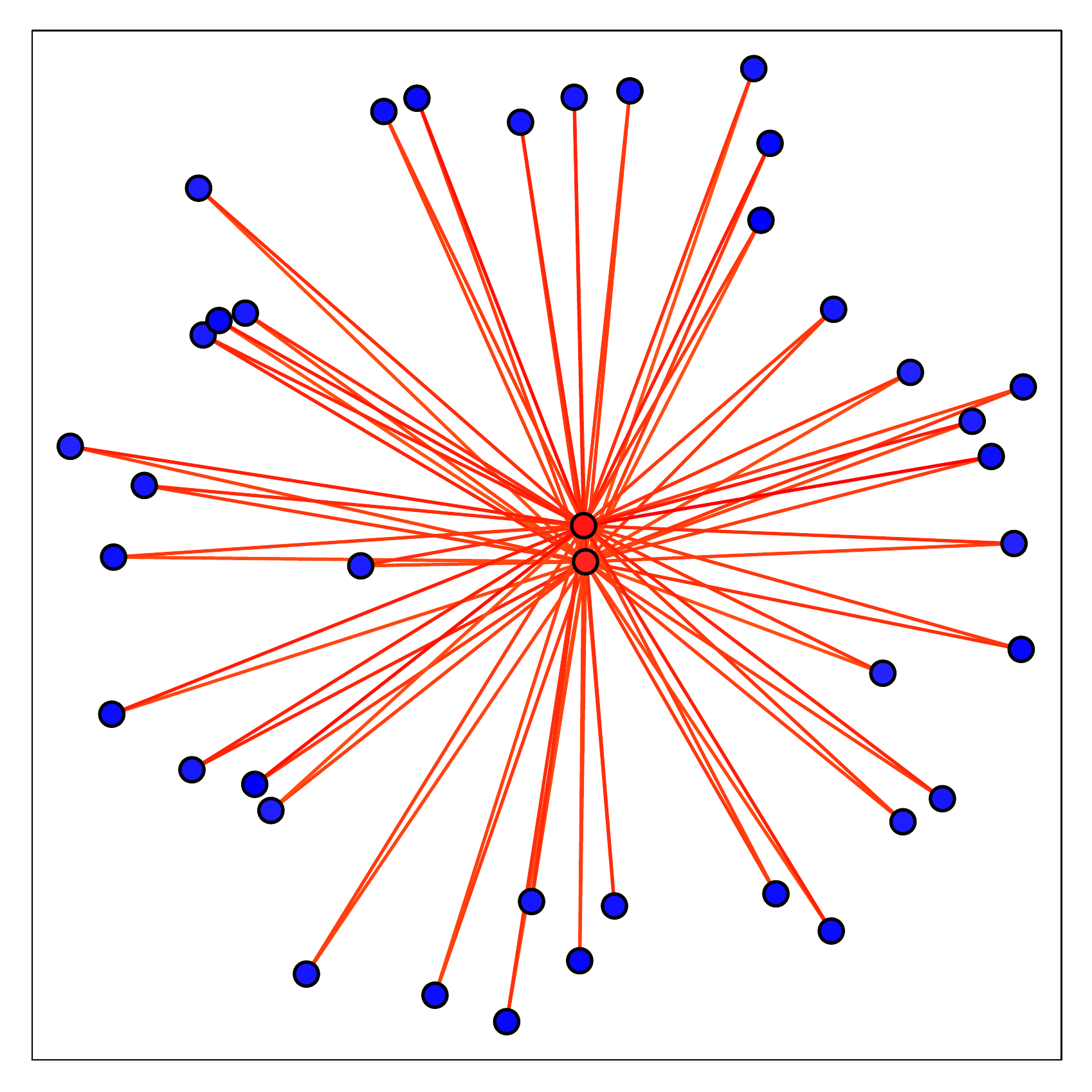} \\
 		$t$ = 5000, $D = 0.08$,  & $t$ = 6000, $D = 0.3$, & $t$ = 6500, $D = 0.83$ \\
 		\includegraphics[width=5 cm]{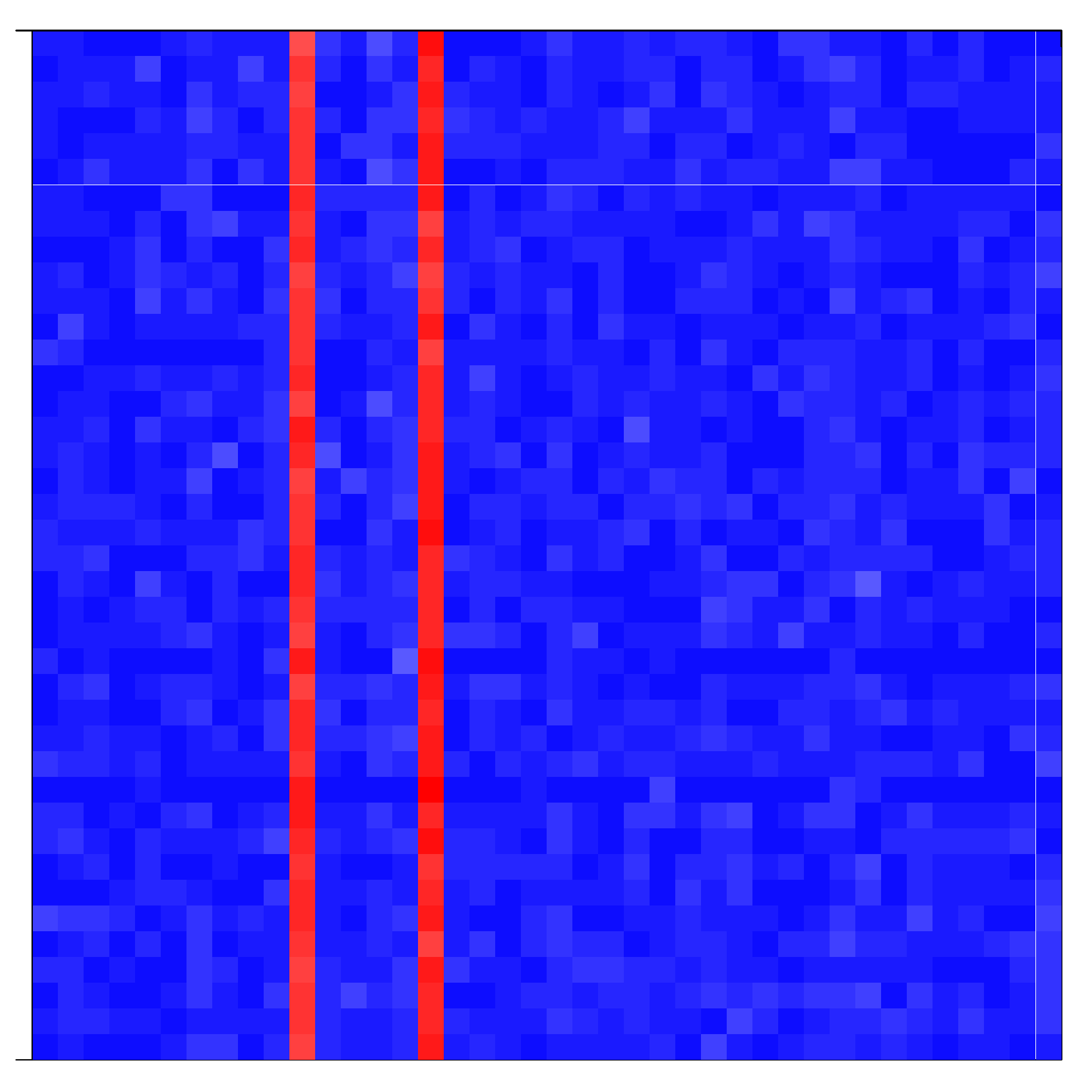} & 	\includegraphics[width= 5 cm]{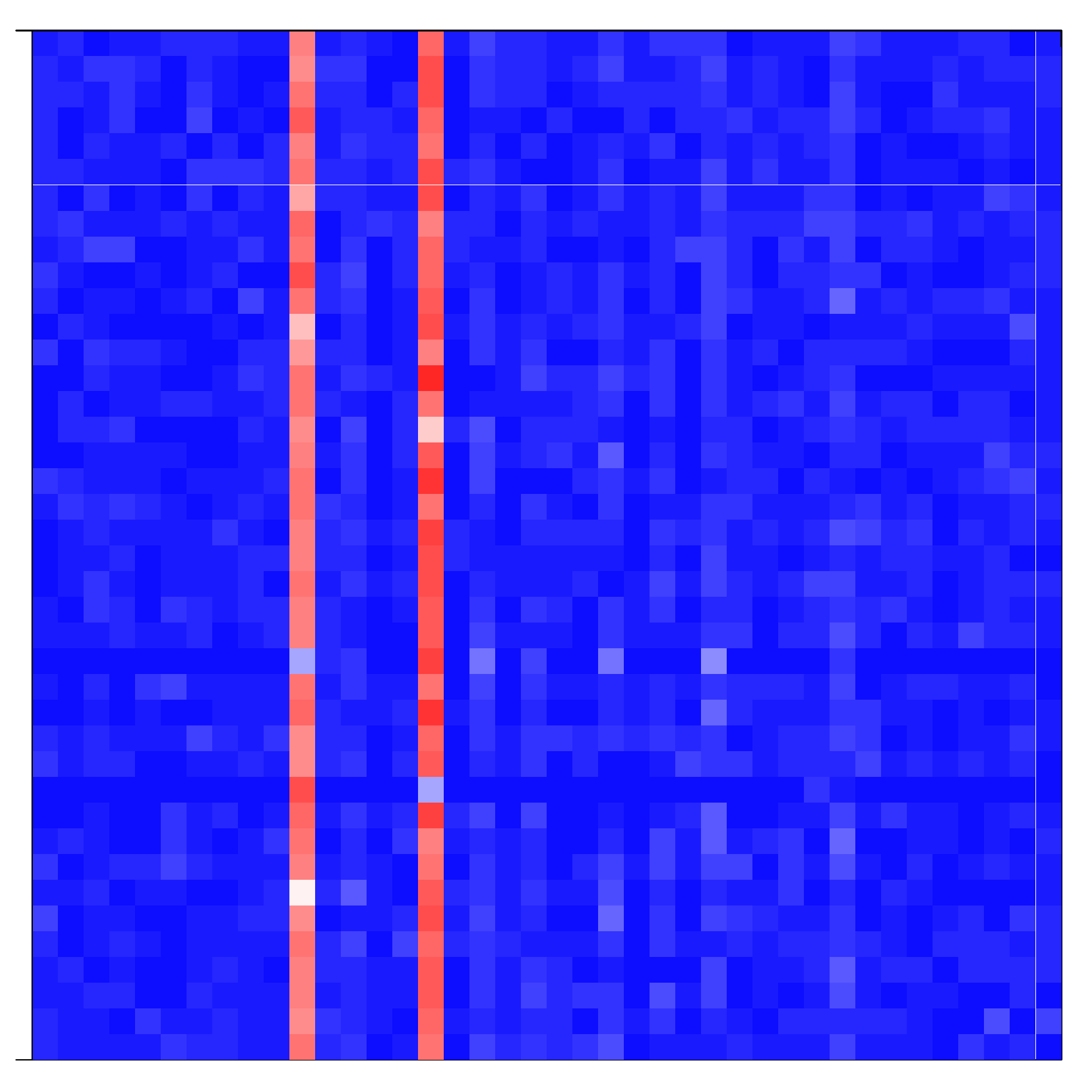}& 	\includegraphics[width= 5 cm]{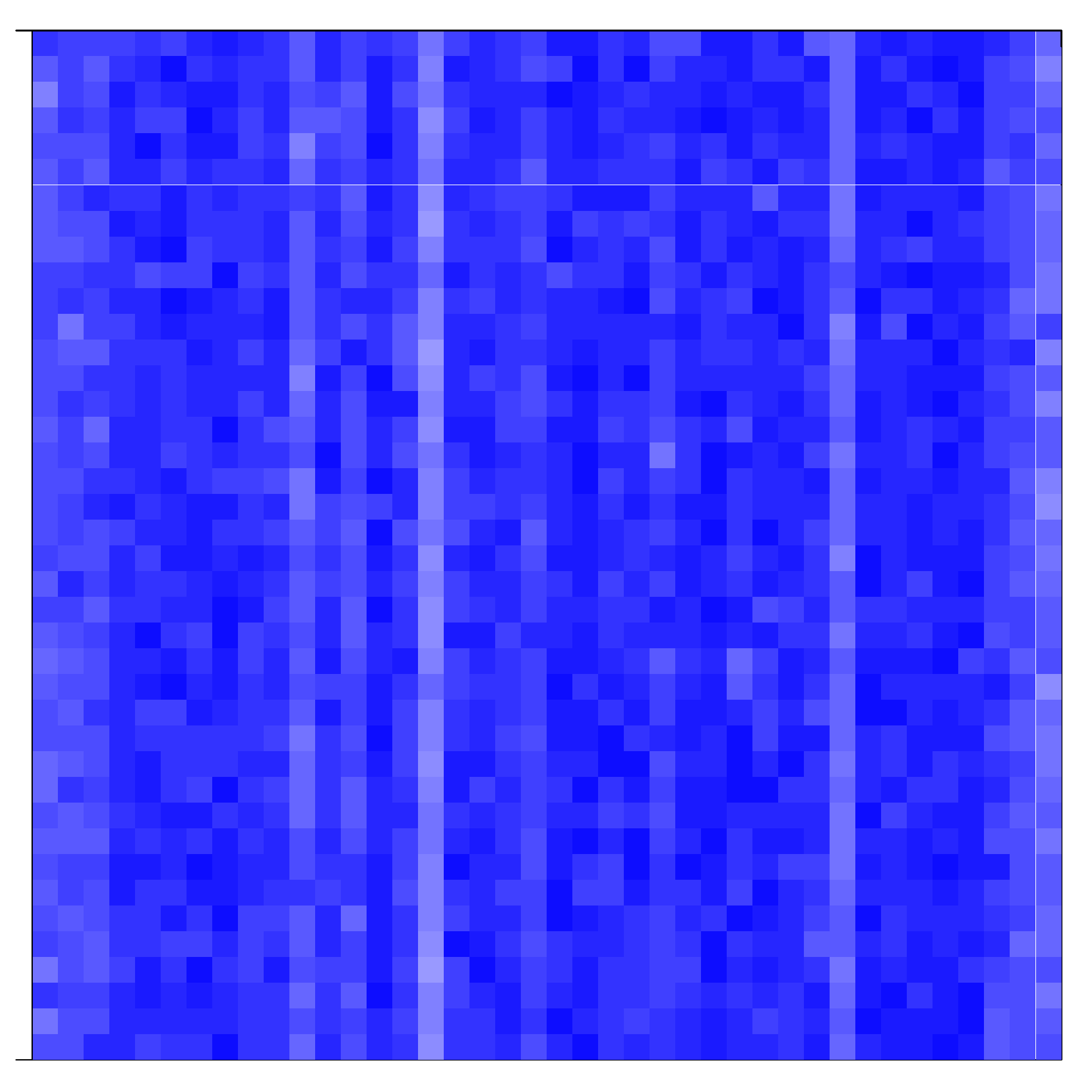} \\
 			\includegraphics[width=5 cm]{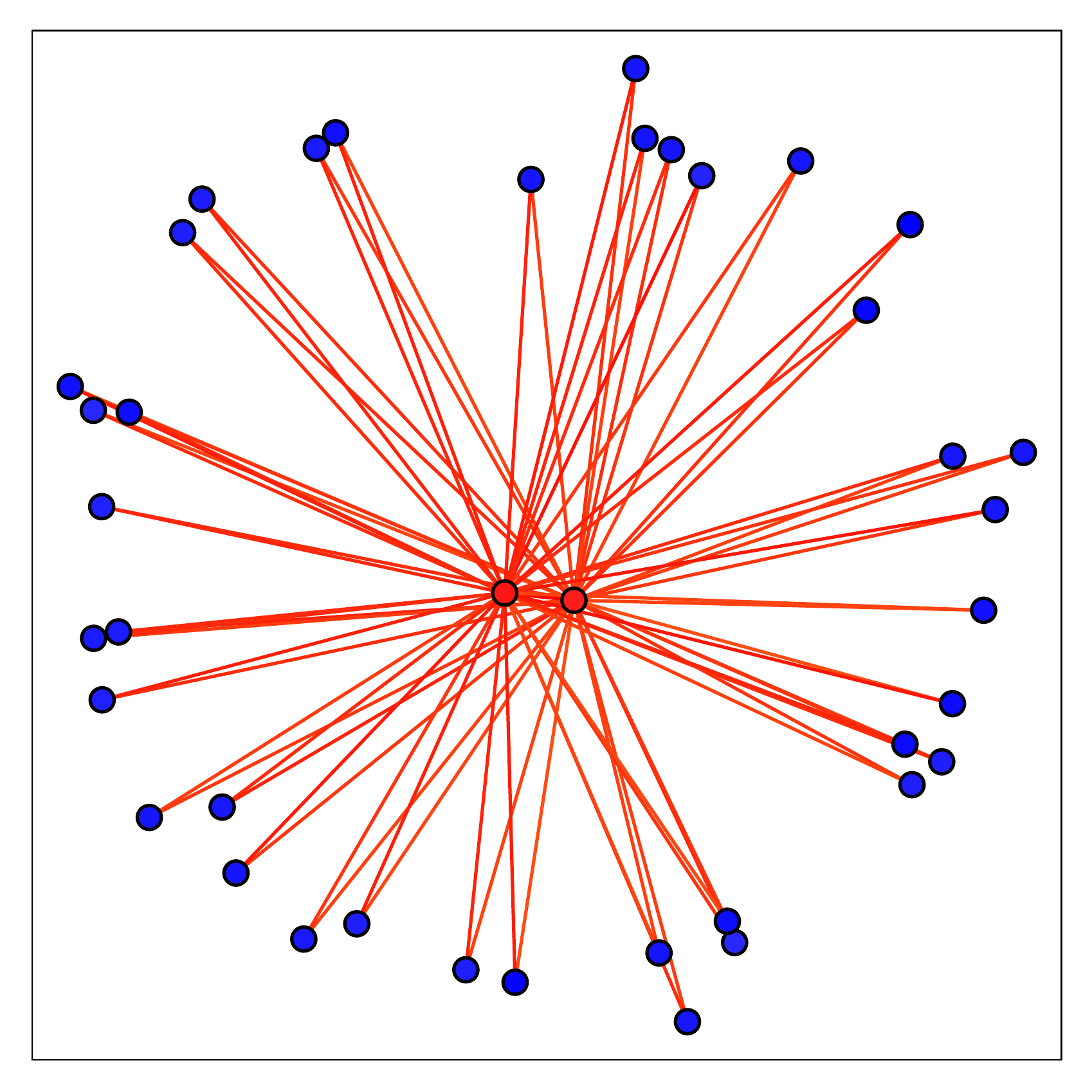} & 	\includegraphics[width= 5 cm]{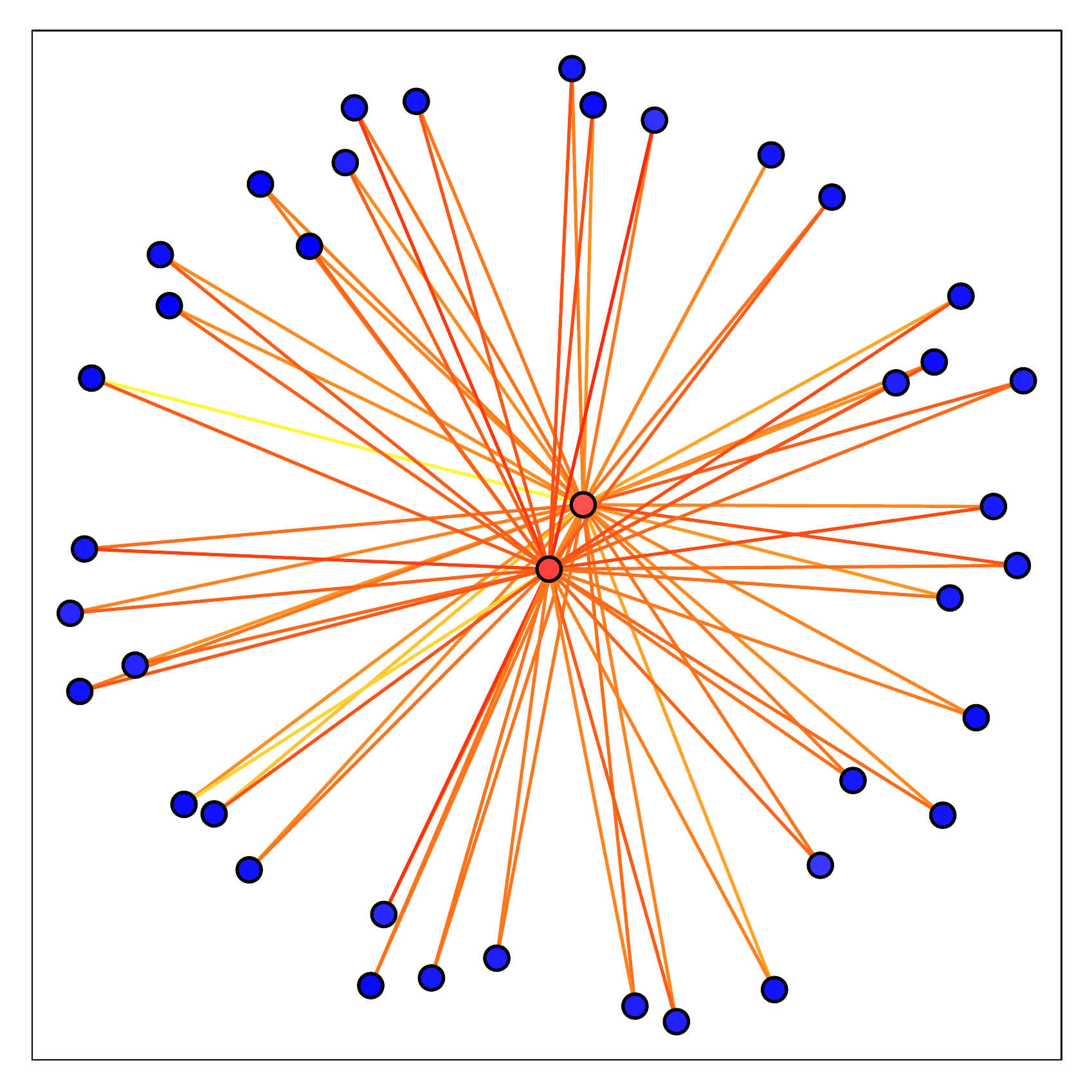}& 	\includegraphics[width= 5 cm]{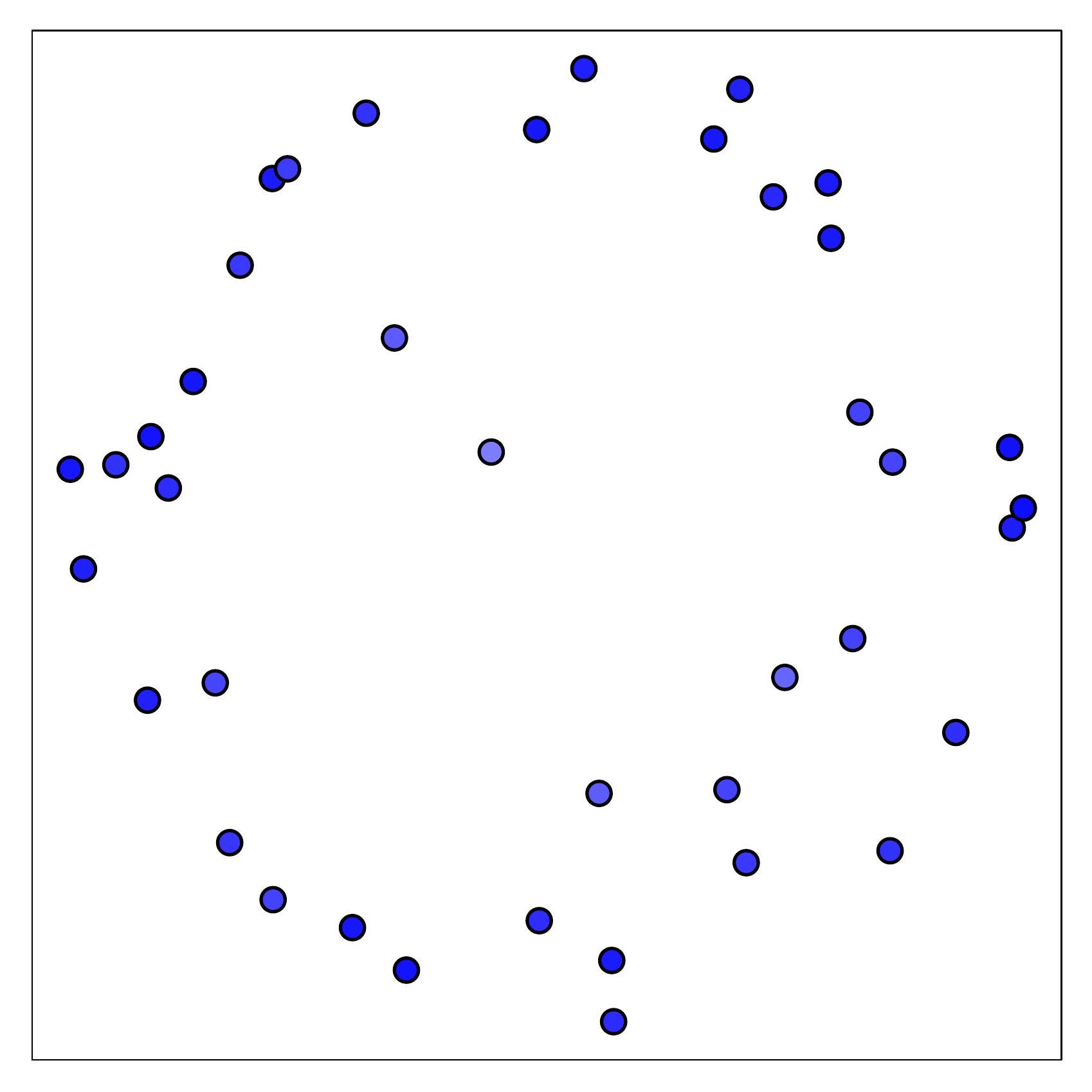} \\
 		$t$ = 7000, $D = 0.81$, & $t$ = 7500, $D = 0.69$, & $t$ = 9000, $D = 0.11$,  \\
 		 	\end{tabular}
 	\caption{Example of mixed pattern between dominance and crisis. Study around the first peak of the dominance indicator for $\omega = 0.5$, $\rho = 0.3$, $N = 40$, $k = 10$, $\sigma = 0.3$, between iterations 5000 and 9000. In a first phase, two agents reinforce each other to dominate the whole population and in the second phase they compete and destroy each other's reputation. See section \ref{representations} for general explanation about the representations. }
 	\label{fig:bigFluct}
 \end{figure}

With this pattern (for instance for $\omega = 0.5$ and $\rho = 0.3$, for $N = 40$, $k = 10$, $\sigma = 0.3$), we observe periods of strong dominance by one or two agents, followed by periods of crisis, all the agents having very negative opinions of all the others and also about themselves.
Figure~\ref{fig:bigFluct} shows the matrix and network representations of six states
around the first peak of dominance. From iteration $6500$ to iteration $7500$,
two individuals, appearing as red columns in the matrix representation whereas all the other columns are blue and in red nodes located in the centre of the network representations, are valued highly positively (red links) by all the others which are colored in blue and located far from the circle center. After the dominance period, there are no more positive links, and all the nodes are blue (negative self value).

Other dominance episodes start with only one dominant individual. In this case, the dominance tends to last longer and it can lead to progressive hierarchy that are similar to the ones observed in the hierarchy pattern. The dominance ends when a second dominant individual emerges and reaches the same level of reputation as the first dominant one. This situation is very instable and it leads rapidly to back to the crisis.

\section{Analytical study}

\subsection{When vanity dominates} 
\label{explanation1-2}

\subsubsection{Case of $\rho = 0$.}

\begin{figure}
 	\centering
 	\begin{tabular}{ccc}
 		\includegraphics[width=5 cm]{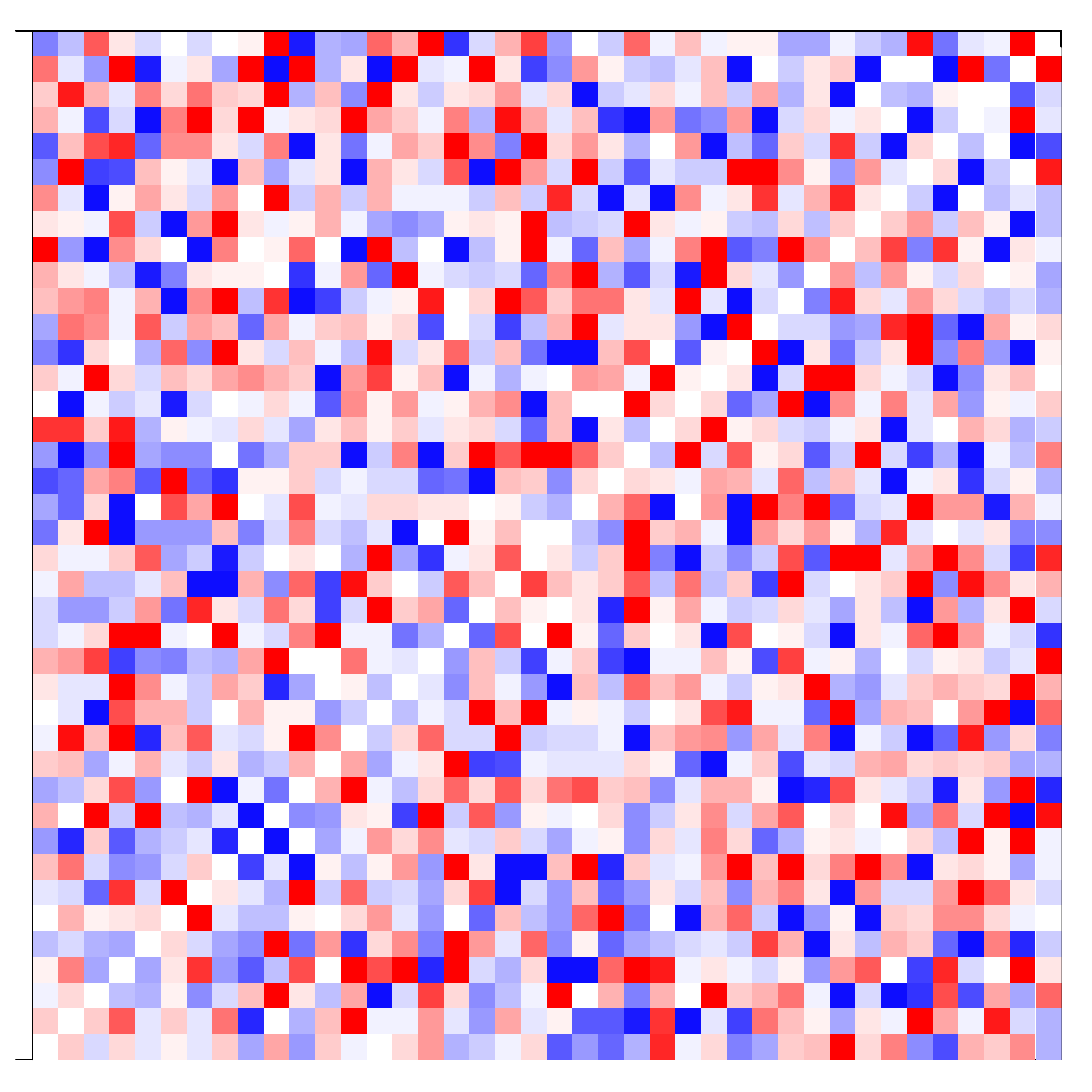} & 	\includegraphics[width= 5 cm]{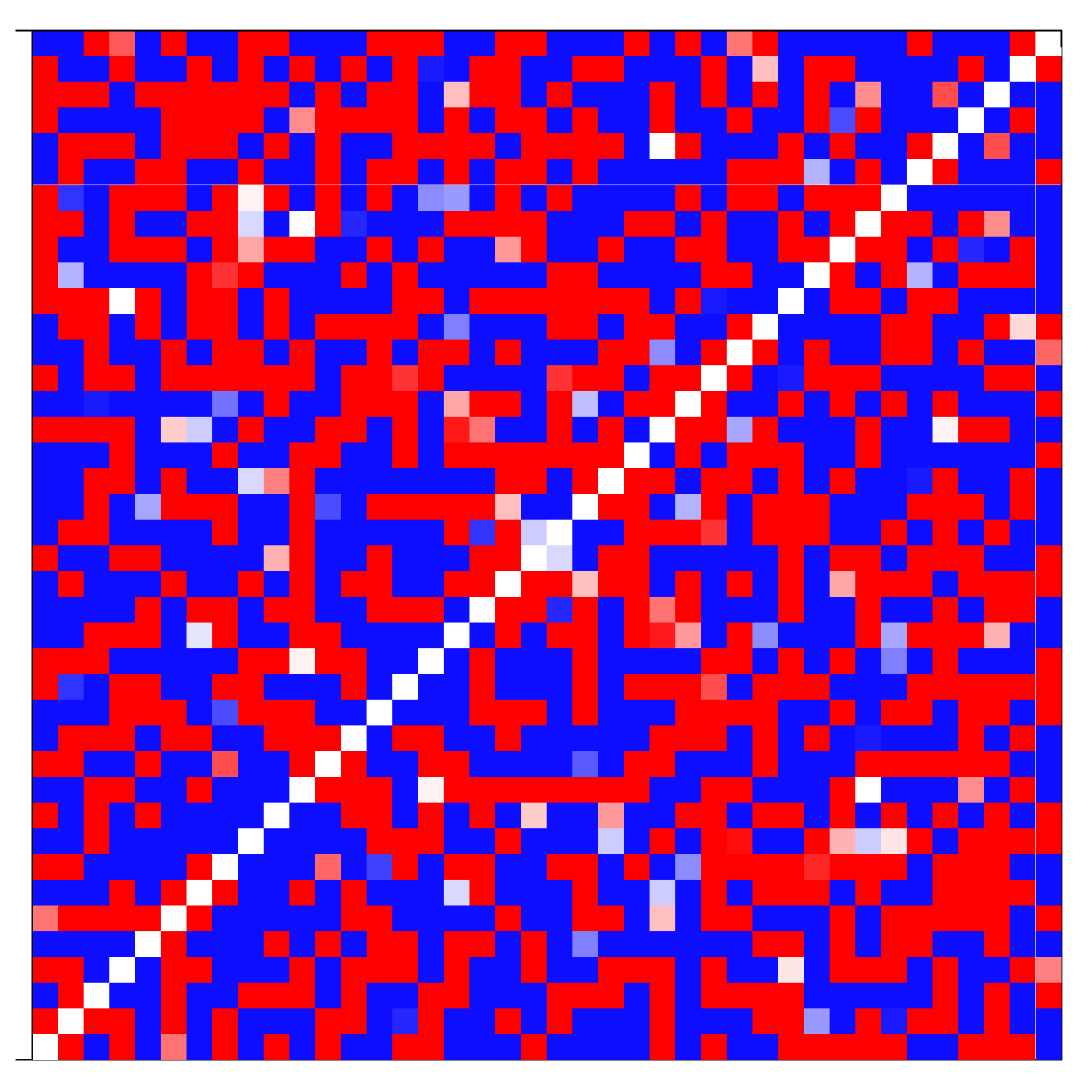}& 	\includegraphics[width= 5 cm]{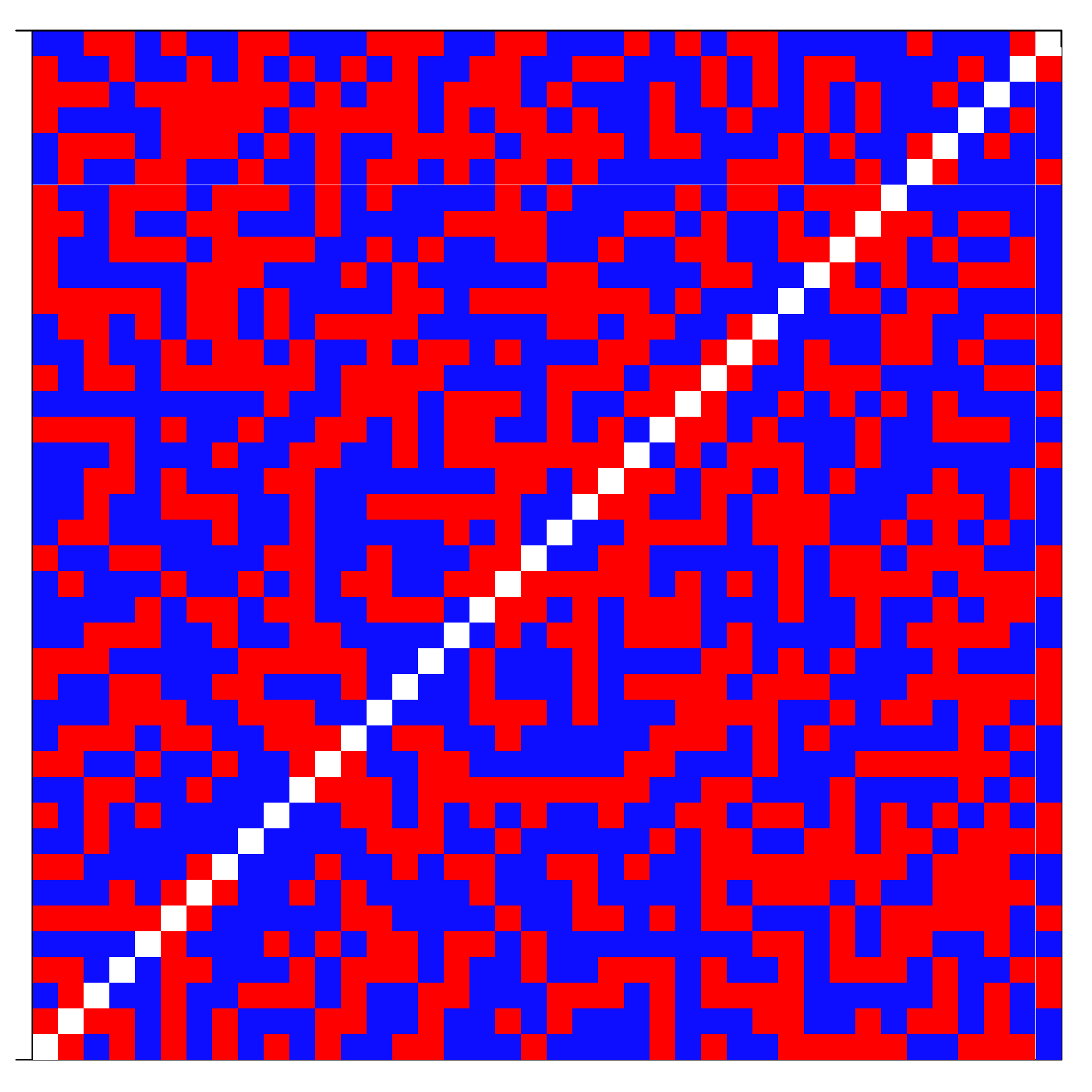} \\
 		\includegraphics[width=5 cm]{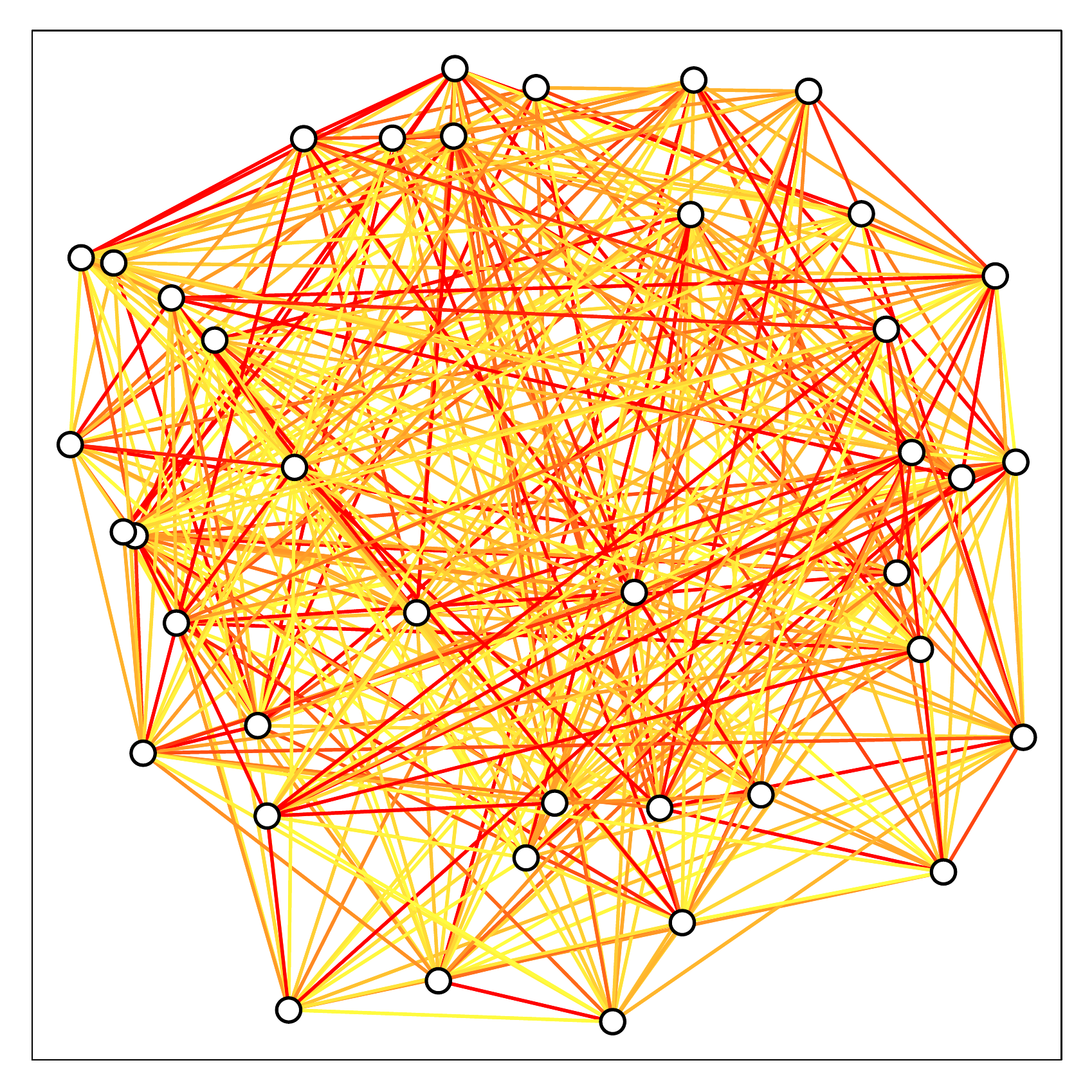} & 	\includegraphics[width= 5 cm]{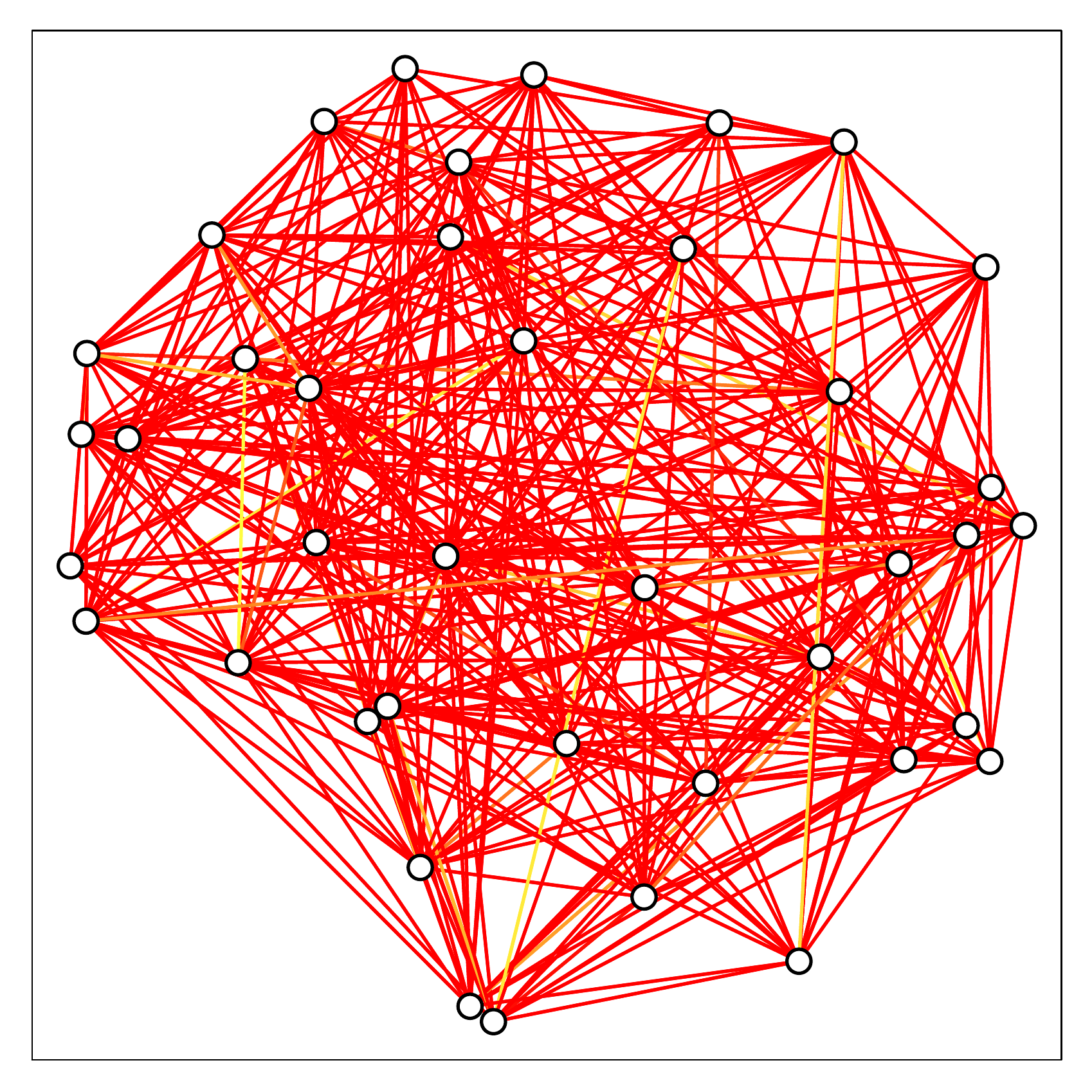}& 	\includegraphics[width= 5 cm]{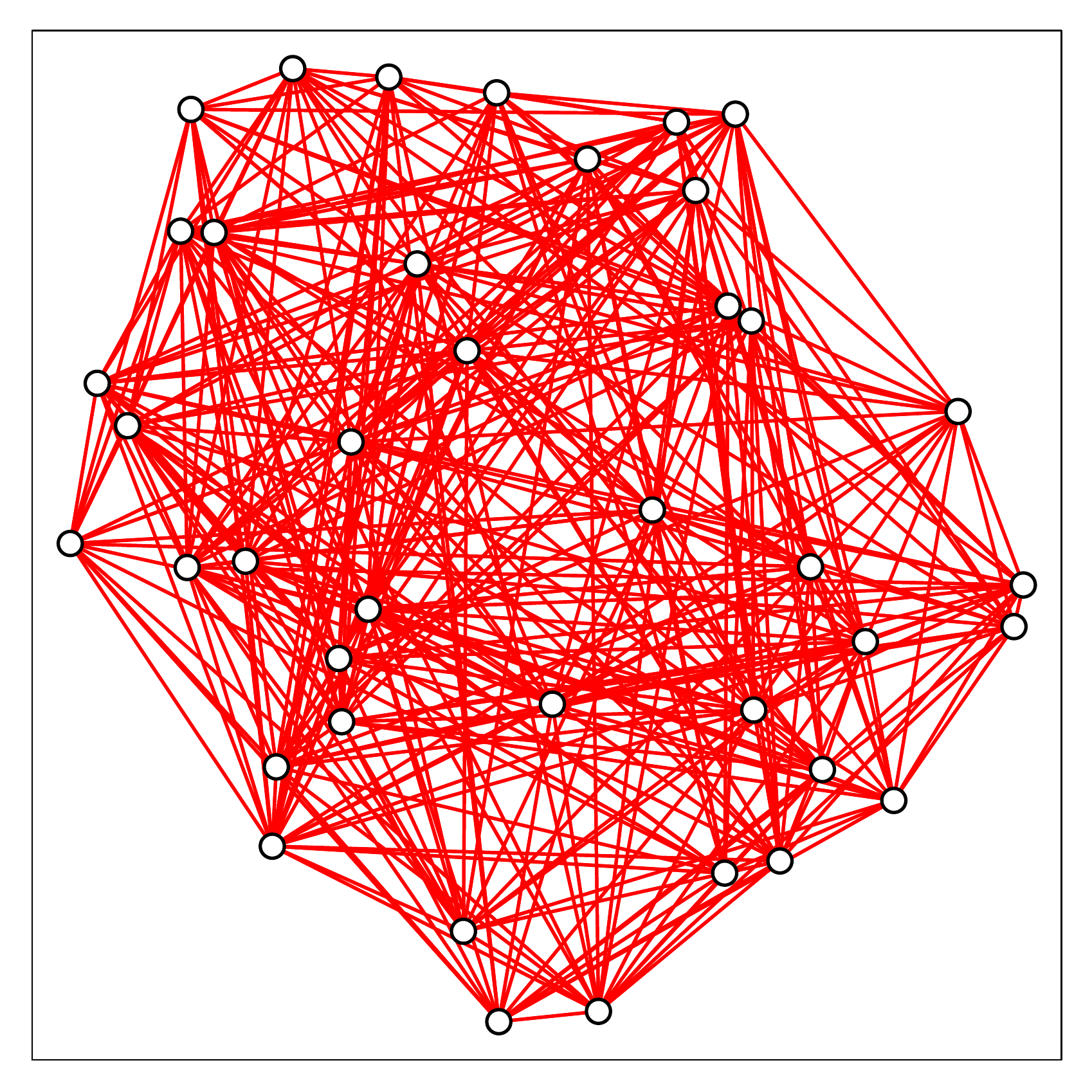} \\
 		$t$ = 200, $\bar{a} = -0.02$ & $t$ = 500, $\bar{a} = -0.07$ & $t$ = 1000, $\bar{a} = -0.07$
 	\end{tabular}
 	\caption{Vanity only. $\rho = 0$, $\omega = 0.4$. The matrix remain symmetric and each agent splits the population between friends and foes. See section \ref{representations} for general explanation about the representations.}
 	\label{fig:vanityOnly}
 \end{figure}

In order to better understand the patterns when $\rho$ is small compared with $\omega$ (i.e. when vanity dominates), we simulate the model with the vanity dynamics alone, i.e. $\rho = 0$, starting
with all agents not knowing each other, hence with all the opinions set to {\em nil}. We get the evolution represented on Figure~\ref{fig:vanityOnly}. After relatively few time steps, all the opinions become either $-1$ or $+1$, except the self opinions which change from {\em nil} to $0$ at the first interactions, and then remain at $0$ because no change occurs on the self opinions in the vanity dynamics (the diagonal of the matrix representation remains white). Each agent separates the population into two groups of almost equal size: the ones she hates and the ones she loves. We also note that the diagonal is a symmetry axis
of the matrix: if I love you, you love me, or if I hate you, you hate me. Note finally that the average of all opinions over the whole population is close to $0$. 
	
We now propose some analytical explanations of the above result. 
By definition of the vanity process, and because $a_{i,i} = a_{j,j} = 0$, when agents $i$ and $j$ meet, the opinions evolve as follows:
\begin{equation}
\begin{cases}
a_{i,j}(t+1) \leftarrow a_{i,j}(t) + \omega ( a_{j,i}(t) + \texttt{Random}(-\delta,\delta)),\\
a_{j,i}(t+1) \leftarrow a_{j,i}(t) + \omega ( a_{i,j}(t) + \texttt{Random}(-\delta,\delta)).
\end{cases}
\end{equation}
From these equations, we can derive the following points:
\begin{itemize}
	\item  If we have, for a couple $(i,j)$ with $i \neq j$, at a given iteration $t$, $a_{i,j}(t) > \delta$ and $a_{j,i}(t) > \delta$, then $a_{j,i}(t) + \texttt{Random}(-\delta,\delta) > 0$ and $a_{i,j}(t) + \texttt{Random}(-\delta,\delta) > 0$. Hence, in this case, after iteration $t$, both $a_{i,j}$ and $a_{j,i}$ keep increasing until they are truncated to 1. 
 \item Similarly, if $a_{i,j}(t) < -\delta$ and $a_{j,i}(t) < -\delta$, then both $a_{i,j}$ and $a_{j,i}$ keep decreasing afterwards until they are truncated to $-1$.
 \item If $a_{i,j}(t) < -\delta$ and $a_{j,i}(t) > \delta$, then at the next steps $a_{i,j}(t)$ will increase and $a_{j,i}(t)$ will decrease. Of course we get the same when interverting $i$ and $j$.
\end{itemize}
This allows us to describe the evolution of the reciprocal opinions of each couple $(i,j)$ in two phases:
\begin{itemize}
	\item In a first phase, when the absolute values of $a_{i,j}$ and $a_{j,i}$ are smaller than $\delta$, the effect of randomness is dominating, and the opinions tend to be close to random walks. When both opinions get close to $\delta$, these random walks are biased towards common growth, when they both get close to $-\delta$ , they are biased towards common decrease, and if their signs are different, the opinions tend both to go to $0$.
	\item As a result of these random processes each opinion couple $(a_{i,j}, a_{j,i})$ necessarily ends up either by being both higher than $\delta$ or both lower than $-\delta$, with an equal probability. This leads to the opinions couples being both at +1 or -1 in the end.
\end{itemize}

\subsubsection{Analysis of the equality pattern}

This analytical study helps formulate qualitative explanations of the equality pattern. 
First, we make the hypothesis that the dynamics of opinion propagation is
negligible except on the diagonal of the matrix, because out of the diagonal
the vanity dynamics are dominating. 
The evolution of the opinions outside the diagonal is mostly driven by the vanity equations:
\begin{equation}
\begin{cases}
a_{i,j}(t+1) \leftarrow a_{i,j}(t) + \omega ( a_{j,i}(t) - a_{i,i}(t) + \texttt{Random}(-\delta,\delta)) \\
a_{j,i}(t+1) \leftarrow a_{j,i}(t) + \omega ( a_{i,j}(t) - a_{j,j}(t) + \texttt{Random}(-\delta,\delta)).
\end{cases}
\end{equation}
If all the values of the diagonal are positive, we see from these equations that this tends to decrease on average the values of $a_{i,j}$ compared with the case of $\rho = 0$, which explains why these values are more frequently negative.
It remains to be explained why the values of the diagonal are positive. Suppose agent $j$ propagates her opinion about $i$ to $i$ herself, we have:
\begin{equation}
a_{i,i}(t+1) \leftarrow a_{i,i}(t) +  p_{i,j} \rho (a_{j,i}(t) - a_{i,i}(t) + \texttt{Random}(-\delta,\delta)).
\label{eq:influence}
\end{equation}
We note that $p_{i,j}$ is higher than 0.5 when $a_{j,i} - a_{i,i}$ is positive. On the contrary, $p_{i,j}$ is lower than 0.5 when $a_{j,i} - a_{i,i}$ is negative. Therefore, the propagation of opinions favours the positive contributions to the diagonal, explaining why this diagonal is positive.

Qualitatively, the agents love those who flatter their ego and believe more those they love, hence they believe those who flatter them. As a result, the agents tend to have a high self-opinion. The positive self opinions tend to shift the other opinions towards negative values. Indeed, having a high self-opinion, the agents feel often undervalued by the others and they decrease their opinion about them by vanity. In return, these agents do the same. 
After a while, the self-opinions fluctuate around an equilibrium value that can be approximated by considering that the sum of the modifications by the interactions with all the other agents is 0:

\begin{equation}
\sum_{j \neq i} p_{i,j} \rho (a_{j,i} - a_{i,i} + \texttt{Random}(-\delta,\delta)) \approx 0.
\end{equation}

Neglecting the random term, this can be rewritten as:

\begin{equation}
a_{i,i} \approx \frac {\sum_{j \neq i}  p_{i,j} a_{j,i}} {\sum_{j \neq i}  p_{i,j}}.
\label{eq:equilibrium}
\end{equation}

If this self-opinion of equilibrium is such that $a_{i,i} > 1 - \delta $, then the random fluctuations can decrease $a_{i,j}$, even if $a_{i,j}=1$, because we can have $a_{j,i}  + \texttt{Random}(-\delta,\delta) < a_{i,i}$. In this case, the couples $(a_{i,j}, a_{j,i})$ that have converged to +1, can enter in negative retaliation loops and finally converge to -1. This leads $a_{i,i}$ to a smaller equilibrium value. The process continues until $a_{i,i} < 1 - \delta $, for all $i$. This explains the two phases that we observe in the simulations. In the first phase, the self-opinions converge to a value which is too high, with a too large number of friends. Then progressively this number of friends decreases, each time one of these friends has her opinion going below the considered self-opinion. This process stops when the self-opinion reaches an equilibrium value for an adequate number of friends that keeps its fluctuations always below $1 - \delta $, preventing any underevaluation by friends. When such a value is reached, the number of friends cannot change anymore and the configuration is stable. Therefore, at the stable state, we must have, for each agent $i$:

\begin{equation}
1 - \delta > \frac {\sum_{j \neq i}  p_{i,j} a_{j,i}} {\sum_{j \neq i}  p_{i,j}}.
\label{eq:stable}
\end{equation}

Putting $a_{i,i}$ at its maximum value in the computation of $p_{i,j}$, and $a_{i,j} \approx 1$ for friends, $a_{i,j} \approx -1$ for foes we get, for friends:

\begin{equation}
p_{i,j}= \frac{1}{1+exp(-\frac{\delta}{\sigma})} = p^+.
\end{equation}

Similarly, for foes:

\begin{equation}
p_{i,j}= \frac{1}{1+exp(\frac{2-\delta}{\sigma})} = p^-.
\end{equation}

If we call $f$ the number of friends, at stability, we can compute the function s(f):
\begin{equation}
s(f) = \frac {fp^+ - (N-f) p^-} {fp^+ + (N-f) p^-}.
\label{eq:stable2}
\end{equation}

An estimate of the number of friends at stability is given by the maximum value of $f$ such that $s(f) < 1 - \delta$. For the case of the equality pattern shown on Figure \ref{fig:smallWorld} ($N = 40$ and $\sigma = 0.35$), we have: for $f = 3$, $s(3) = 0.79$, $f = 4$, $s(4) = 0.85$ and $1-\delta = 0.8$. Therefore, an evaluation of the number of friends at stability is $f = 3$, corresponding to the number of friends observed on average (3.4). 

\begin{figure}
 	\centering
 		\includegraphics[width=8 cm]{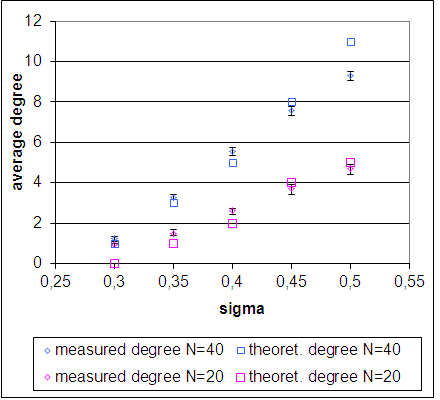} 
 	\caption{Variations of the average number of friends of each agent in the equality pattern when $\sigma$ varies, compared with the theoretical prediction for $N=40$ and $N=20$. The measured values are averaged on 10 replicas. The error bars correspond to the minimum and maximum over the 10 replicas. They show that the average number of friends does not vary significantly with the replicas. The other parameter values are: $\omega=0.3$, $\rho=0.01$, $\delta = 0.2$, $k=5$.}
 	\label{fig:averageDegree}
 \end{figure}

This provides an evaluation of the number of friends that each agent has. The tests shown on Figure \ref{fig:averageDegree} confirm that this approximation is correct. However, it should remain in a range where $\rho$ is small compared with $\omega$, and increasing $k$ increases also the effect of the opinion propagation, therefore it can also modify the validity of the approximation.

This analysis allows us to predict the effect of the main parameters on this pattern. For instance:
\begin{itemize}
	\item Increasing $\delta$ (noise parameter) decreases the final equilibrium value of the self opinion, and hence the final number of friends of each agent;
	\item Increasing $\sigma$ (ruling the slope of the sigmoid function of the influence parameter) will decrease the difference of weights between friends and foes in equation \ref{eq:equilibrium}, and therefore decrease the equilibrium value for a given number of friends. This will thus result in a larger number of friends at the stable state;
	\item Increasing the number of agents in the population decreases the value of the self-opinions for the same number of friends (see eq. \ref{eq:equilibrium}). Therefore, it will result in a larger number of friends at the stable state.
	\item Increasing $\rho$ increases the fluctuations around the equilibrium self-opinion and hence will decrease the number of friends at the stable state.
\end{itemize}

\subsubsection{First analysis of the small world properties in the equality pattern}

In the presentation of the equality pattern, we mentioned that the network of friends shows in some cases the properties of small worlds. We did not perform a systematic study to identify the limits of these properties in the parameter space. From our first investigations however, it seems that the small world properties require very small values of parameter $\rho$ compared with parameter $\omega$. For instance, the values of $\rho$ that we used in the experiments shown on Figures \ref{} and \ref{} and leading to the equality pattern are too high to provide small worlds. 

We a small value of $\rho$, we are in the domain of validity of the approximation made in the previous paragraph, and the general scheme applies: a fast establishment of to a too large network of friends, leading to too high self-opinions, leading to the progressive elimination of some friends until reaching a sustainable equilibrium with more moderate self-opinions. A reasonable hypothesis is that during this phase of progressive elimination of friends, the friends who have positive links between them tend to be more protected than friends who have negative links between them. To be more precise, consider agent $i$, who has two friends $j_1$ and $j_2$. If $j_1$ and $j_2$ are friends, they will tend to support each other when they discuss with $i$. Therefore, if there a vanity struggle between $i$ and $j_1$ starts, $j_2$ will tend to influence $i$ positively. The influence of $j_2$ on $i$ is important because $j_2$ is a friend of $i$. On the contrary, if $j_1$ and $j_2$ are foes, each one tend to influence negatively $i$ about the other. This influence favours in general a to get a higher clustering coefficient than in random networks.

This rapid analysis explains why the smal-world properties can take place but not not systematically, because:
\begin{itemize}
	\item they require an adequate level of opinion propagation: if it is too high, the phase where the friends are too numerous does not take place, if it is too small, it is not sufficient to influence the elimination of friends of friends during the elimination phase.
	\item the parameter $k$ (number of acquaintances talked about) should not be too small compared with $N$ (size of the poulation), because otherwise the influence of the friends is very limited.
\end{itemize}
  
Table 1 shows the results of a few experiments that confirm this analysis: In these experiments, $k = 10$, favouring the propagation of opinion, gives a higher clustering coefficient than $k = 2$. We also observe that, in these examples, decreasing the value of $\rho$ tends to decrease the clustering coefficient, but also the mean shortest path. Of course, more systematic explorations would be necessary to come to confirm and refine these conclusions.

\begin{table}[h]
 \centering
	\begin{tabular}{|c|c|c|c|c|c|c|c|c|}
	\hline
 	\multicolumn{4}{|c|}{}	 & 	average & clustering & random  & mean  & random \\
 	\multicolumn{4}{|c|}{parameters} & degree  & coefficient & clustering &  shortest  & mean sh. \\
  $\omega$ & $\rho$  & $\sigma$ & $k$  &  &  & coefficient & path & path \\
 		 \hline
0.3& $ 0.01$ & 0.35 & 2 &	3.83  & 0.09 & 0.08	& 2.88  & 3.36\\
   &      &  &  10 &	3.02  & 0.23 & 0.07	& 4.82 & 7.08\\
   \hline
0.6 & 0.05 &  0.5 & 2  & 7.19 & 0.18 & 0.19& 2.03 & 2.04 \\
	& & & 10 & 3.51 & 0.25 & 0.09 & 3.43 & 4.58 \\
 \hline
 	\end{tabular}
 	 	\label{table:SW}
 	 	\caption{Comparison of clustering coefficient and shortest path between simulations of the model and random networks of same number of nodes and average degree. Each result is the average of 10 replicas. Note that in these experiments when $k = 10$, the network shows the small world properties, this is not the case for $k = 2$.}
\end{table}

\subsubsection{Analysis of the Elite pattern}

The elite pattern occurs when the opinion propagation is stronger than in the previous case which leads to the situation where the number of friends is less than one on average. In this case, with the random fluctuations from the initialisation, the model converges to a situation where there are two types of agents:
\begin{itemize}
	\item the elite agents who have one friend supporting them and who have a positive self opinion, and they have a very negative opinion of all the others (except themselves and their friend);
	\item the second category agents who did not manage to have a friend and have a negative self-opinion, and a very negative opinion of all the other second category agents.
\end{itemize}

The second category agents are sensitive to the opinion propagation from the elite agents. Indeed, the second category agents having a strongly negative self-opinion, they are influenced even by agents for whom they have a negative opinion (for an opinion which is the same as the self opinion, the propagation coefficient is 0.5, see Figure \ref{fig:propaCoef}). Let us suppose that there are $m$ elite agents, each having a single friend with an opinion of 1 about them, and having a self opinion of about $1 - \delta$, and $m-2$ other elite agents having an opinion $-1$ about them. At a stationary state, the opinion $a_{s,e}$ of the second category agents about the elite agents would respect an equilibrium between the influences of the different elite agents:
\begin{equation}
		 (a_{s,e} - 1 + \delta) + \frac{k}{N-2}((a_{s,e} - 1) + (m-2)(a_{s,e} + 1)) \approx 0.
		 \label{eq:elite}
\end{equation} 

In equation \ref{eq:elite}, the influence of the self-opinion is multiplied by 1, because the considered elite agent always influences the second category agents about herself, whereas the influence of the others's opinion takes place when they talk about her, having a probability $\frac{k}{N-2}$ to be talked about (because the random drawing excludes the discussing pair of agents). Moreover, there is one elite agent having an opinion +1, and $m-2$ having opinion -1. Finally, we neglect the influence of the second category agents because they all have a similar opinion about the elite agents. We get:

\begin{equation}
		a_{s,e} \approx \frac{(N-2) (1 - \delta) - k(m-3)}{N-2 + k(m-1)}.
\end{equation} 

We verified on two examples with $N = 40$, $k = 2$, $\delta = 0.2$, $\sigma = 0.3$ that this approximation gives the right order of magnitude:
\begin{itemize}
	\item For $\omega = 0.3$ and $\rho = 0.15$, the number of members of the elite is 24. The measured average opinion of the second category agents about the elite agent is -0.15 whereas the approximation yields -0.14.
	\item For $\omega = 0.5$ and $\rho =0.4$, the number of members of the elite is 4. The measured average opinion of the second category agents about the elite agent is 0.61 whereas the approximation yields 0.64.
\end{itemize}

This result seems satisfactory, given the level of approximation made in the reasoning. We can deduce from this that the higher the number of the elite, the lower will be the opinion of the second category for the elite. But when the opinion of the second category agents is high, this tends to increase the self-opinion of the elite agents, and this increases the risk of instability. This corresponds to the observations we made.

The second category agents are more stable if the opinion propagation coefficient $\rho$ is relatively high because in this case, the influence of the very negative opinion of elite agents tends to prevent them to grow with possible positive loops of vanity among second category agents. This rise of couples of second category agents allowing them to come back the elite is possible though, especially in the mixed pattern equality and elite.

\subsubsection{Analysis of the mixed pattern between equality and elite}

In the mixed pattern between equality and elite (see Figure \ref{fig:eliteSW}) we have some elite agents that become temporarily second category and then become elite again. We are also in a case where the number of friends per agent is around 1, and with the fluctuations the agents have chances to loose their only friend. Once the agent becomes isolated, she progressively looses her high self-opinion because all the other agents are sending her messages of negative value. This explains why these agents form a second ring in the network representation. It can happen that the agent looses her good self opinion first and then looses her friend. This can be explained with the random fluctuations of the model, where they can interact for a long time only with their foes, and their self-opinion can reach a threshold below which their foes become more influent than her single friend (for which the propagation coefficient does not increase much). With this negative self-opinion, the agent propagates a negative opinion of herself, which progressively influences her friend.

When her self-opinion becomes very negative, she becomes more influenced by the agents of the elite, who have a high self opinion. This increases the opinion of the second category agents about the elite and, this opinion cannot increase too much because each agent of the elite is viewed very negatively by all the other elite agents except her single friend (see the reasoning made in the previous paragraph).

The main difference with the previous pattern is that, the vanity being stronger, there is a high chance that the second category agents enter into positive loops of friendship reinforcement, despite the opinion propagation of the elite. When this takes place, the second category agents reinforce each other, they restore their positive self-opinion and come back to the elite. This explains why there can be fluctuations between equality and elite patterns.

\subsection{When opinion propagation dominates}

\subsubsection{Study of the case $\omega = 0$}

\begin{figure}
 	\centering
 	\begin{tabular}{ccc}
 		\includegraphics[width=5 cm]{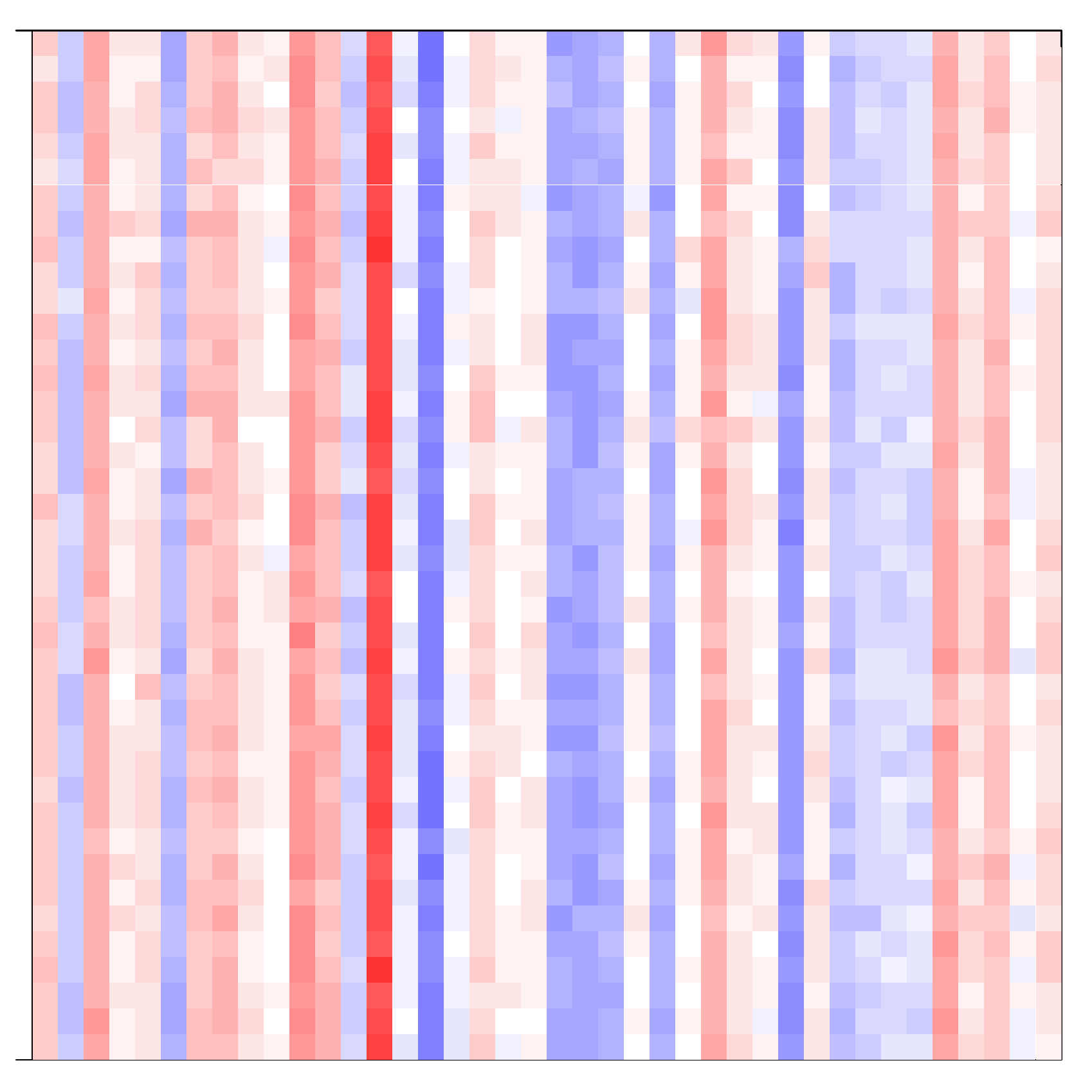} & 	\includegraphics[width= 5 cm]{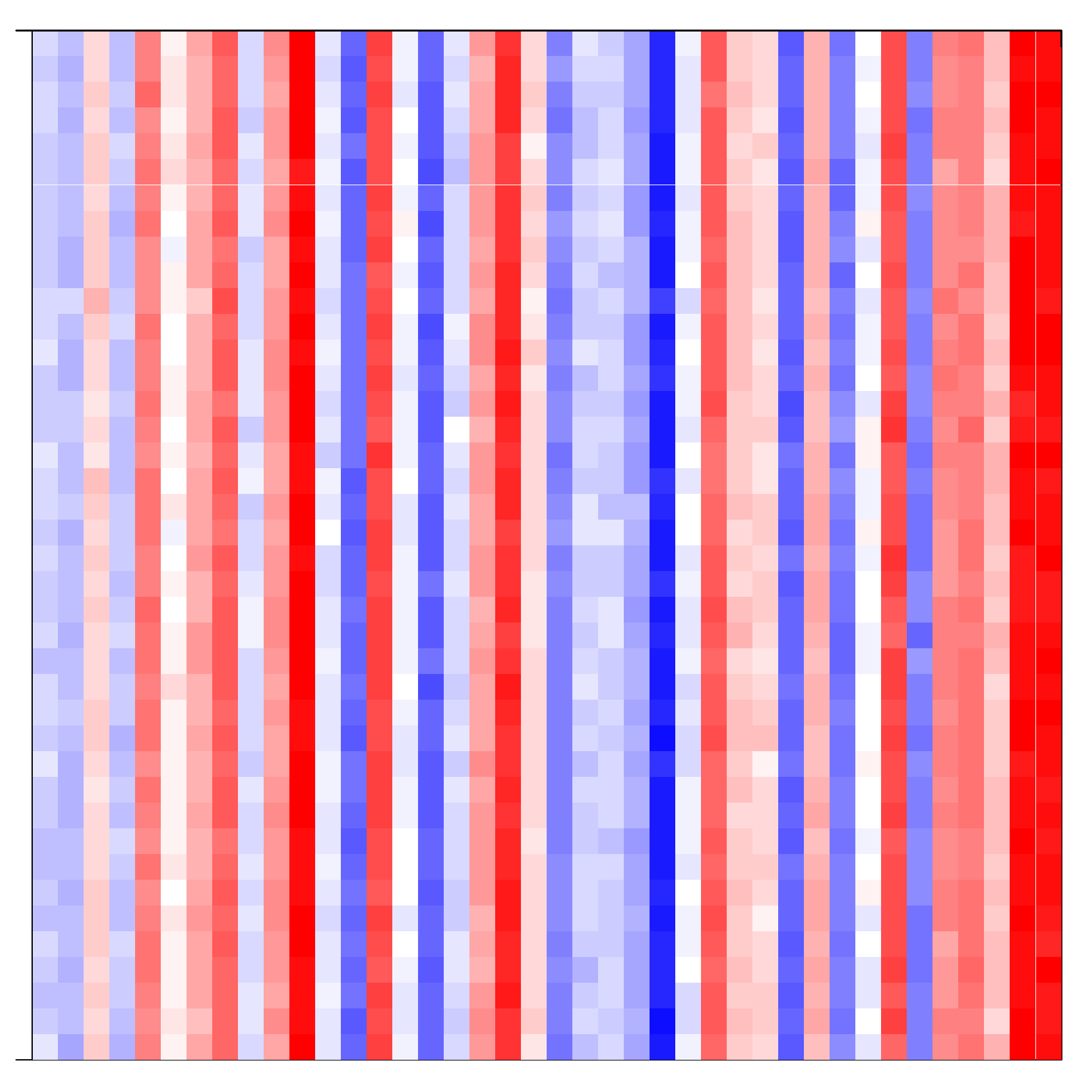}& 	\includegraphics[width= 5 cm]{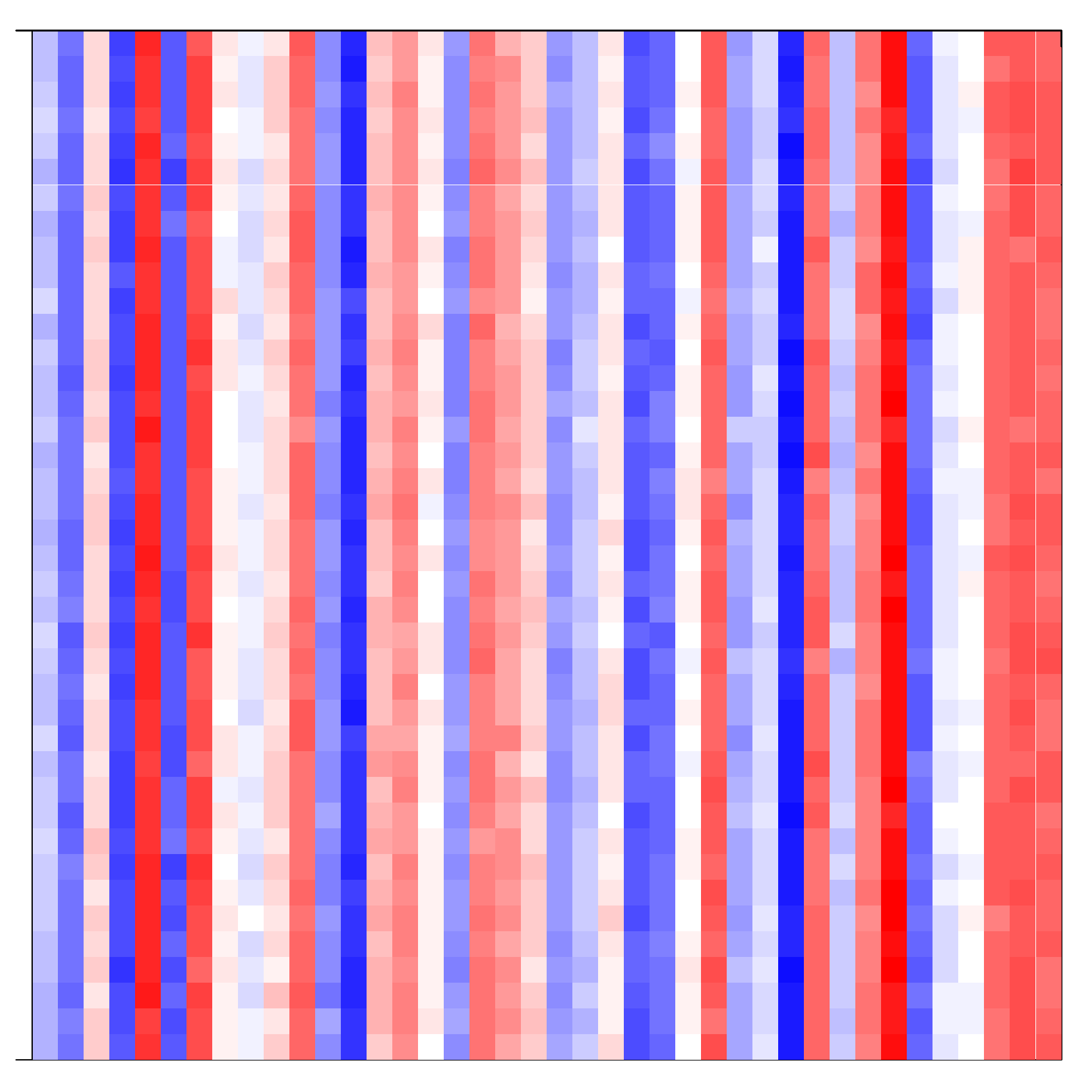} \\
 		\includegraphics[width=5 cm]{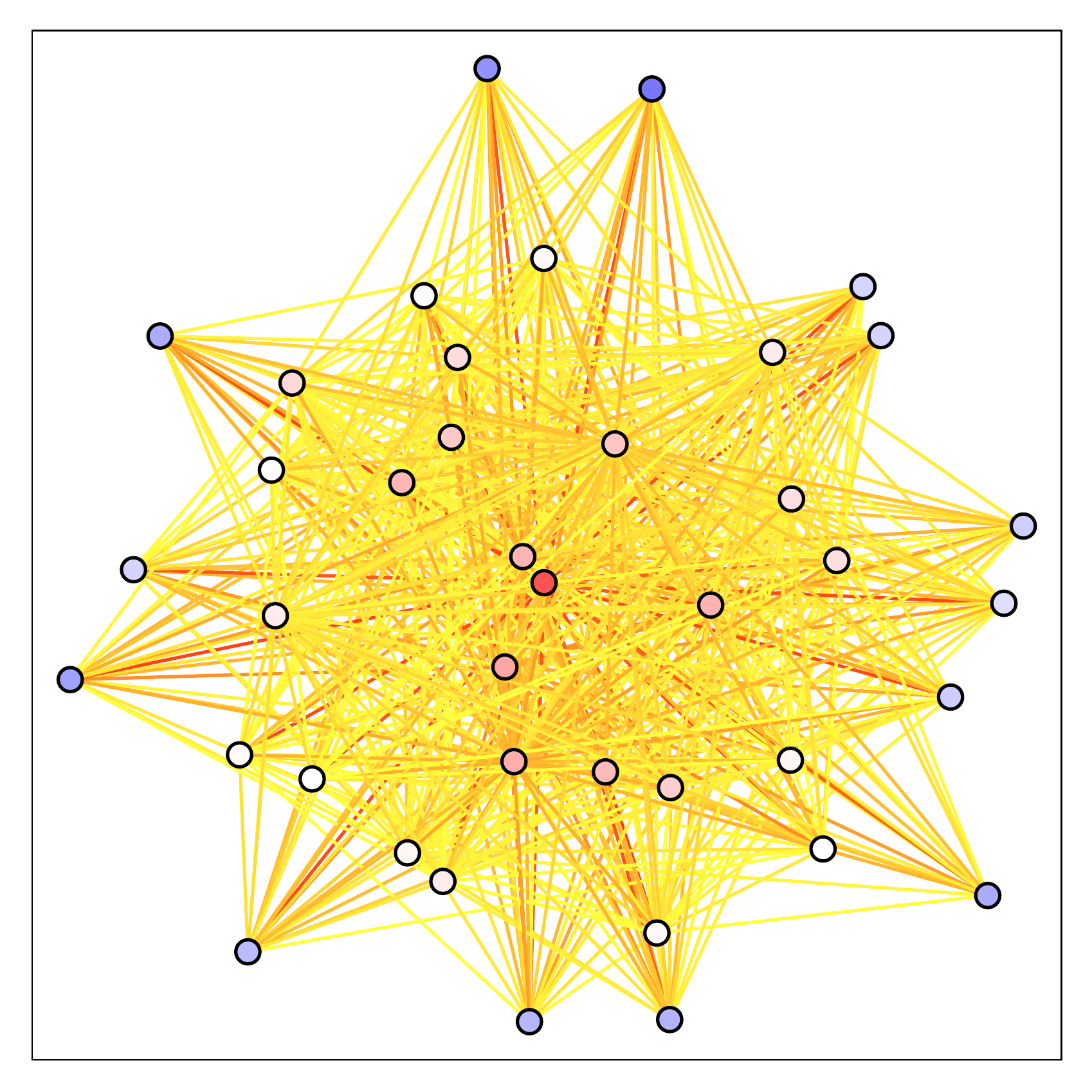} & 	\includegraphics[width= 5 cm]{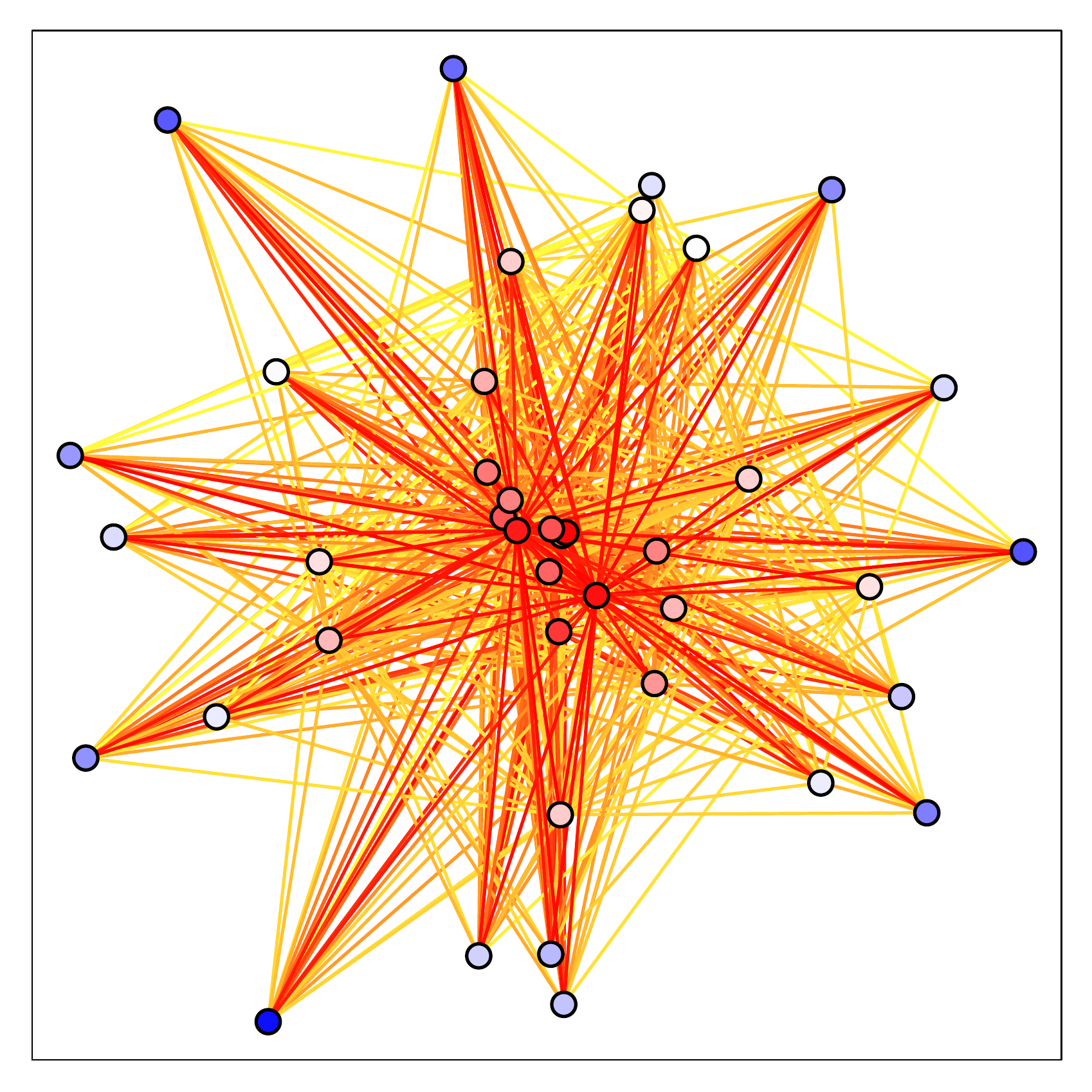}& 	\includegraphics[width= 5 cm]{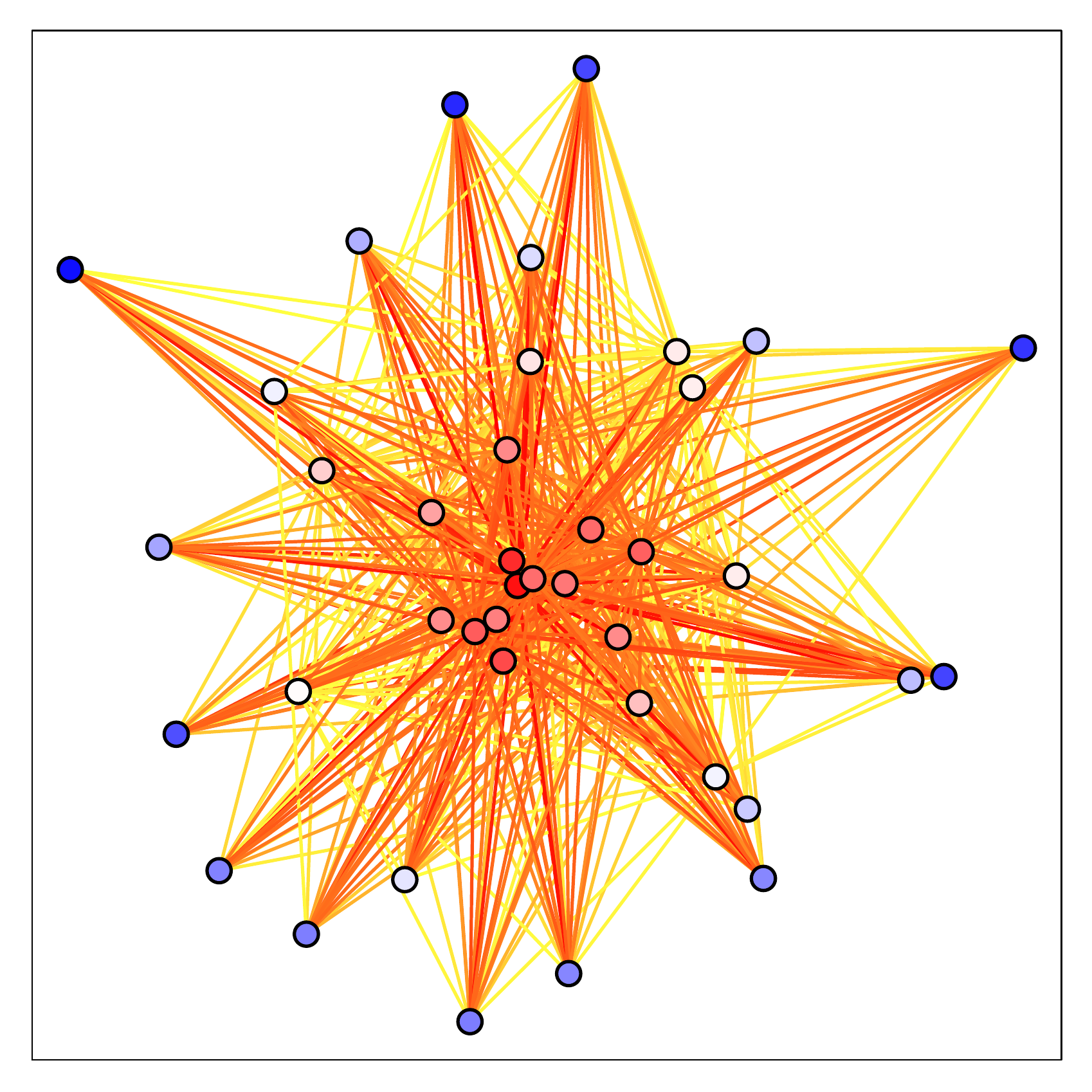} \\
 		$t$ = 10000, $\bar{a} = -0.03$ & $t$ = 50 000, $\bar{a} = -0.16 $ & $t$ = 100 000, $\bar{a} = 0.03$
 	\end{tabular}
 	\caption{Evolution of the opinions with the dynamics of opinion propagation only ($N = 40$, $\omega = 0$, $\rho = 0.3$). }
 	\label{fig:repOnly}
 \end{figure}

Studying the model with the opinion influence only ($\omega = 0$) can give us
some clues for understanding the model when $\omega$ is small. Figure~\ref{fig:repOnly} shows the corresponding pattern, that suggests the following observations: 
\begin{itemize}
\item Like in the hierarchy and dominance patterns,  all the agents have similar opinions about each agent, leading to matrix representations with columns of almost homogeneous colours (the reputation $r_i$ of agent $i$). 
\item Unlike in the hierarchy pattern, after a while, the values of reputations $r_i$ are almost uniformly distributed on the segment $[-1,+1]$. The values $r_i$ are not stable over time, they can take any value in the segment $[-1,+1]$ if the simulation is long enough.
\end{itemize}

In the case $\omega = 0$, it is easier to show analytically why the columns of the matrix tend to be homogeneous.

Let us assume $i$ and $j$ meet, the propagation of opinion process implies:
\begin{equation}
  \begin{cases}
 a_{i,j}(t+1) &= a_{i,j}(t) + \rho p_{i,j}\left(a_{j\, j}(t)-a_{i,j}(t)+\delta_{j,j}\right)\\
 a_{j,j}(t+1) &= a_{j,j}(t) + \rho p_{j\, i}\left(a_{i,j}(t)-a_{j,j}(t)+\delta_{i,j}\right)
  \end{cases}\, .
\end{equation}
Let us introduce the variables $D_{i,j}(t)=a_{i,j}(t)-a_{j,j}(t)$, hence
\begin{equation}
D_{i,j}(t+1) = D_{i\,j}(t)\left(1-\rho (p_{i\,j}+p_{j\,i})\right) +\Delta\, ,
\end{equation}
where $\Delta=\rho p_{i\, j}\delta_{j,j}(t)-\rho p_{j\,i}\delta_{i,j}(t)$ is the stochastic contribution. By definition
$p_{i\,j}+p_{j\,i}\in [0,2]$ and it can be equal to zero only $a_{i\,i}-a_{i\,j}>>1$ and $a_{j\,j}-a_{j\,i}>>1$. So neglecting for a while
the stochastic term and defining $\theta=\left(1-\rho (p_{i\,j}+p_{j\,i})\right)$ we have:
\begin{equation}
|D_{i,j}(t+1)| \leq \theta |D_{i\,j}(t)|\, .
\end{equation}
Therefore, if we suppose $\rho < 0.5$, we have $\theta <1$ and then, as $t$ increases
$D_{i,j}(t+1)$ goes to zero. This is the required result.

\begin{figure}
 	\centering
 	\begin{tabular}{cc}
 		\includegraphics[width=8 cm]{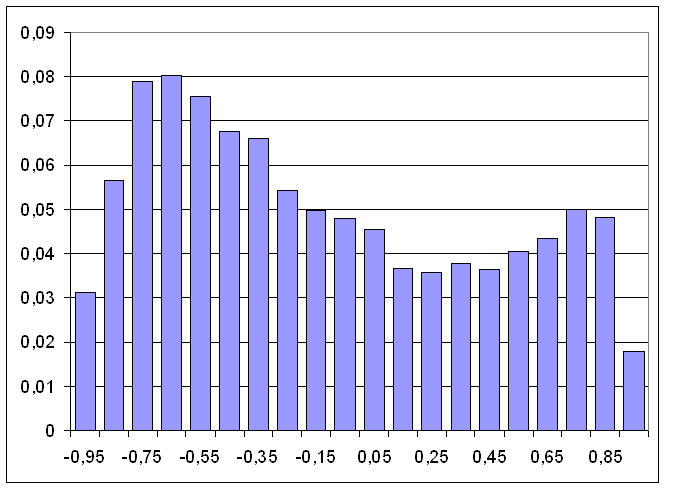} & 	\includegraphics[width= 8 cm]{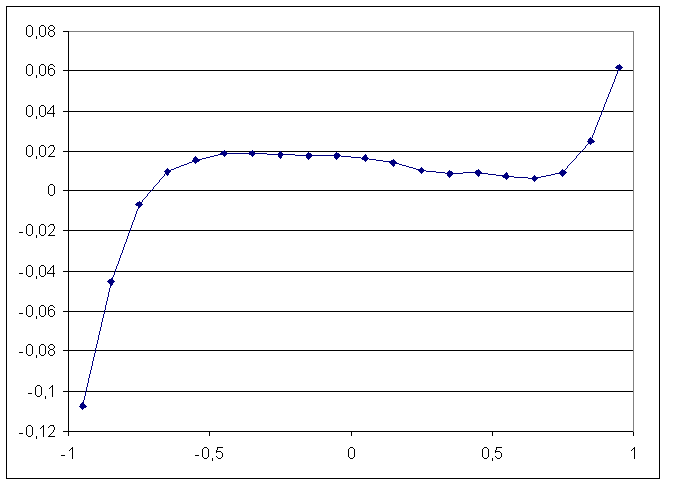} 
 	\end{tabular}
 	\caption{Graphs for $\omega$ = 0, $\rho$ = 0.3. On the left: the density of self opinion averaged on 50000 iterations. On the right: the average difference between the self opinion and the reputation ($a_{i,i} - r_i$).}
 	\label{fig:GraphsOmega0}
 \end{figure}

Now, with the addition of the stochastic part, the dynamics of the reputations seems to be a random walk between $-1$ and
$+1$. However, when computing the distribution of the reputations over 50000
iterations in a sliding window, we observe that this distribution is not
perfectly uniform. Figure \ref{fig:GraphsOmega0} on the left shows the distribution of the self opinions (which are close to the reputations) and we observe that the average value tends to be slightly negative, and the distribution has two maxima close to its extremes (around -0.7 and around +0.8), with a minimum between these maxima around 0.3.

Moreover, we observe on Figure \ref{fig:GraphsOmega0} on the right that the average self-opinion tends to be slightly higher
than the reputation, except when the self-opinion of the agents are close to
the extremes. The values of the extremes are due to side effects: the
reputation tends to fluctuate more slowly than the self opinion. The average
positive bias for the self opinion is more interesting. It is due to the
propagation coefficient which tends to be higher when the self-opinion gets
higher. Indeed, because of this difference, when an agent self-opinion is higher
than her reputation, the others have less influence on the self-opinion than when the self-opinion is lower than the reputation (everything else being equal). However, the effect of this average difference between the self-opinion and the reputation depends on the value of the agent's reputation:
\begin{itemize}
	\item If the agent's reputation (and hence her self-opinion) is among the lowest, then she is very sensitive to the influence of most of the others that she values more than herself (they have thus a high propagation coefficient). Therefore, when the agent self-opinion is higher than her reputation (which statistically takes place more often than the opposite), the agent's self-opinion tends to follow her reputation (the average opinion of the others about herself), thus it decreases. This explains a tendency towards negative values.
	\item If the agent's reputation (and hence her self-opinion) is among the highest, then on average she tends to impose her self-opinion to the others, because her propagation coefficient is high in the exchanges. Thus, on the contrary, when her self-opinion is higher than her reputation (statistically the most frequent), she tends to increase her reputation towards her self-opinion.
\end{itemize}

To summarise, the highly valued individuals tend to lead the other's opinions
and, with the statistical bias for a self-opinion higher than the reputation,
they tend to increase their reputation. This is the contrary for the badly
valued individuals who tend to naturally decrease their self-opinion, only by
the effect of the propagation coefficient. Nevertheless, these are statistical
tendencies, which take place stochastically and there are random
movements in the opposite directions. This explains the shape of the
reputation distribution. Below a threshold of the self opinion, the
reputations and self-opinions tend to be biased towards negative values, and
above towards positive values. The value of this threshold depends on
parameter $\sigma$ determining the propagation coefficient and also on $k$ the
number of agents about whom the opinions are propagated during the
encounters. Indeed, this number has an impact on how the agents propagate
their opinion about the others. 

\subsubsection{Analysis of hierarchy pattern} 

These observations are useful to understand the processes behind the emergence of hierarchy. First note that the vanity process enhances the tendency of self-opinions to be higher than the reputations. Indeed, the small statistical positive bias for self-opinion that is due to the opinion propagation leads, on average, the agents to consider themselves as (slightly) undervalued by the others, thus they devalue them by vanity. This is very similar to the process that we observed in the equality pattern, but it is slower because of the averaging effect of the opinion propagation. This explains the tendency to get negative opinions in hierarchy pattern. 

\subsection{Analysis of mixed pattern between crisis and dominance}

\begin{figure}
 	\centering
 	\begin{tabular}{ccc}
 		\includegraphics[width=5 cm]{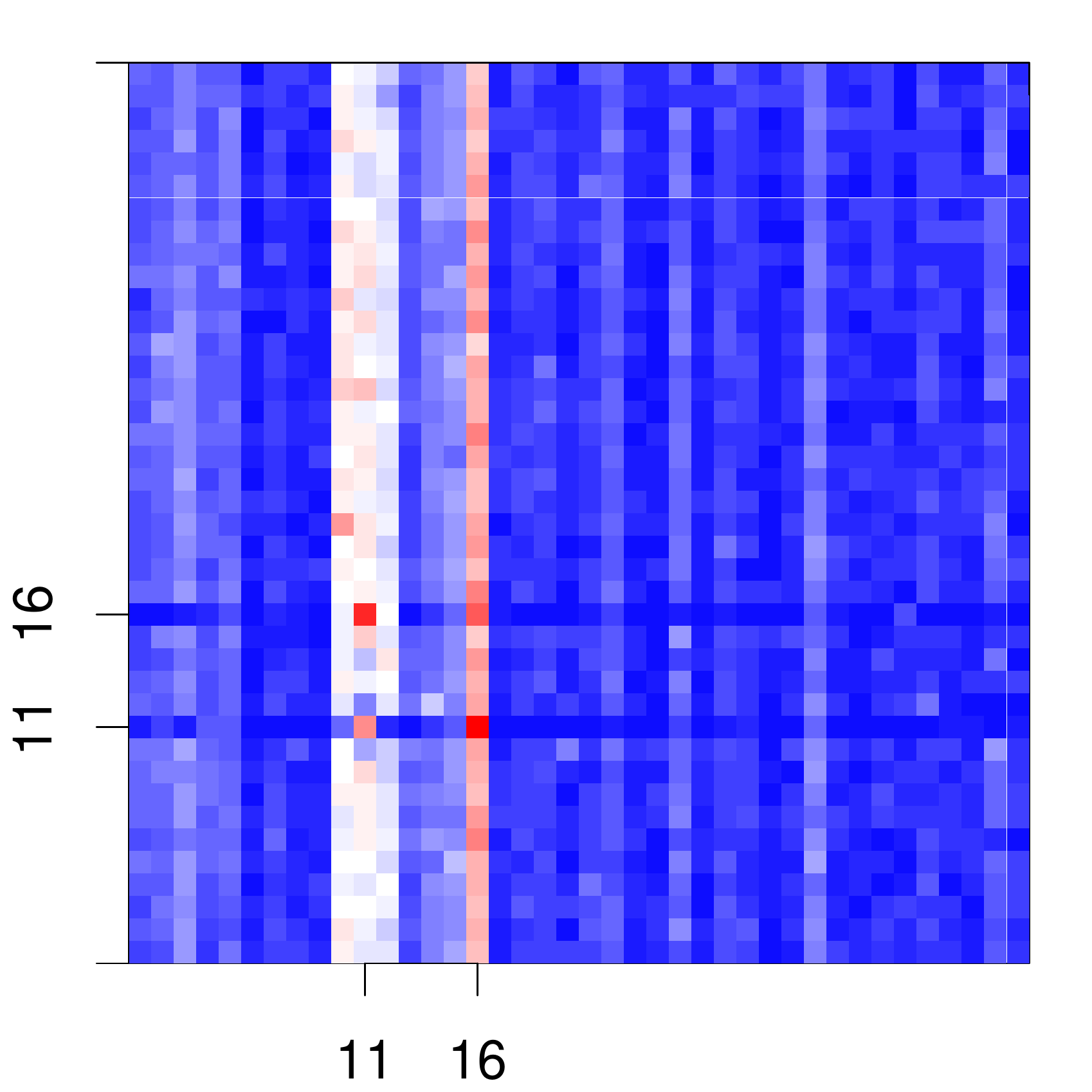} & 	\includegraphics[width= 5 cm]{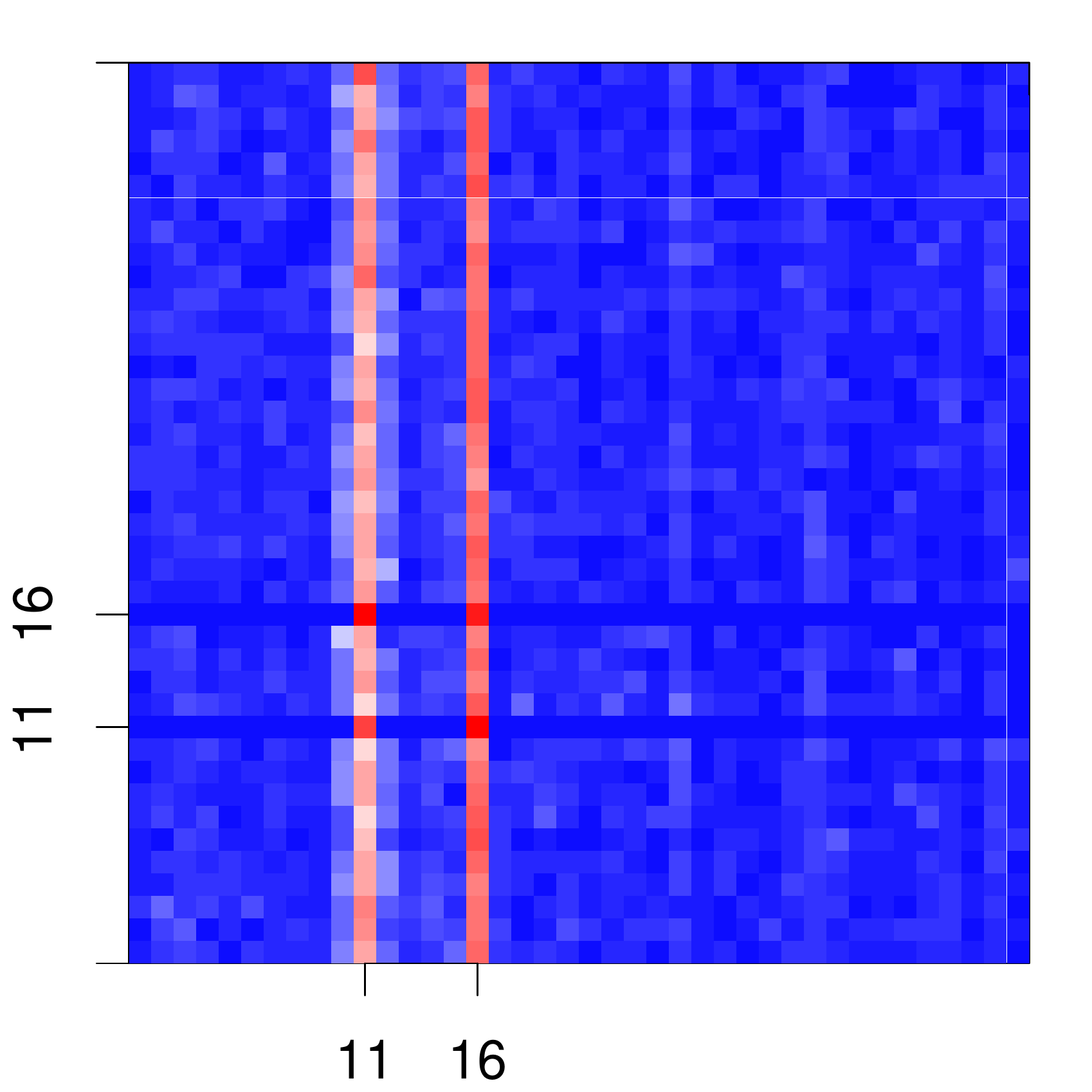}& 	\includegraphics[width= 5 cm]{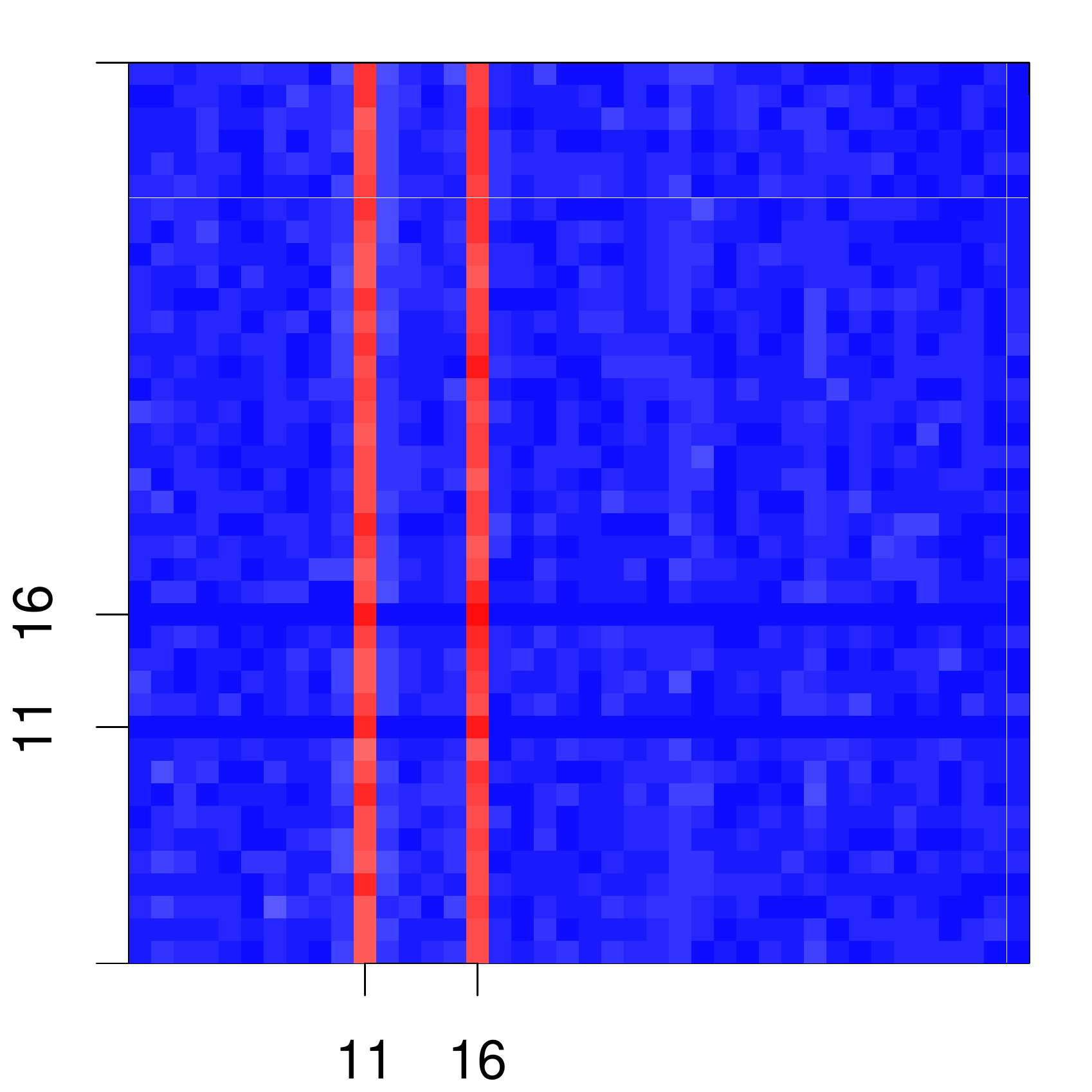} \\
 	 		$t$ = 6050 & $t$ = 6200 & $t$ = 6300 \\
 	 		 		\includegraphics[width=5 cm]{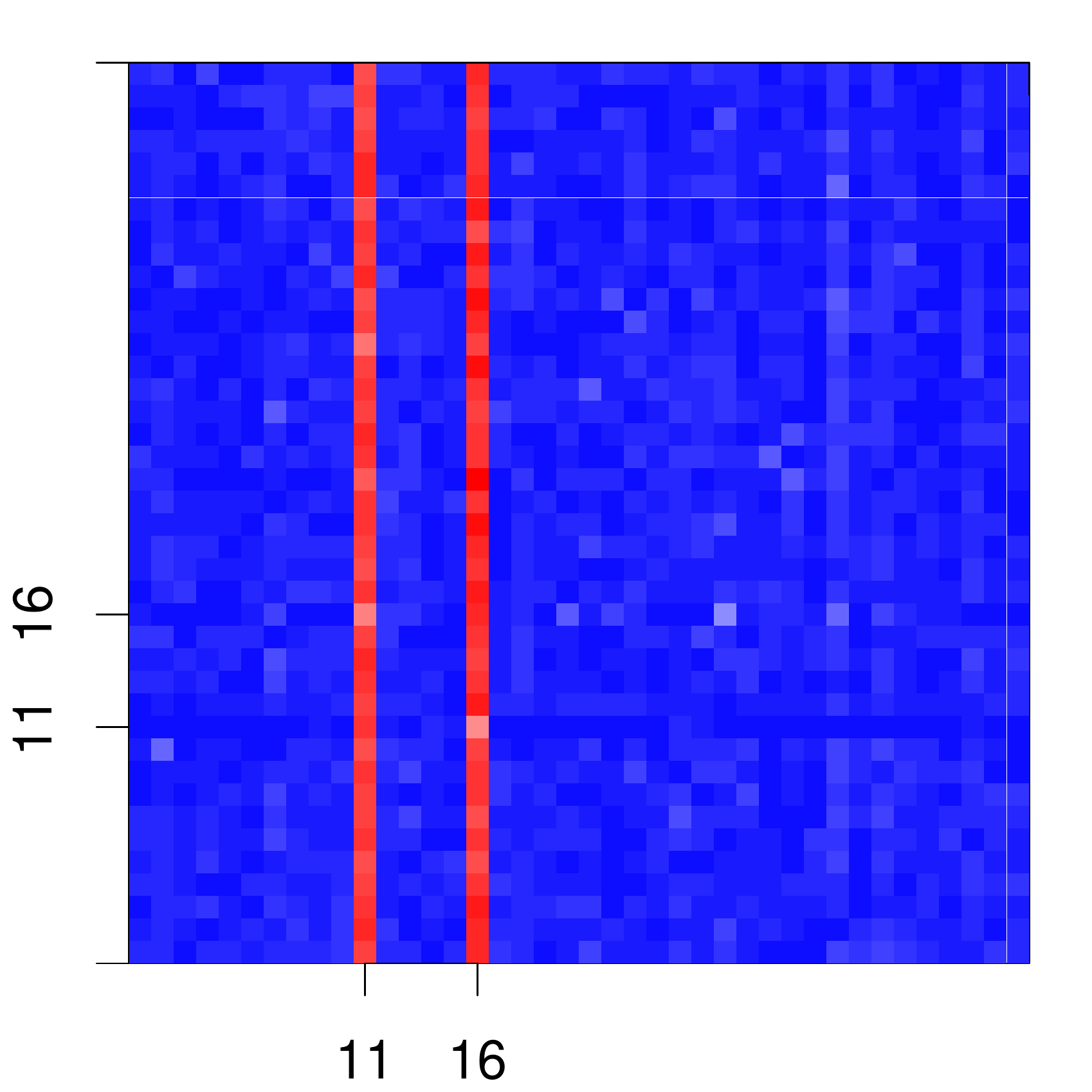} & 	\includegraphics[width= 5 cm]{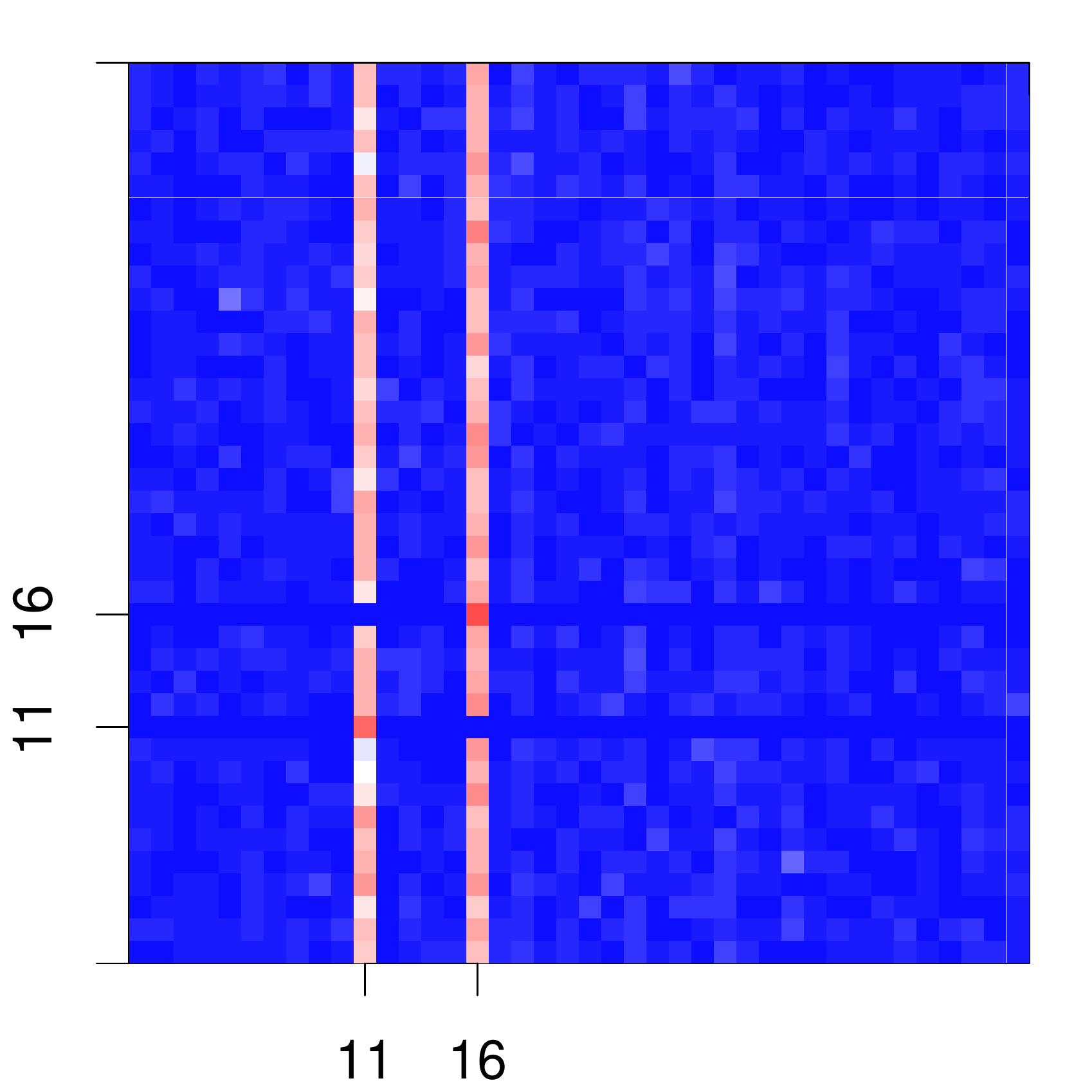}& 	\includegraphics[width= 5 cm]{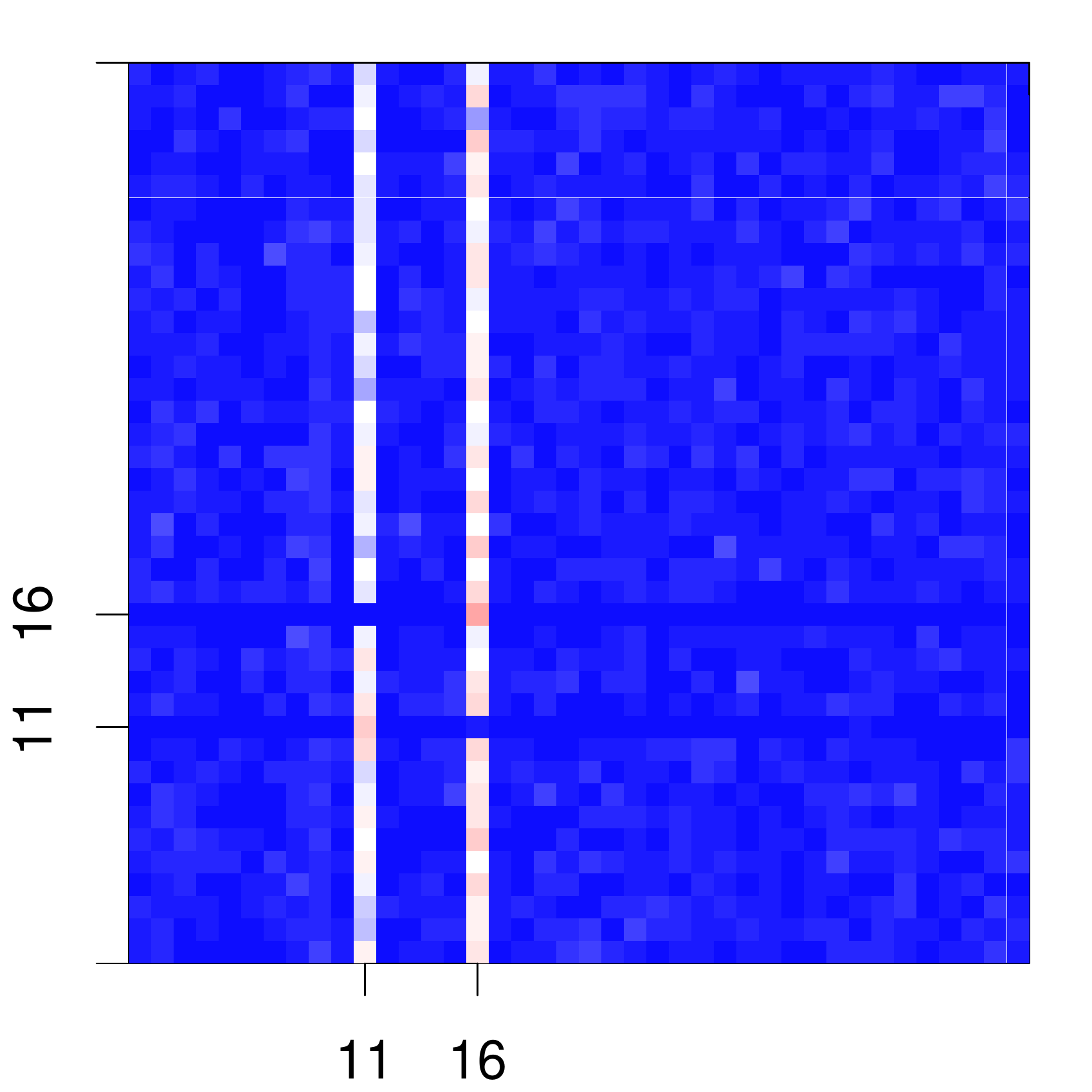} \\
 	 		$t$ = 7400 & $t$ = 7600 & $t$ = 7800
 	\end{tabular}
 	\caption{Crisis-dominance pattern: Rise (first line of matrices) and fall (second line of matrices) of double leadership. At $t$ = 6050, we observe that two agents ($i$ = 11 and $i$ = 16) support strongly each other (their reciprocal opinions are represented by the two red squares symmetric with respect to the diagonal), and both have positive self opinions. Moreover both tend to have very negative opinions of the rest of the population. At $t$=6 200, 6300, we observe that this process reinforces the reputations of both agents and decrease all the other opinions that become close to -1. At $t$ = 7400, we observe that the two leaders do not support each other anymore, on the contrary, their opinion about each other (lighter squares) are lower than their respective reputations. Hence they enter in a negative loop of vanity in which they lead the rest of the population ($t$ = 7600, 7800). The crisis pattern will take place after a few hundred iterations. }
 	\label{fig:P4DoubleLeaderRiseFall}
 \end{figure}

\begin{figure}
 	\centering
 	\begin{tabular}{ccc}
 		\includegraphics[width=5 cm]{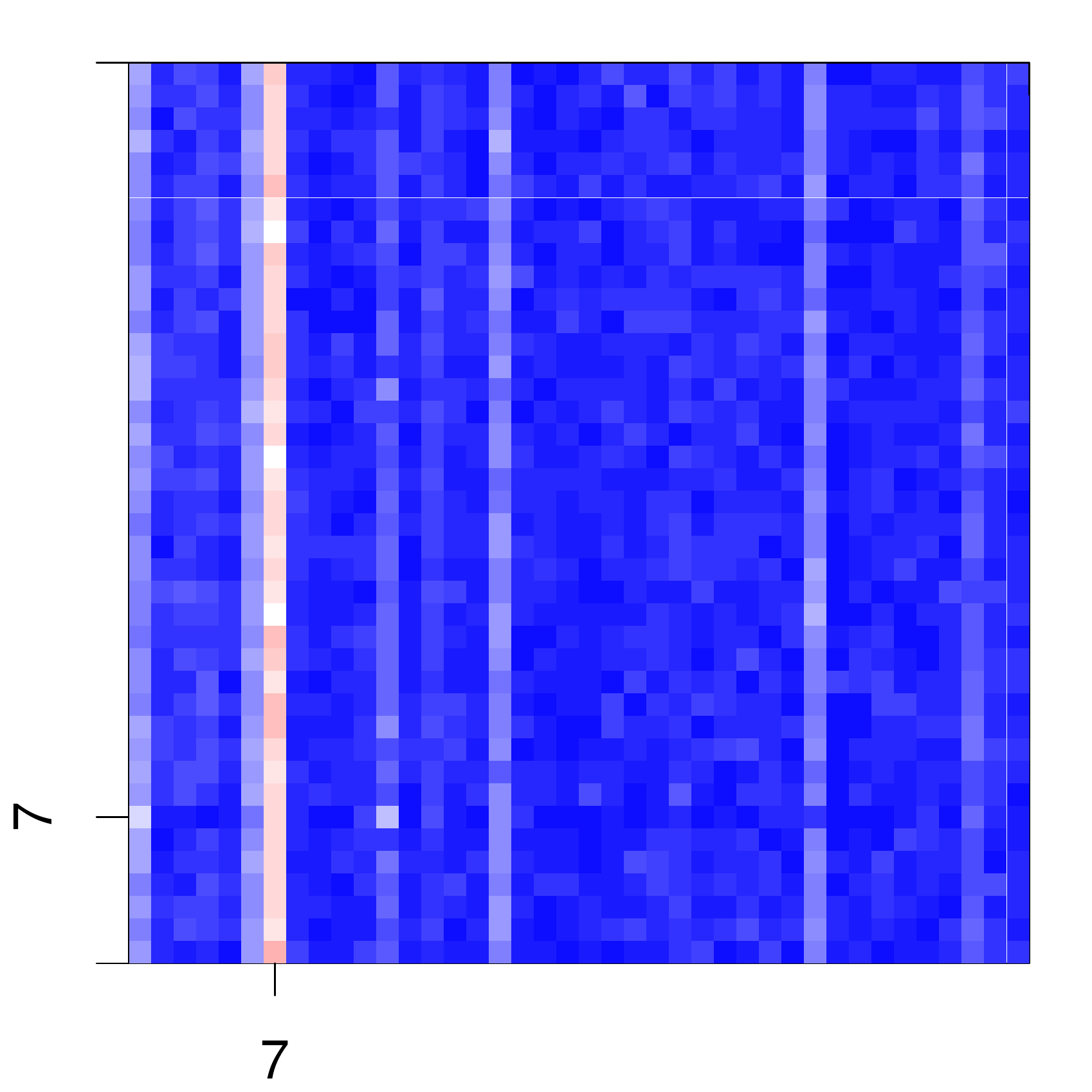} & 	\includegraphics[width= 5 cm]{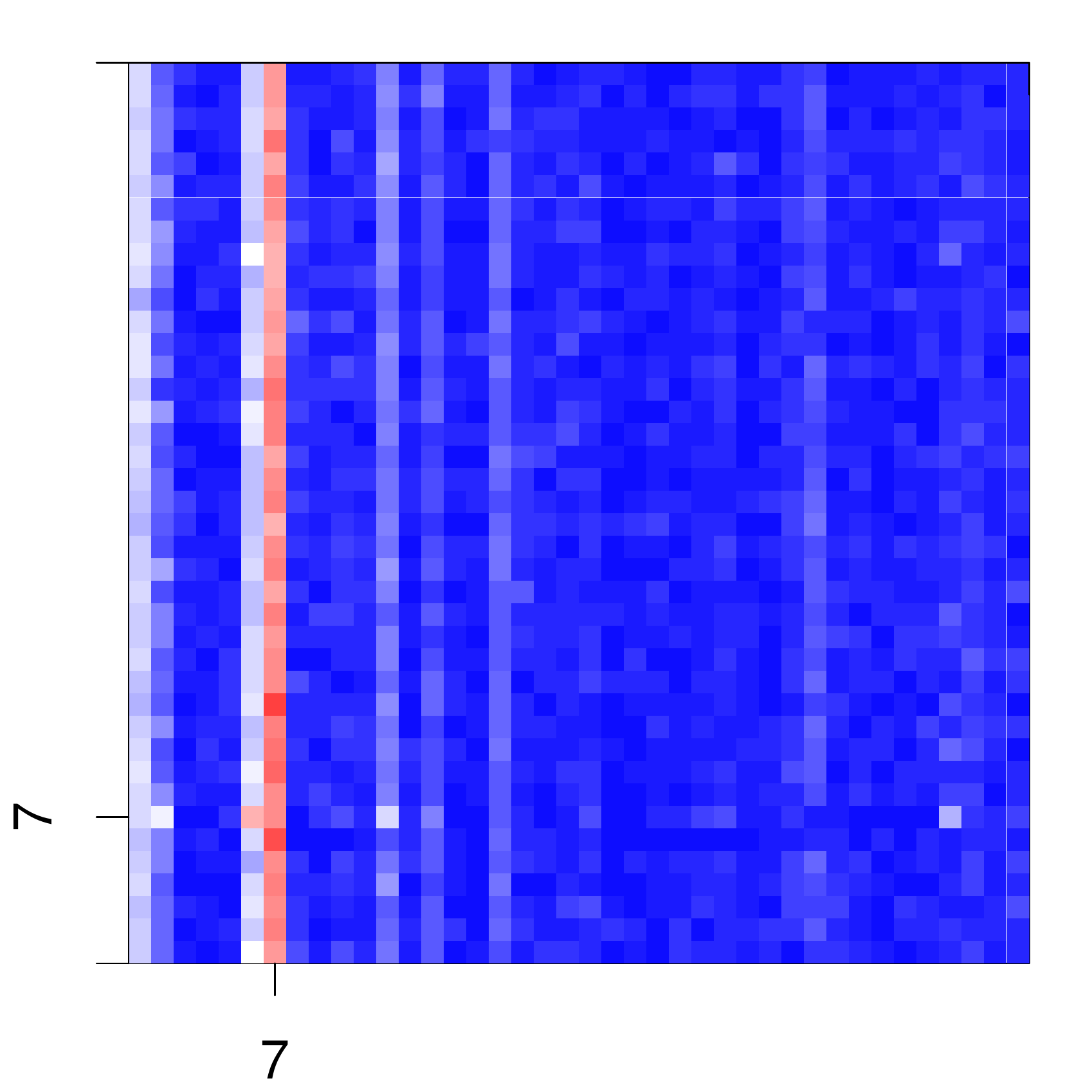}& 	\includegraphics[width= 5 cm]{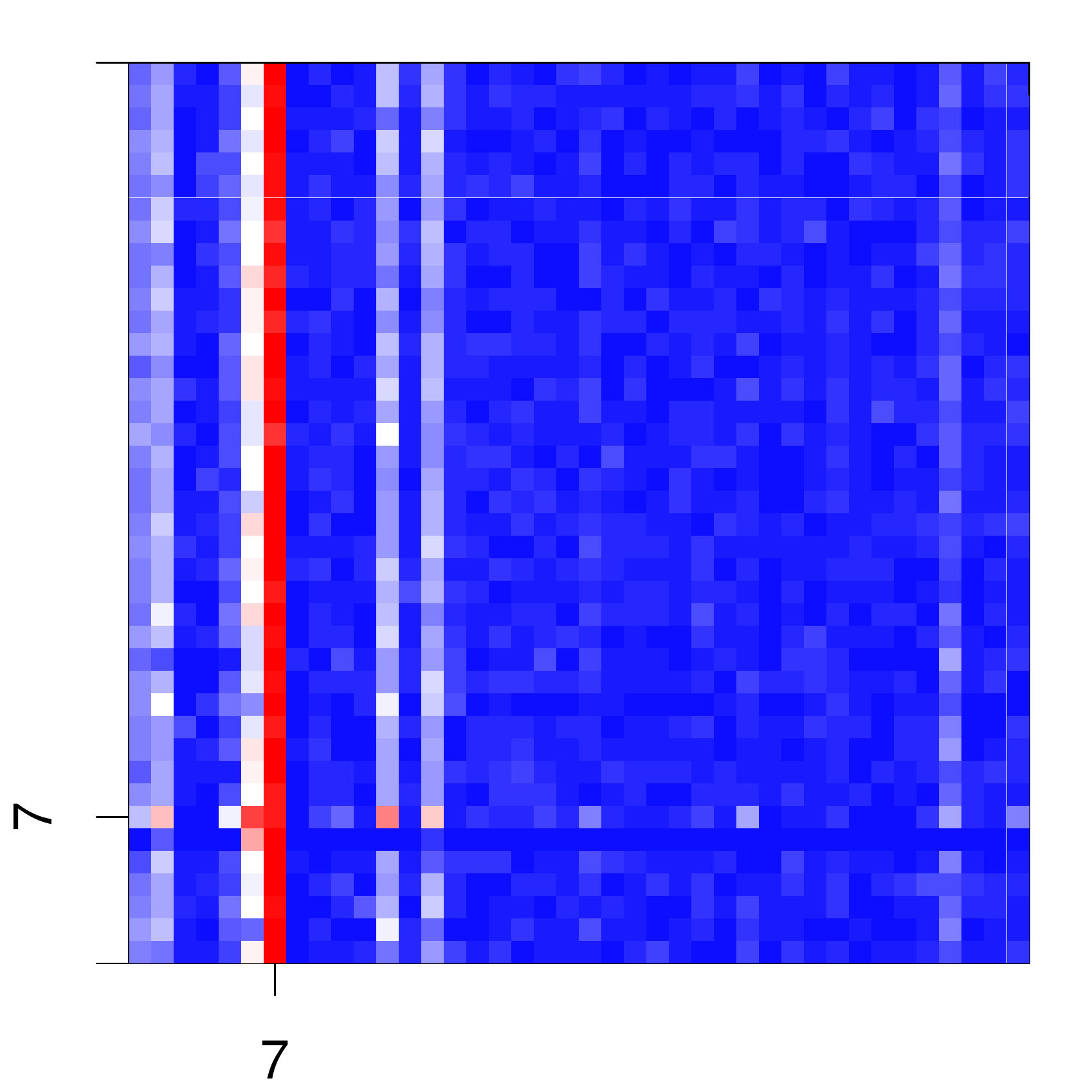} \\
 	 		$t$ = 200100 & $t$ = 200300 & $t$ = 200500
 	\end{tabular}
 	\caption{Crisis-dominance pattern: Rising of a single leader. At $t$ = 200100, we see that an agent (column with light pink cells at $i$ = 7) has a higher reputation than the others, and she supports 3 other agents more than the average (white squares at $j$ = 1, 12, 17). At $t$ = 200300; we observe that her reputation significantly increased, that she supports other agents (particularly the pink square at $j$ = 6). At $t$ = 200500, the dominance of agent 7 is complete, and she supports strongly several agents that have their reputation rising (columns with several white or pink squares). }
 	\label{fig:P4SingleLeaderRising}
 \end{figure}

\begin{figure}
 	\centering
 	\begin{tabular}{ccc}
 		\includegraphics[width=5 cm]{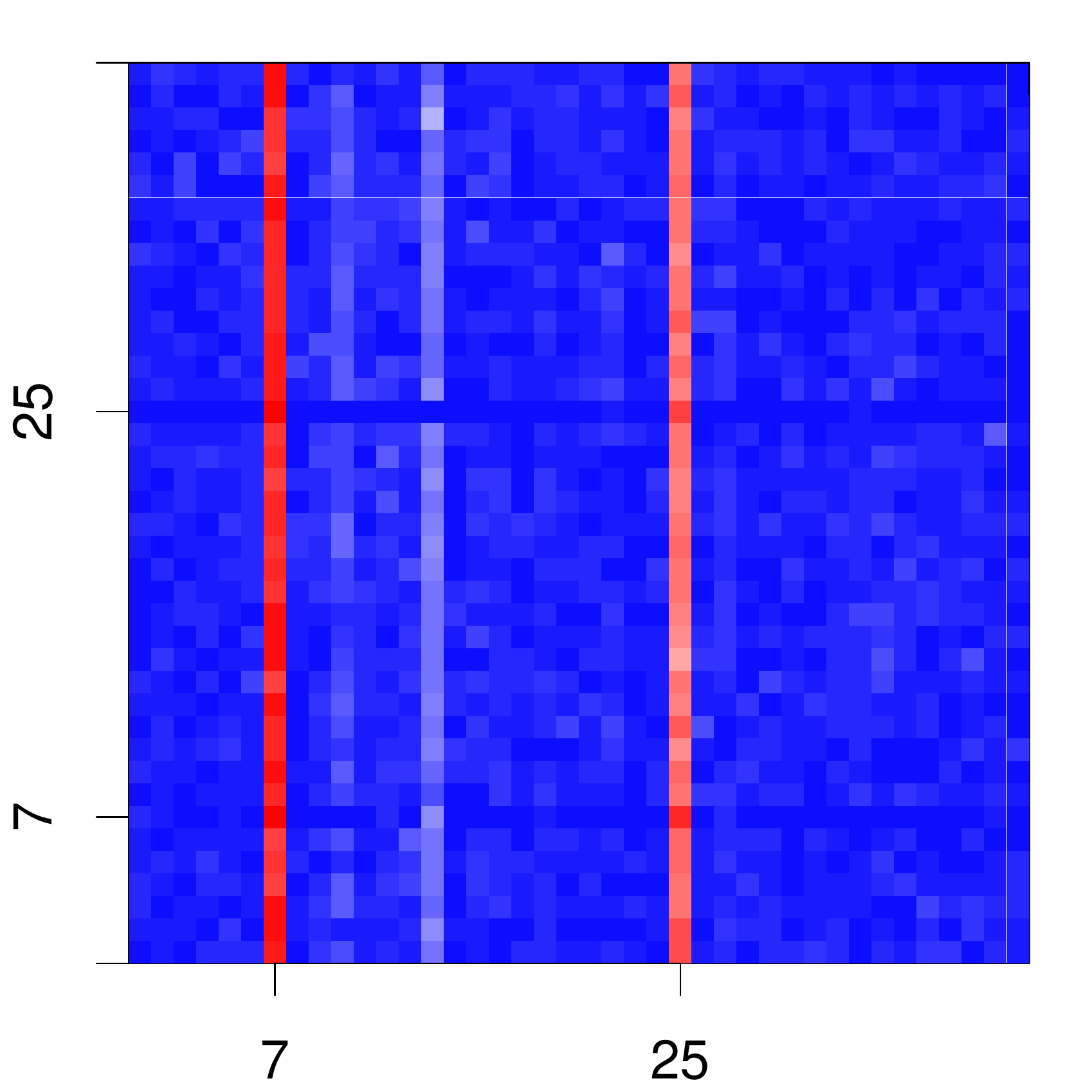} & 	\includegraphics[width= 5 cm]{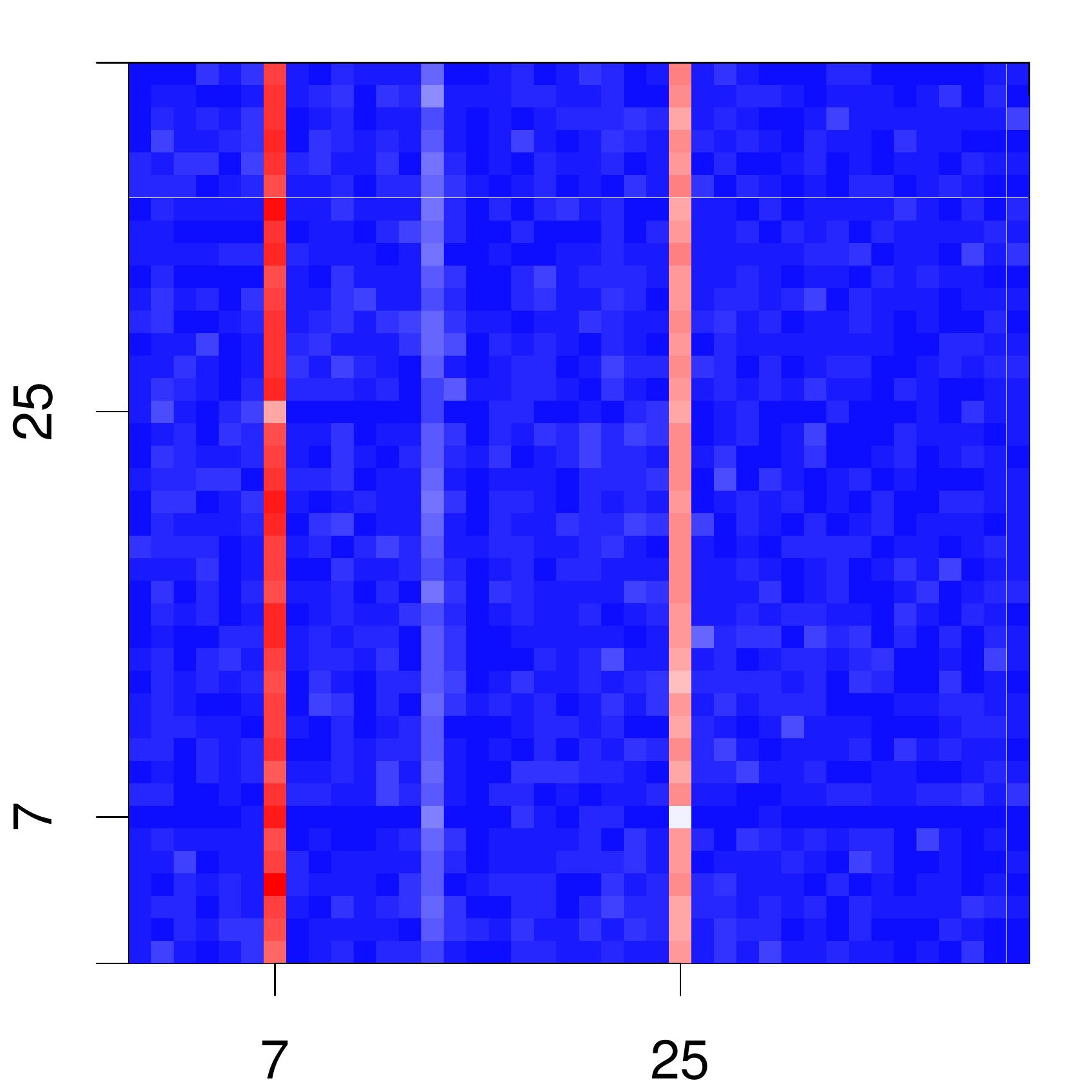}& 	\includegraphics[width= 5 cm]{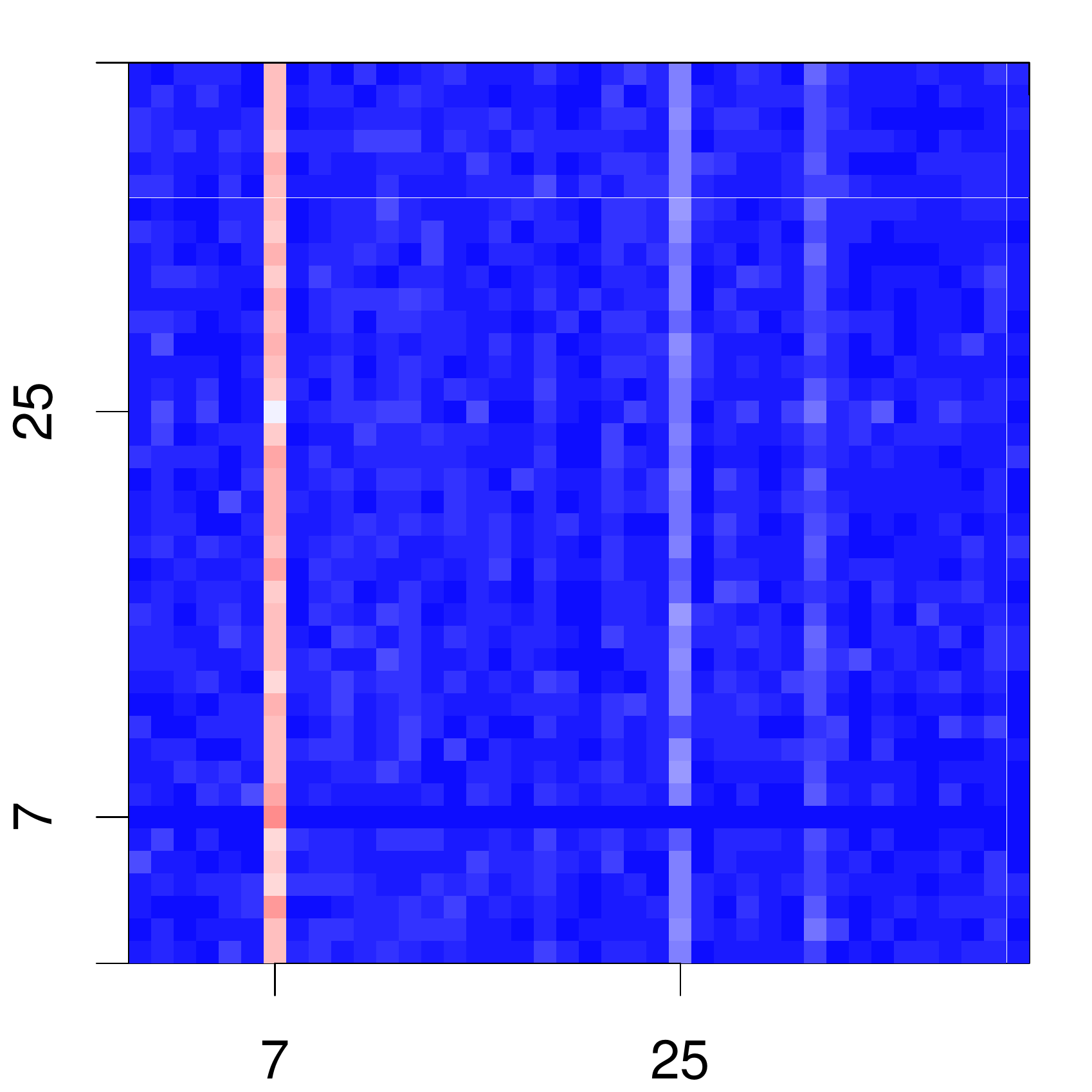} \\
 	 		$t$ = 204300 & $t$ = 204500 & $t$ = 205000 \\
 		\includegraphics[width=5 cm]{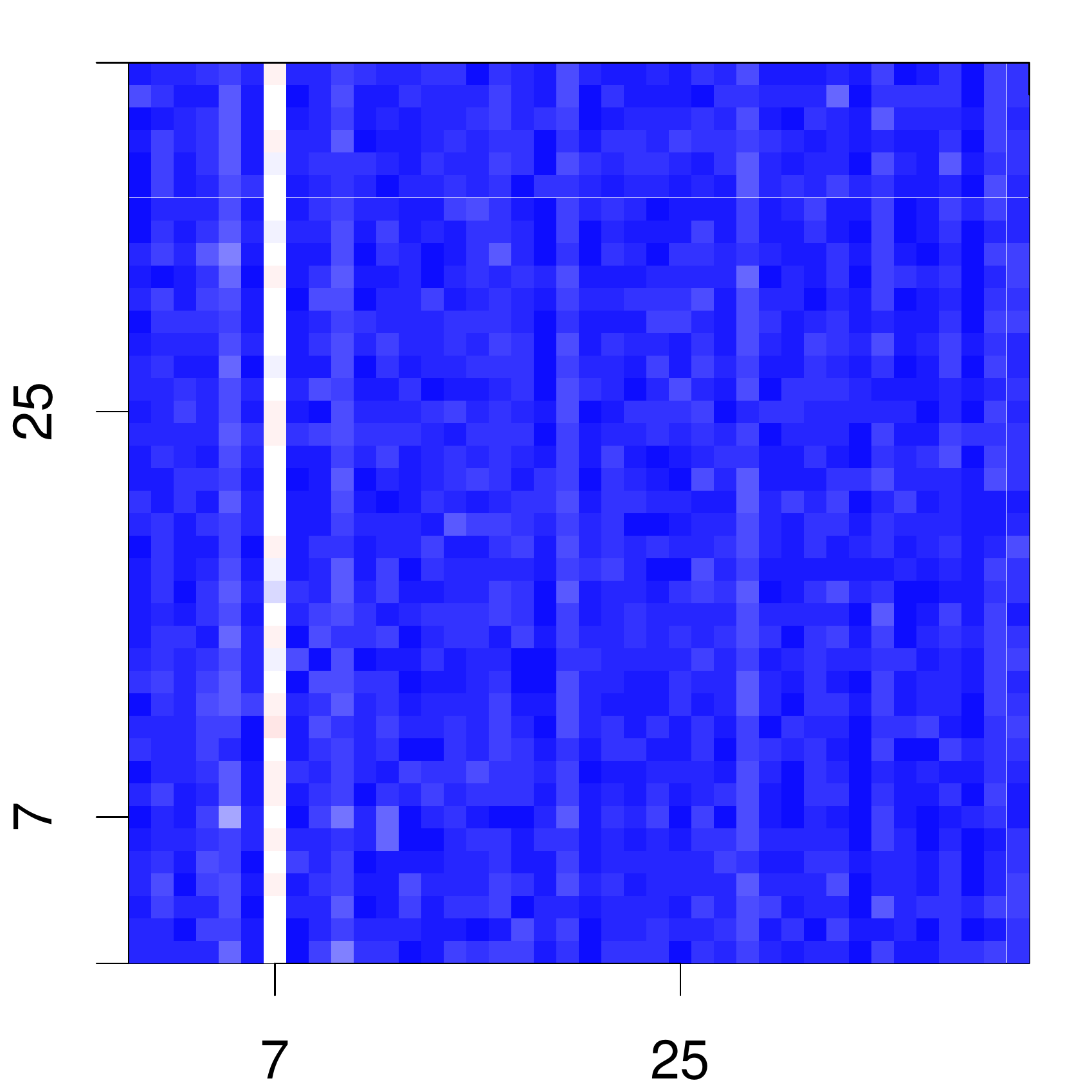} & 	\includegraphics[width= 5 cm]{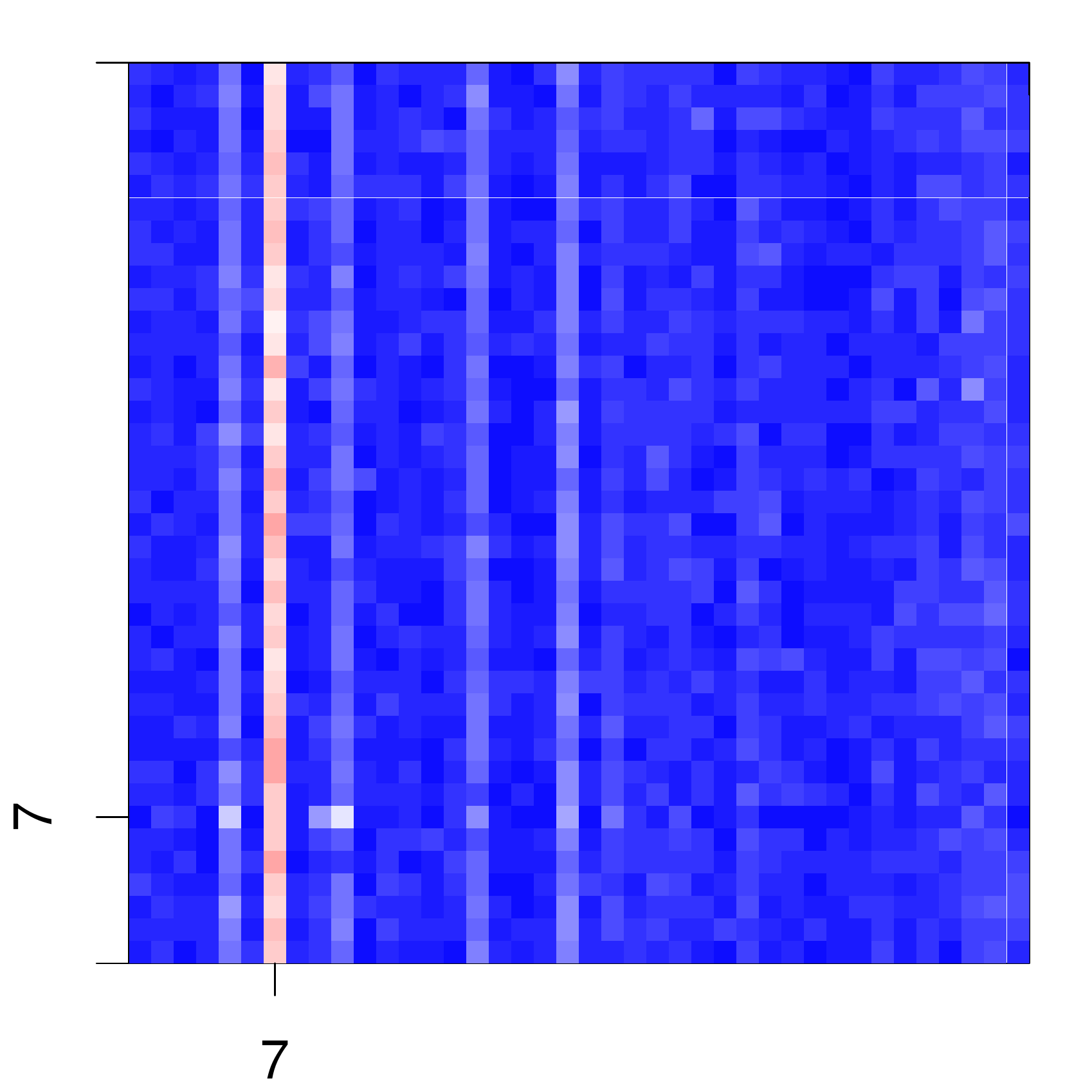}& 	\includegraphics[width= 5 cm]{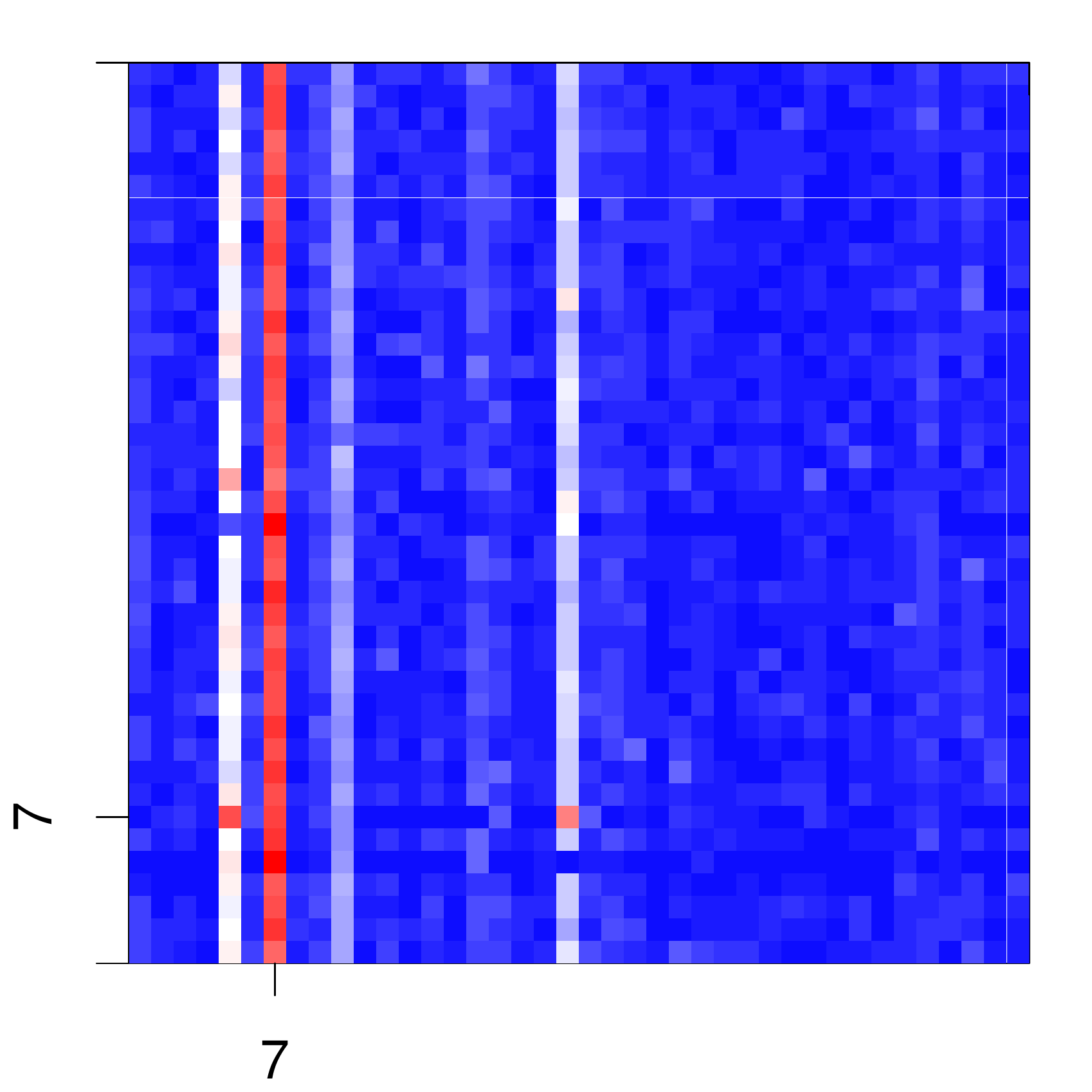} \\
 		$t$ = 205700 & $t$ = 206000 & $t$ = 206300
 	\end{tabular}
 	\caption{Crisis-dominance pattern: the leader manages to get through the competition with a challenger. At $t$ = 204 500, the opinion of the challenger ($i$ = 25) for the leader ($i$ = 7) and of the challenger for the leader are significantly lower than their respective self opinions. They enter in a negative loop of vanity that ends up at $t$ = 205700 with the challenger returning into the majority of very low reputations, and the leader having still some limited support from one other agent. The leader's reputation decreased much, but it is high enough to rise again ($t$ = 206000, 206300) and a new episode of strong dominance will take place.  }
 	\label{fig:P4Challenger}
 \end{figure}

The crisis pattern appears as an extreme result of the tendency for a positive self bias that is amplified by increasing the vanity.
But in the considered mixed pattern, the state of crisis is not stable, it generates episodes of dominance that we now try to explain qualitatively: 
\begin{itemize}
	\item There are fluctuations of the reputations during the crisis. When an agent's reputation (and thus self-opinion) becomes higher enough than the others, it becomes less sensitive to the opinion propagation from the others. In this case, the opinions of this leading agent become more strongly driven by the vanity than by the opinion propagation. In particular, this agent can establish loops of positive reinforcements with a few other agents that are not averaged immediately by opinion propagation. This increases the self-opinion of the leader who propagates this good opinion with a strong influence. Moreover, the leader propagates her good opinion about the agents that have a good opinion of her, which reinforces these agents in the population and their propagation of their good opinion about the leader. In summary, when an agent reaches some level of reputation (with all the others having very low reputations) she tends to lead the opinions, reinforcing herself and those who propagate a good opinion of her.
	\item When this leader's reputation becomes close to 1, it cannot grow anymore. Then a struggle with her potential rivals takes place. Indeed, in general, her allies, with the fluctuations, end up by having a lower opinion of the leader than the leader's self-opinion. In this case, the leader begins to decrease her opinion about this challenger. The leader is followed by the rest of the population with limited damage for her reputation if the challenger is not too strong. On the contrary, when the challenger has reached a reputation which is as strong as the one of the leader, then the fight is fatal for both of them and leads to come back to the generalised distrust.
	\item The rival that reaches the level of the leader and causes her loss can appear more or less rapidly. In some cases, the leader and her rival grow together right at the beginning and reach the top reputation almost at the same time. In this case, the dominance episode is short (first peak in the graph of dominance around iteration 7000). Moreover, the rapid growth of the two leaders self-opinions leads them to decrease their opinions about all the rest of the population, by vanity. Figure \ref{fig:P4DoubleLeaderRiseFall} illustrates the rise and fall of a couple of leaders.
	\item On the contrary, when the leader manages to reach the top
          reputation with a significant difference with her followers, she may
          keep her dominance for more than 10 thousand time steps, with
          several attempts from challengers that she manages to stop soon
          enough. In this dominance episodes, there are periods where a
          progressive hierarchy is established with several agents having
          intermediate reputations. Figure \ref{fig:P4SingleLeaderRising} illustrates the rise of a single leader and figure \ref{fig:P4Challenger} her successful struggle with a challenger.
\end{itemize}

\subsection{Conclusions of the theoretical analysis}

Globally, our analysis points out the main drivers of the model behaviour when vanity or opinion propagation dominates. In both cases, the relation between the self-opinion and the reputation is important. It is responsible for the general tendency of the model to yield negative opinions. We propose some quantitative approximations correctly predicting some measures of the model behaviour in the equality and elite patterns. For the patterns where opinion propagation dominates, our explanations are more qualitative. Nevertheless, this analytical work is far from complete and many questions remain, like for instance explaining the stability of the dominance pattern, the quantification of the dominance periods in the crisis-dominance pattern, etc...

\section{Discussion - conclusion}

In the last decades, the researches in the field of complexity showed that the collective behaviour of a population of interacting agents can be very difficult to predict because of cascade effects or global structure emergence from local interactions. Computer simulations give the possibility to simulate and study these effects. We argue that this approach can be a source of new arguments in philosophical debates making hypotheses about individual interactions and we believe that this paper illustrates this point. We translated Hobbes hypothesis about the role of vanity in human relations into a computer model and explored by simulation the consequences of this hypothesis on the emergence of patterns. Our work can thus be seen as using computer simulation as a tool for philosophical reflection: it allows one to make thought experiments with idealised populations of interacting agents. 

We think that our results can shed a new light on Hobbes theses. Indeed, it is striking that our model leads to two situations on which Hobbes focuses particularly: the general distrust (crisis) where men are all enemies of each other (and of themselves), and the absolute dominance of one agent (the Leviathan). In this respect, the model confirms the main intuitions of Hobbes. Moreover, in the crisis-dominance pattern, these two situations alternate dynamically, as if they were the two sides of the same coin. The main point is that they take place spontaneously as an effect of the individual interactions. Moreover, we show that the same hypotheses can lead to very different patterns that were not considered by Hobbes (equality, elite).

The model can also be related with more recent general theories of social interactions, in particular the ones of Ren\'e Girard (\citep{Girard1972,Girard1982}). Indeed, with his thesis about mimetism as being the main driver of social interactions, Girard puts forward a mixture of the ingredients that are present in our model: imitation (opinion propagation) and rivalry (vanity). Girard considers that these ingredients can lead to a state of generalised crisis, where all hierarchies are abolished, and also to the absolute dominance of one agent. Moreover, he insists on the intrinsic instability of the leadership, the book of Job in the Bible being for him a prototypical illustration of such an instability. Our model is in accordance with these general views: the same mechanisms can lead to the general crisis and to the dominance, and the dominance is always unstable. However, Girard assumes also that the same mechanisms lead to scape goat structures where an agent is universally dispised. We have not observed this last pattern with our model. If this is confirmed that the model cannot generate such a pattern, it would mean that our model does not implement correctly Girard's hypotheses about individual interactions, or that Girard is wrong about such interactions leading to scape goat patterns.

Another philosophical debate in which our model could contribute is about how to define a human subject. Indeed, in the model, the agents have a perception of themselves which is always mediated by the others. Their self-opinion derives from the opinion that the others express about them. This is an important option which differs strongly from the Cartesian subject who is able to found herself. The subject who builds herself in the interaction with the others corresponds better to the more recent philosophical traditions which give a particular importance to language (after the "linguistic turn"). 

These remarks show that this model could help contribute to some philosophical debates. The question is then: can it be related to more concrete social observations? 
 
One could answer positively to this question, arguing that the patterns emerging from the model show major common features with familiar real social situations. For instance:
\begin{itemize}
	\item When vanity dominates strongly, the agents are in priority careful to reward a good image of themselves or to punish a bad one, and otherwise they are not influenced much by other opinions, then the model predicts that agents do not establish a hierarchy and establish strong links with a small set of friends, all the others being enemies. In some cases, the structure of this friendship network shows the properties of small worlds. This equality pattern could be related to the rejection of hierarchies and the tendency to priviledge privatized lives in modern societies, turning exclusively to the small circle of family and close friends. It can be argued that the corresponding dynamics of the model (strong vanity, low opinion propagation) are coherent with the tendencies of the modern individual. The fact that the network has a small world property for some parameter values is an interesting emerging feature, coming from the dynamics of interactions where people talk about the people they know and value their friends. Some more detailed properties of the model behaviour related to the equality pattern can also be interesting to interpret in a sociological perspective: 
\begin{itemize}
\item The number of friends is related to the self-opinion of the agent. If this self-opinion is too high, then there is a high probability of being offended by a friend and entering a loop of negative vanity retaliations, ending up by loosing this friend. 
	\item The number of friends tends to decrease when the value of parameter $\sigma$ increases. This parameter can be interpreted as ruling the importance given to differences of values between individuals. In a society where the differences count much (small $\sigma$) the number of friends tends to be low because the individuals do not change their self-opinion when criticised by agents they value lower than themselves. As a consequence, they get easily a high opinion of themselves which increases their probability to loose friends. On the contrary, when $\sigma$ is high, the difference of values counts less and the agents tend decrease their self-opinion with the critical opinions of the agents they value lower than themselves. This lower self-opinion is more compatible with a higher number of friends. 
	\item Finally, we observe that in a larger population, the agents tend to have a larger number of friends, because these friends compensate the hostility of a larger number of agents.  
\end{itemize}
 \item When opinion proagation is more important and vanity dominates less strongly, there is a transition towards the elite pattern, where two types of agents appear: the elite and the second category agents. The process leading to this differentiation can also be interpreted in socio-psychological terms: 

\begin{itemize}
	\item The second category agents are those who have lost all their friends, and had their self-opinion dramatically decreased consequently. This situation can be related to depression due to isolation.
	\item The elite agents have only one friend and they hate the other elite agents as well as the second category agents. Being in the elite is a constant competition where each has a low number of friends and many foes.
	\item The second category agents have a moderately good opinion of the elite agents, and a strongly negative opinion of the other second category agents. The elite agents can count on some support from the second category agents.
	\item When the elite agents are many, they tend to be less valued by the second category agents and more stable than when they are a low number. 
\end{itemize}
 
	\item The second family of patterns takes place when the opinion propagation dominates. This situation can be interpreted as a larger influence of the population opinion on each individual, corresponding to a traditional society where the group dominates the individual. In this case, the first striking feature is that the population has almost always a consensus on its opinion about each individual. Therefore, there is a very small difference between the self-opinions and the reputation of the individuals (whereas this difference is very strong in the vanity dominating patterns). Hence, this is curiously when the agents care less about their image (low vanity) that the reputation becomes meaningful and shared by the whole population.
This family of patterns	suggests the following remarks related to psycho-sociological interpretations:
\begin{itemize}
	\item The hierarchy pattern presents a pyramidal figure of the group, with a number of agents increasing when the reputation is decreasing. It could be related to traditional, hierarchical societies (as opposed to modern individualist societies), and it emerges from hypotheses corresponding to a more traditional view of the influence of the group.
	\item A positive bias for the agent self-opinion is responsible for a global tendency to adopt negative opinions about the others, when some vanity is introduced. This positive self bias is generally observed in social psychology \citep{Hoorens1993}. The studies of links between self-esteem, leadership and groups (see e.g. \citet{Hogg20031}) could be related to the mechanisms that we analysed in our model.
	\item In the crisis-dominance pattern, we observe that the leadership is given to agents who tend to have a more positive view of the others than average and who are able to resist to the general opinion. This corresponds to the charismatic leader as defined in \citep{vanKnippenberg2004825}. In our model, the dominant agents are able to make the others change strongly their opinions, which also corresponds to a recognised feature of leadership in the literature \citep{Mary2006654,Hogg20031}.
	\item The emergence of absolute leadership in totalitarian regimes has been interpreted as an attempt to restore a traditional order in a modern society by M. \cite{Gauchet2010}. This can be related with our results where this absolute leadership takes place when mixing the vanity (that can be related to the modern individualism) and opinion propagation (that can be related to the traditional power of the society on the individual).
\end{itemize}	
\end{itemize}

However, a closer analysis would lead to remain careful about some of these interpretations: 
\begin{itemize}
	\item Experiments show that a low self-esteem is very rarely expressed by subjects, whereas the majority of agents have a low self-opinion in several patterns of the model (hierarchy, dominance and crisis). A possible interpretation is that the usual subjects submitted to experiments are more likely to be closer to the equality pattern where everybody has a high self esteem. Moreover, as underlined previously, our model yields the usually observed positive self bias, compared with the other's opinions, hence this failure to match psycho-sociology observations about self-esteem should probably be discussed further.
	\item A major problem is that interpretations considering large societies are certainly too demanding  to the model. In its present version, it is indeed limited to small populations, because it is supposed that everybody can meet with everybody and know everybody. Therefore, it is closer to group dynamics. 
\end{itemize}

In order to extend the study to larger sets of agents (thousands to millions for instance), it will be necessary to make new assumptions: to limit the number of agents that an agent can have in mind and discuss about, to limit the agents she can discuss with through different assumptions of a priori networks of interactions (possibly evolving with the opinions). Moreover, in the present setting, we considered identical agents at the beginning. It could be interesting to introduce some diversity, for instance in the propagation coefficient. However, we think that there is still a lot of work to complete the experimental and theoretical study of this first version of the model, particularly for the patterns of dominance, hierarchy and crisis where our analysis remains qualitative.

\section{Acknowledgment} 
The work of T. Carletti has been partially supported by the FNRS
grant \lq\lq Mission Scientifique 2010-2011\rq\rq  allowing him to visit the
Irstea - LISC during the fall 2010.

\bibliographystyle{plain}
\bibliography{GO_refs}

\begin{thebibliography}{26}
\providecommand{\natexlab}[1]{#1}
\providecommand{\url}[1]{\texttt{#1}}
\expandafter\ifx\csname urlstyle\endcsname\relax
  \providecommand{\doi}[1]{doi: #1}\else
  \providecommand{\doi}{doi: \begingroup \urlstyle{rm}\Url}\fi

\bibitem[Bagnoli et~al.(2007)Bagnoli, Carletti, Fanelli, Guarino, and
  Guazzini]{PhysRevE.76.066105}
Franco Bagnoli, Timoteo Carletti, Duccio Fanelli, Alessio Guarino, and Andrea
  Guazzini.
\newblock Dynamical affinity in opinion dynamics modeling.
\newblock \emph{Phys. Rev. E}, 76:\penalty0 066105, Dec 2007.
\newblock \doi{10.1103/PhysRevE.76.066105}.
\newblock URL \url{http://link.aps.org/doi/10.1103/PhysRevE.76.066105}.

\bibitem[Buckley et~al.(2004)Buckley, Winkel, and Leary]{Buckley200414}
Katherine~E. Buckley, Rachel~E. Winkel, and Mark~R. Leary.
\newblock Reactions to acceptance and rejection: Effects of level and sequence
  of relational evaluation.
\newblock \emph{Journal of Experimental Social Psychology}, 40\penalty0
  (1):\penalty0 14 -- 28, 2004.
\newblock ISSN 0022-1031.
\newblock \doi{10.1016/S0022-1031(03)00064-7}.
\newblock URL
  \url{http://www.sciencedirect.com/science/article/pii/S0022103103000647}.

\bibitem[Carletti et~al.(2011)Carletti, Fanelli, and Simone]{ACS.14.1}
Timoteo Carletti, Duccio Fanelli, and Righi Simone.
\newblock Emerging structures in social networks guided by opinions' exchanges.
\newblock \emph{ACS}, 14:\penalty0 28, 2011.
\newblock \doi{10.1142/S021952591100286X}.
\newblock URL
  \url{http://www.worldscinet.com/acs/14/1401/S021952591100286X.html}.

\bibitem[Castellano et~al.(2009)Castellano, Fortunato, and
  Loreto]{castellano2009statistical}
C.~Castellano, S.~Fortunato, and V.~Loreto.
\newblock {Statistical physics of social dynamics}.
\newblock \emph{Reviews of Modern Physics}, 81\penalty0 (2):\penalty0 591--646,
  2009.

\bibitem[Deffuant(2006)]{Deffuant_2006}
G.~Deffuant.
\newblock Comparing extremism propagation patterns in continuous opinion
  models.
\newblock \emph{Journal of Artificial Societies and Social Simulation},
  9\penalty0 (3):\penalty0 1--24, 2006.
\newblock URL \url{http://jasss.soc.surrey.ac.uk/9/3/8.html}.

\bibitem[Deffuant et~al.(2000)Deffuant, Neau, Amblard, and
  Weisbuch]{deffuant2000mixing}
G.~Deffuant, D.~Neau, F.~Amblard, and G.~Weisbuch.
\newblock {Mixing beliefs among interacting agents}.
\newblock \emph{Advances in Complex Systems}, 3\penalty0 (4):\penalty0 87--98,
  2000.

\bibitem[Fein(1997)]{FeinSpencer1997}
Steven~J. Fein, Steven;Spencer.
\newblock Prejudice as self-image maintenance: Affirming the self through
  derogating others.
\newblock \emph{Journal of Personality and Social Psychology}, 73\penalty0
  (1):\penalty0 31--44, 1997.
\newblock \doi{10.1037/0022-3514.73.1.31}.

\bibitem[Fortunato(2004)]{fortunato2004universality}
S.~Fortunato.
\newblock {Universality of the Threshold for Complete Consensus for the Opinion
  Dynamics of Deffuant et al.}
\newblock \emph{International Journal of Modern Physics C}, 15:\penalty0
  1301--1307, 2004.

\bibitem[{Galam}(2008)]{GalamReview2008}
S.~{Galam}.
\newblock {Sociophysics:. a Review of Galam Models}.
\newblock \emph{International Journal of Modern Physics C}, 19:\penalty0
  409--440, 2008.
\newblock \doi{10.1142/S0129183108012297}.

\bibitem[Gargiulo and Huet(2010)]{Gargiulo2010EPL}
F.~Gargiulo and S.~Huet.
\newblock {Opinion dynamics in a group-based society }.
\newblock \emph{EPL}, 91\penalty0 (58004):\penalty0 2--6, 2010.

\bibitem[Gauchet(2010)]{Gauchet2010}
Marcel Gauchet.
\newblock \emph{A l'épreuve des totalitarismes, 1914-1974}.
\newblock Gallimard, 2010.

\bibitem[Girard(1972)]{Girard1972}
René Girard.
\newblock \emph{La violence et le sacré}.
\newblock Grasset, 1972.

\bibitem[Girard(1982)]{Girard1982}
René Girard.
\newblock \emph{Le bouc émissaire}.
\newblock Grasset, 1982.

\bibitem[Hogg and van Knippenberg(2003)]{Hogg20031}
Michael~A Hogg and Daan van Knippenberg.
\newblock Social identity and leadership processes in groups.
\newblock volume~35 of \emph{Advances in Experimental Social Psychology}, pages
  1 -- 52. Academic Press, 2003.
\newblock \doi{10.1016/S0065-2601(03)01001-3}.
\newblock URL
  \url{http://www.sciencedirect.com/science/article/pii/S0065260103010013}.

\bibitem[Hoorens(1993)]{Hoorens1993}
Vera Hoorens.
\newblock Self-enhancement and superiority biases in social comparison.
\newblock \emph{European Review of Social Psychology}, 4\penalty0 (1):\penalty0
  113--139, 1993.
\newblock \doi{10.1080/14792779343000040}.
\newblock URL
  \url{http://www.tandfonline.com/doi/abs/10.1080/14792779343000040}.

\bibitem[Huet and Deffuant(2010)]{PUB00029104}
S.~Huet and G.~Deffuant.
\newblock Openness leads to opinion stability and narrowness to volatility.
\newblock \emph{Advances in Complex Systems}, 13\penalty0 (3):\penalty0
  405--423, 2010.
\newblock \doi{http://dx.doi.org/10.1142/10.1142/S0219525910002633}.
\newblock URL \url{http://cemadoc.cemagref.fr/cemoa/PUB00029104}.

\bibitem[Huet et~al.(2008)Huet, Deffuant, and Jager]{huet2008rejection}
S.~Huet, G.~Deffuant, and W.~Jager.
\newblock {A Rejection Mechanism In 2d Bounded Confidence Provides More
  Conformity}.
\newblock \emph{Advances in Complex Systems}, 11\penalty0 (4):\penalty0
  529--549, 2008.

\bibitem[Leary et~al.(2006)Leary, Twenge, and Quinlivan]{Leary01052006}
Mark~R. Leary, Jean~M. Twenge, and Erin Quinlivan.
\newblock Interpersonal rejection as a determinant of anger and aggression.
\newblock \emph{Personality and Social Psychology Review}, 10\penalty0
  (2):\penalty0 111--132, 2006.
\newblock \doi{10.1207/s15327957pspr1002_2}.
\newblock URL \url{http://psr.sagepub.com/content/10/2/111.abstract}.

\bibitem[Lorenz(2007)]{lorenz2007continuous}
J.~Lorenz.
\newblock {Continuous Opinion Dynamics Under Bounded Confidence:. a Survey}.
\newblock \emph{International Journal of Modern Physics C}, 18:\penalty0
  1819--1838, 2007.

\bibitem[Mary and Uhl-Bien(2006)]{Mary2006654}
Mary and Uhl-Bien.
\newblock Relational leadership theory: Exploring the social processes of
  leadership and organizing.
\newblock \emph{The Leadership Quarterly}, 17\penalty0 (6):\penalty0 654 --
  676, 2006.
\newblock ISSN 1048-9843.
\newblock \doi{10.1016/j.leaqua.2006.10.007}.
\newblock URL
  \url{http://www.sciencedirect.com/science/article/pii/S1048984306001135}.
\newblock <ce:title>The Leadership Quarterly Yearly Review of
  Leadership</ce:title>.

\bibitem[Srivastava and Beer(2005)]{Srivastava2005}
Sanjay Srivastava and Jennifer~S. Beer.
\newblock How self-evaluations relate to being liked by others: Integrating
  sociometer and attachment perspectives.
\newblock \emph{Journal of Personality and Social Psychology}, 89\penalty0
  (6):\penalty0 966 -- 977, 2005.
\newblock ISSN 2005-16185-010.
\newblock \doi{10.1037/0022-3514.89.6.966}.
\newblock URL
  \url{http://www.sciencedirect.com/science/article/pii/S1048984304000864}.

\bibitem[Stephan and Maiano(2007)]{StephanMaiano2007}
Yannick Stephan and Christophe Maiano.
\newblock On the social nature of global self-esteem: A replication study.
\newblock \emph{The Journal of Social Psychology}, 147\penalty0 (5):\penalty0
  573--575, 2007.
\newblock \doi{10.3200/SOCP.147.5.573-576}.
\newblock URL
  \url{http://www.tandfonline.com/doi/abs/10.3200/SOCP.147.5.573-576}.

\bibitem[Sznadj-Weron(2005)]{Sznajd-Weron2005}
K~Sznadj-Weron.
\newblock {Sznajd model and its applications}.
\newblock \emph{Acta Physica Polonica B}, 36\penalty0 (8), 2005.

\bibitem[Urbig et~al.(2008)Urbig, Lorenz, and Herzberg]{Urbig2008}
D.~Urbig, J.~Lorenz, and H.~Herzberg.
\newblock Opinion dynamics: The effect of number of peers met at once.
\newblock \emph{Journal of Artificial Societies and Social Simulation},
  11(2)\penalty0 (3):\penalty0 4, 2008.
\newblock URL \url{http://jasss.soc.surrey.ac.uk/9/3/8.html}.

\bibitem[van Knippenberg et~al.(2004)van Knippenberg, van Knippenberg, Cremer,
  and Hogg]{vanKnippenberg2004825}
Daan van Knippenberg, Barbara van Knippenberg, David~De Cremer, and Michael~A.
  Hogg.
\newblock Leadership, self, and identity: A review and research agenda.
\newblock \emph{The Leadership Quarterly}, 15\penalty0 (6):\penalty0 825 --
  856, 2004.
\newblock ISSN 1048-9843.
\newblock \doi{10.1016/j.leaqua.2004.09.002}.
\newblock URL
  \url{http://www.sciencedirect.com/science/article/pii/S1048984304000864}.

\bibitem[Wood and Forest(2011)]{WoodForest2011}
J.~V. Wood and A.~L. Forest.
\newblock Seeking pleasure and avoiding pain in interpersonal relationships.
\newblock In M.~D. Alicke and C.~Sedikides, editors, \emph{Handbook of
  Self-Enhancement and Self-Protection}, pages 258--78. New York, London: The
  Guilford Press, 2011.

\end{thebibliography}

\end{document}